\documentclass[journal]{IEEEtran}
\usepackage[pdftex]{graphicx}
\usepackage{enumitem}
\usepackage{algpseudocode}                   
\usepackage{algorithmicx}                   
\usepackage{caption}
\usepackage{amsbsy}
\usepackage{datetime2}

\newcommand{\CatphanTM}{{Catphan\textsuperscript{\textregistered}}}
\newcommand{\CTP}{{CTP404\space}}
\newcommand{\CatphanTMlong}{{\CatphanTM}\CTP}

\begin{document}
\title{An Improved Method of Total Variation Superiorization Applied to Reconstruction in Proton Computed Tomography}
\author{\IEEEauthorblockN{Blake~Schultze\IEEEauthorrefmark{1},
Yair~Censor\IEEEauthorrefmark{2},
Paniz~Karbasi\IEEEauthorrefmark{1},
Keith~E.~Schubert\IEEEauthorrefmark{1}~\IEEEmembership{Senior Member,~IEEE}, and
Reinhard~W.~Schulte\IEEEauthorrefmark{3}~\IEEEmembership{Member,~IEEE}}

\IEEEauthorblockA{\IEEEauthorrefmark{1}Department of Electrical and Computer Engineering, Baylor University, Waco, TX 76798, USA, email: Blake\_Schultze@baylor.edu, Paniz\_Karbasi@baylor.edu, Keith\_Schubert@baylor.edu}

\IEEEauthorblockA{\IEEEauthorrefmark{2}Department of Mathematics, University of Haifa, Haifa 3498838, Israel, email: yair@math.haifa.ac.il}

\IEEEauthorblockA{\IEEEauthorrefmark{3}Department of Basic Sciences, Division of Biomedical Engineering Sciences, Loma Linda University, Loma Linda, CA 92350 USA, email: rschulte@llu.edu}%
\\\vspace{0.5cm}\flushleft\textbf{NOTE: This work has been submitted to the IEEE for possible publication. Copyright may be transferred without notice, after which this version may no longer be accessible.}\vspace{-0.35cm}}
\IEEEtitleabstractindextext{%
\begin{abstract}
Previous work has shown that total variation superiorization (TVS) improves reconstructed image quality in proton computed tomography (pCT).  The structure of the TVS algorithm has evolved since then and this work investigated if this new algorithmic structure provides additional benefits to pCT image quality.  Structural and parametric changes introduced to the original TVS algorithm included: (1) inclusion or exclusion of TV reduction requirement, (2) a variable number, $N$, of TV perturbation steps per feasibility-seeking iteration, and (3) introduction of a perturbation kernel $0<\alpha<1$.  The structural change of excluding the TV reduction requirement check tended to have a beneficial effect for $3\le N\le 6$ and allows full parallelization of the TVS algorithm.  Repeated perturbations per feasibility-seeking iterations reduced total variation (TV) and material dependent standard deviations for $3\le N\le 6$.  The perturbation kernel $\alpha$, equivalent to $\alpha=0.5$ in the original TVS algorithm, reduced TV and standard deviations as $\alpha$ was increased beyond $\alpha=0.5$, but negatively impacted reconstructed relative stopping power (RSP) values for $\alpha>0.75$.  The reductions in TV and standard deviations allowed feasibility-seeking with a larger relaxation parameter $\lambda$ than previously used, without the corresponding increases in standard deviations experienced with the original TVS algorithm.  This work demonstrates that the modifications related to the evolution of the original TVS algorithm provide benefits in terms of both pCT image quality and computational efficiency for appropriately chosen parameter values.
\end{abstract}
}
\maketitle
\IEEEdisplaynontitleabstractindextext
\IEEEpeerreviewmaketitle
\section{Introduction}
\IEEEPARstart{P}{roton} %
computed tomography (pCT) is a relatively new imaging modality that has been developed from early beginnings~\cite{Cormack76,Hanson78,Hanson79,Hanson81} towards a recent preclinical realization of a pCT scanner~\cite{JBCGK16,BJSS16,BCSBB17}; a comprehensive review of pCT development can be found in~\cite{Johnson18}. The main motivation of pCT has been to improve the accuracy of proton therapy dose planning due to more accurate maps of relative stopping power (RSP) with respect to water, which determines how protons lose energy in human tissues in reference to water as a medium. The same method can also be used to image the patient immediately before treatment to verify the accuracy of the treatment plan about to be delivered.  Proton therapy, like therapy with other heavy charged particles, e.g., carbon ions, is very susceptible to changes in tissue RSP, and small differences of a few percent, both random and systematic, can lead to range errors exceeding the desired limit of 1-2 mm~\cite{Paganetti12}. Thus, the planner of proton and ion therapy must increase margins around the target, which leads to unwanted exposure of normal tissues to high dose.

The faithful reconstruction of proton RSP maps, in terms of accuracy and reproducibility, is an important part of the successful clinical implementation of pCT. The approach that has been selected as the most promising in recent years, although technologically demanding, is to track individual protons through the patient and to predict their most likely path (MLP)~\cite{SPTS08,FDDBS15} in addition to measuring the energy loss of each proton and converting it to water-equivalent pathlength (WEPL). This has led to pCT reconstruction algorithms that are based on solving large and sparse linear equation systems, where each equation has the linear combination of intersection lengths of tracked protons through individual object voxels and the unknown RSP of those voxels on the left-hand side of the equation and the measured WEPL on the right-hand side of the equation. A solution of such large systems can be found with algorithms using projections onto convex sets and solving them iteratively as shown previously~\cite{PSCBM10}. The noise content of the reconstructed images depends on many factors, such as the thickness of the object, the number of protons used in the image formation, and the details of the iterative algorithm, such as the number of iterations performed and the relaxation parameter chosen.

The superiorization method (SM) is another relatively recent development that has found its place between feasibility-seeking and constrained optimization in medical physics applications~\cite{HGDC12}. The superiorization method has also been tested as a technique to improve the image quality, in particular the noise properties, of pCT images when combined with the diagonally-relaxed orthogonal projections (DROP) algorithm~\cite{PSCR10}.  Superiorization reduces, not necessarily minimizes, the value of a target function while seeking constraints-compatibility. This is done by taking a solely feasibility-seeking algorithm, analyzing its perturbation resilience, and proactively perturbing its iterates accordingly to steer them toward a feasible point with reduced value of the target function. When the perturbation steps are computationally efficient, this enables generation of a superior result with approximately the same computation time and efficiency (computational cost) as that of the original feasibility-seeking algorithm.

The mathematical principles of the SM over general consistent \textquotedblleft problem structures\textquotedblright\ with the notion of bounded perturbation resilience were formulated in~\cite{CDH10}. The framework of the SM was extended to the inconsistent case by using the notion of strong perturbation resilience in~\cite{Censor14,CDHST14}. In~\cite{CDHST14}, the efficacy of the SM was also demonstrated by comparing it with the performance of the projected subgradient method for constrained minimization problems.

A comprehensive overview of the state of the art and current research on superiorization appears in our continuously updated bibliography Internet page, which currently contains 82 items~\cite{website:Censor}. Research works in this bibliography include a variety of reports ranging from new applications to new mathematical results for the foundation of superiorization. A special issue entitled: \textquotedblleft Superiorization: Theory and Applications\textquotedblright{} of the journal Inverse Problems appeared in~\cite{CHJ17}.

Recently published works also attest to the advantages of the superiorization methodology in x-ray CT image reconstruction. These include reconstruction of CT images from sparse-view and limited-angle polyenergetic data~\cite{HWF17}, statistical tomographic image reconstruction~\cite{HZM17}, CT with total variation and with shearlets~\cite{GH17}, and superiorization-based multi-energy CT image reconstruction~\cite{YCW17}.

In this work, we report on improvements in noise properties (total variation, standard deviation of regions of interest in different materials) and computational efficiency when applying novel modifications of superiorization to pCT reconstruction.
\section{Motivation}
Iterative projection methods seeking feasible solutions have been shown to be an effective image reconstruction technique for pCT~\cite{PC15}, but reconstructed images exhibit local RSP fluctuations that cannot be removed by the reconstruction process alone.  Inelastic electronic and nuclear events result in a statistical distribution of energy loss and, consequently, of the WEPL values calculated from measurements.  These statistical variations in WEPL manifest in the reconstructed image as correlated localized fluctuations in the reconstructed RSP values.  Although iterative reconstruction algorithms~\cite{PSCBM10} are less sensitive to these variations than reconstruction transform methods, such as filtered backprojection (FBP)~\cite{BR67,RL71}, there is a propagation and amplification of WEPL uncertainty with successive iterations.  Hence, although accuracy tends to increase with each iteration, as reconstruction nears convergence, updates of the solution from subsequent iterations are increasingly dominated by growing fluctuations.  Thus, beyond a certain number of iterations, image quality begins to degrade, placing a limit on the maximum number of useful iterations and preventing steady-state convergence.  WEPL uncertainty is inherent in the physical process and cannot be avoided, but techniques have been developed to reduce fluctuations and limit their propagation during iterative reconstruction.  Given the amplification of uncertainty in the iterative process, any reduction in local RSP variations may lead to improved convergence behavior and, therefore, increase the accuracy of reconstructed RSP values. Therefore, the current work focused on further reduction of the noise content in the overall image as well as in certain regions of defined RSP.

The two measures that were adopted to quantify the prevalence and magnitude of these RSP fluctuations are total variation (TV) and standard deviation.  For an introduction to TV for image analysis see, e.g.,~\cite{CCCNP10}.  Standard deviation is a commonly used measure of variability around the mean in statistics. In image analysis, it is often employed to characterize the amount of fluctuation present in a region of interest that is known to present a homogeneous material.  These measures provide a basis for comparing the effectiveness of techniques developed to address the noise problem in iterative image reconstruction.  Total variation superiorization (TVS) is a technique for reducing image noise content without reducing the sharpness of edges between boundaries of materials. TV superiorization consists of repeated steepest descent steps of TV interlaced between iterations of a feasibility-seeking algorithm.

In pCT reconstruction, feasibility-seeking tends to accentuate RSP variations present due to WEPL uncertainty.  Whereas this sharpens edges between different material regions, it also results in an amplification of RSP fluctuations during iterative image reconstruction.  Performing TV reduction steps between consecutive feasibility-seeking iterations slows the growth of RSP variations. This permits more feasibility-seeking iterations before fluctuations grow to dominate updates of the solution.  Hence, although the reduction in TV is itself an important aspect of TVS, another important aspect is the increased number of useful iterations made possible by the reduction in the amplification of RSP fluctuations.
\section{Methods}
\subsection{TVS Algorithms}
The efficacy of TVS for image reconstruction in pCT has been demonstrated in previous work~\cite{PSCR10}.  In recent years, the algorithmic structure of the superiorization method has undergone some evolution in ways that offer several potential benefits in pCT.  The details of this evolution can be found in the Appendix of \cite{Censor17}, titled ``The algorithmic evolution of superiorization''.  In addition, there were certain aspects of the original TVS algorithm, here referred to as OTVS, that had been proposed but were not previously investigated in its application to pCT.

With the new version of the TVS algorithm, here referred to as NTVS, we investigated both the structural changes and aspects previously not investigated of the OTVS algorithm.  The notation and other algorithmic details of the NTVS algorithm can be found in Appendices~\ref{app:defterms} and ~\ref{app:ntvs}. The definition of the OTVS algorithm investigated here and in previous work is provided in this notation in Appendix~\ref{app:otvs}.
\subsection{NTVS Algorithm}
The NTVS algorithm investigated in this work combines properties that were scattered among previous works on TVS in x-ray CT, see the Appendix of~\cite{Censor17}. These properties, listed next, were never combined in a single algorithm, as we do here, neither for x-ray CT nor for pCT.
\begin{enumerate}[label=(\arabic*)]
    \item Exclusion of the TV reduction verification step (step (\ref{alg:TVcheck}) of the OTVS algorithm in Appendix~\ref{app:otvs}).
    \item Usage of powers of the perturbation kernel $\alpha$ to control the step-sizes $\beta_k$ in the TV perturbation steps.
    \item Incorporation of the user-chosen integer $N$ (step (\ref{alg:N}) of the NTVS algorithm in Appendix~\ref{app:ntvs}) that specifies the number of TV perturbation steps between consecutive feasibility-seeking iterations.
    \item Incorporation of a new formula for calculating the power $\ell_k$, $\ell_k=\textrm{rand}(k, \ell_{k-1})$, used to calculate the step-size $\beta_k=\alpha^{\ell_k}$ at iteration $k$ of feasibility-seeking (step (\ref{alg:lk}) of the NTVS algorithm in Appendix~\ref{app:ntvs}).
\end{enumerate}
The step verifying the reduction of TV (step (\ref{alg:TVcheck}) of the OTVS algorithm in Appendix~\ref{app:otvs}) is not time consuming, but such decision-controlled branches present their own challenges with respect to computational efficiency.  Although there are technically a few computations with data dependencies (e.g., norm calculations), in each case, these can either be rearranged/reformulated or simply repeated separately to generate data independent calculations, making parallel computation of the algorithm possible. Hence, if the branching introduced by the TV reduction verification can be removed without compromising image quality, the NTVS algorithm can be incorporated into the existing parallelization scheme, providing up to a 30\% reduction in sequential operation count (computation time) and eliminating the repeated perturbations until a reduced TV is achieved (computation time and efficiency).  This change is similar to, but distinct from, the investigations performed by Penfold et al~\cite{PSCR10} in developing the OTVS algorithm for application in pCT reconstruction in which he found that the computationally expensive feasibility proximity check step of the classical TVS algorithm~\cite{HD08,DHC09,BDHK07} could safely be removed. Inclusion or exclusion of such checks not only affects computational efficiency, but these can also have a significant impact on image quality. Hence, the formulation of the NTVS algorithm presented in this work permits exclusion of the TV reduction check by demonstrating the surprising result that its removal has a positive impact on reconstructed image quality in addition to its computational benefits. The algorithm representing NTVS with the TV reduction requirement included is defined in Appendix~\ref{app:ntvs}), but this is only provided for reference purposes and is not intended for use.

The OTVS algorithm, initializing TVS with $\beta_0=1$ and simply halving the perturbation magnitude each time through the TV perturbation loop, prevented access to one of the most influential variables of TVS: the perturbation kernel $\alpha$.  With the magnitude of the perturbations given by $\beta_{k}=\alpha^{\ell_k}$, convergence is maintained by requiring $0<\alpha<1$.  The primary purpose of $\alpha$ is to control the rate at which $\beta_k$ converges to zero.  In OTVS, $\beta_0=1$ and $\alpha=0.5$ results in a relatively modest initial perturbation and a rapidly decreasing $\beta$ such that little to no perturbation is applied after the first few feasibility-seeking iterations.  Hence, OTVS perturbations applied after subsequent feasibility-seeking iterations are unlikely to have a meaningful impact on the amplification of RSP variations. This results in an overall under-utilization of TV perturbations.  Thus, NTVS provides direct control of $\alpha$, and its performance for various values of $\alpha$ was investigated in this work.

With the ability to increase the perturbation kernel $\alpha$, larger reductions in TV can be generated; this also produces slower-decaying perturbations, which may not be desired.  Alternatively, larger reductions in TV can also be generated by applying perturbations multiple times per feasibility-seeking iteration without increasing the magnitude of individual perturbations.  Hence, NTVS introduces a variable $N$ controlling the number of repetitions of TV perturbations between feasibility-seeking iterations.

Since the exponent $\ell$ increases after each of the $N$ applied perturbations, reducing the perturbation coefficient $\beta_{k}=\alpha^{\ell}$, an increase in $N$ causes the perturbation magnitude to converge to zero earlier in reconstruction.  To preserve meaningful perturbations in later iterations, the exponent $\ell$ is adjusted between feasibility-seeking iterations by decreasing it to a random integer between its current (potentially large) value and the (potentially much smaller) iteration number $k$, i.e., $\ell_k=\textrm{rand}(k, \ell_{k-1})$.  This update was suggested and justified in \cite[page 38]{Langthaler14}, \cite[page 36]{Prommegger14} and subsequently used in~\cite{Havas16} for maximum likelihood expectation maximization (MLEM) algorithms and in the linear superiorization (LinSup) algorithm~\cite[Algorithm 4]{Censor17}.  Although the decrease of $\ell_k$ is random within a bounded range, on average, the corresponding perturbation coefficient $\beta$ experiences a sizeable increase. For $N=1$, the difference between $\ell_{k-1}$ and $k$ is only nonzero when the TV reduction requirement is included and at least one perturbation did not reduce TV; when the TV reduction requirement is excluded, $\ell_{k-1}$ and $k$ are always equal and, therefore, $\ell_k$ is never decreased. Since $\ell_k$ is incremented after each of the $N$ perturbations, the random decrease in $\ell_k$ becomes increasingly important as $N$ increases. This random decrease slows the rate at which $\beta_k$ converges towards zero while preserving the convergence property, given that the iteration number $k$, which increases sequentially, is set as the lower limit.
\subsection{Input Data Sets}
The preliminary investigations of the NTVS algorithm were performed using a simulated pCT data set to quantify the variations generated by the random increase in $\ell_k$ between feasibility-seeking iterations (step ~\ref{alg:lk} of Algorithm~\ref{alg:NTVS}). The simulated data set of the \CatphanTMlong phantom module (The Phantom Laboratory Incorporated, Salem, NY, USA) was generated using the simulation toolkit Geant4~\cite{geant4} and contained approximately $120$ million proton histories. The definitive investigations were then performed for two experimental data sets: (1) a scan obtained with an experimental pCT scanner~\cite{JBCGK16} containing approximately $250$ million proton histories of the same \CatphanTMlong phantom and (2) an experimental pCT scan of a pediatric anthropomorphic head phantom (model HN715, CIRS, Norfolk, VA, USA) containing approximately the same number of proton histories. All pCT data sets were generated with the phantom rotating on a fixed horizontal proton beam line producing a cone beam (simulated data set) or a rectangular field using a magnetically wobbled beam spot (experimental data sets). The simulated data set was generated with 90 fixed angular step intervals of 4 degrees ranging from 0 to 356 degrees, and the experimental data sets were generated from a continuous range of projection angles between 0 and 360 degrees.

\begin{figure}[h!]
  \centering
     \includegraphics[width=0.85\linewidth]{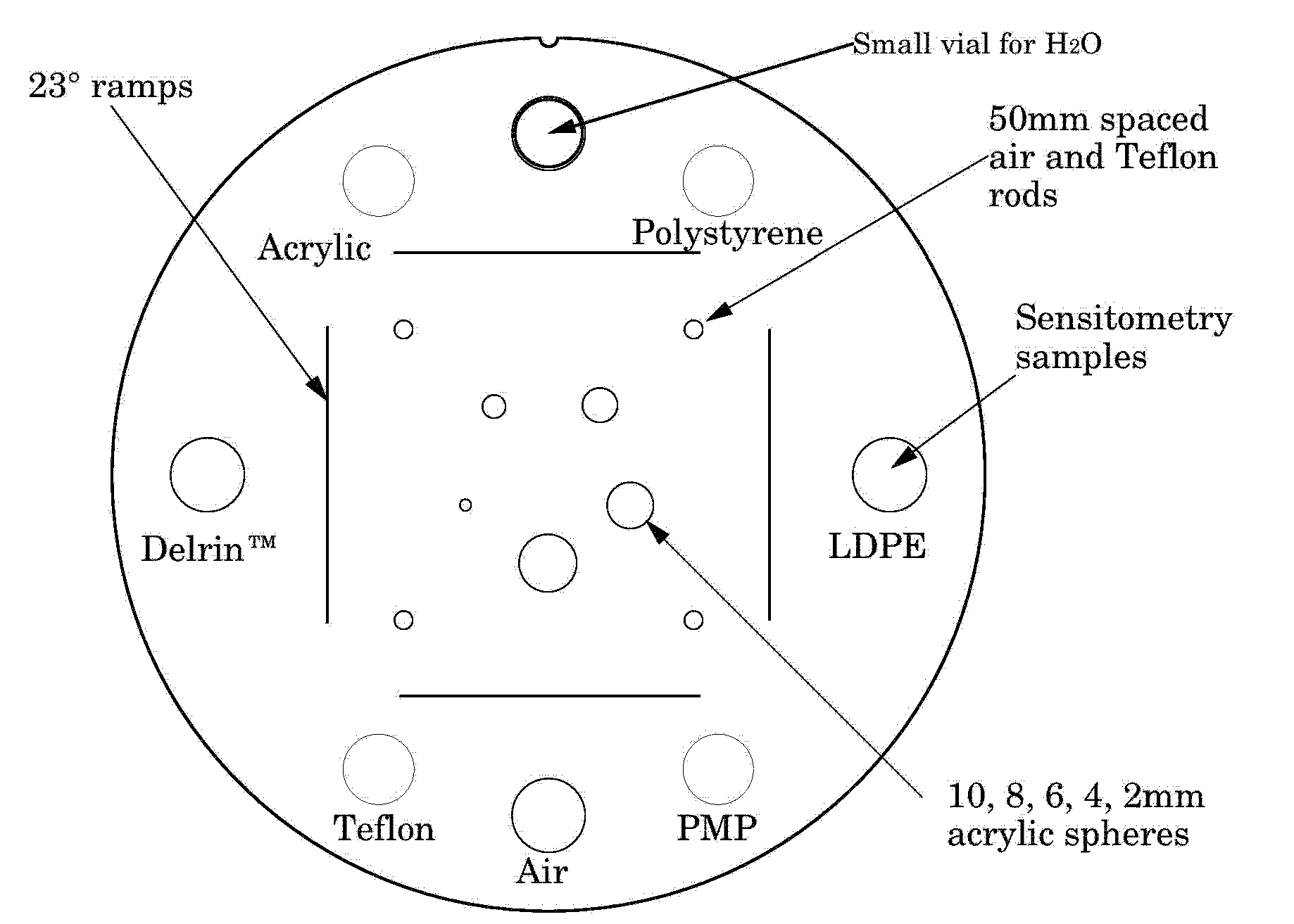}
\caption{\CatphanTMlong phantom composition and geometry of the material inserts.}
\label{fig:CTP404}
\end{figure}

The \CatphanTMlong phantom is a 15~cm diameter by 2.5~cm tall cylinder composed of an epoxy material with an $\textrm{RSP}\approx1.144$; because this value was not known at the time when the simulated data set was created, it was explicitly set to $\textrm{RSP}=1.0$ (water) in the Geant4 simulation. The theoretical and experimentally measured RSP of the materials of the phantom are listed in Table~\ref{tab:CTP404-RSP}. The phantom (see Figure~\ref{fig:CTP404}) has three geometric types of contrasting material inserts embedded with the centers of each arranged in evenly spaced circular patterns of varying diameter $d$ as follows:
\begin{enumerate}
    \item $d=30\textrm{ mm}$: five acrylic spheres of diameter $2,4,6,8,$ and $10$mm, with the center of each lying on a circular cross section midway along the phantom's axis.
    \item $d=50\sqrt{2}\textrm{ mm}$: four 3 mm diameter rods, $3\;\times$ air and $1\;\times$ Teflon, running the length of the phantom.
    \item $d=120\textrm{ mm}$: eight 12.2 mm diameter cylindrical holes, six filled with materials of known composition\footnote{$2\times$air, acrylic, polymethylpentene (PMP), low density polyethylene (LDPE), Teflon\textsuperscript{\textregistered}, Delrin\textsuperscript{\textregistered}, and polystyrene} and two left empty (air-filled), running the length of the phantom.
\end{enumerate}
Since the acrylic spheres have an $\textrm{RSP}\approx1.160$, in the case of the experimental data, these cannot be discerned from the surrounding epoxy material ($\textrm{RSP}\approx1.144$) of nearly the same RSP.
\begin{table}[h!]
\centering
\begin{tabular}{|l|c|c|c|c|c|c|c|c|}\hline
                & Air       &PMP  & LDPE    &Epoxy  \\\hline
Simulated       & 0.0013    &0.877 &0.9973  &1.024       \\
Experimental    & 0.0013    &0.883  &0.979  & 1.144     \\\hline\hline
                &Polystyrene &Acrylic &Delrin   &   Teflon\\\hline
Simulated       &1.0386     &   1.155 &  1.356   &  1.828 \\
Experimental    &1.024      &   1.160  &  1.359   &  1.79\\\hline
\end{tabular}
\caption{RSP of the material inserts for the simulated and experimental \CatphanTMlong data sets}
\label{tab:CTP404-RSP}
\end{table}
\begin{table}[h!]
\centering
\begin{tabular}{|l|c|c|c|c|c|c|c|c|}\hline
                & Soft Tissue       &Brain Tissue  & Trabecular Bone\\\hline
Experimental    & 1.037 & 1.047 & 1.108  \\\hline
\end{tabular}
\caption{RSP of the tissue/bone regions of interest analyzed in the pediatric head phantom.}
\label{tab:HN715-RSP}
\end{table}
\subsection{Data Preprocessing and Implementation Details of Image Reconstruction}
Details of the pCT data preprocessing, calibration, and image reconstruction have been presented previously~\cite{SKGPSS15,GBPGP17}.  For the purposes of this work, feasibility-seeking was performed using the DROP algorithm of \cite{CEHN08} with blocks containing 3200 (simulated data set) and 25,600 (experimental data set) proton histories. The smaller block-size was chosen for the simulated data set, which had only half of the histories as the larger experimental data sets and thus more noise. In general, smaller block sizes further accentuate noise during the reconstruction, and would potentially benefit more from NTVS. The intent of the preliminary investigation with the simulated data set was twofold: (1) to provide a larger opportunity for improvement with NTVS to better assess its benefits for more noisy data sets and (2) to quantify the magnitude of random variations in performance arising from the random increases in $\ell_k$. The random number generator used to determine the random increase in $\ell_k$ between feasibility-seeking iterations was assigned a random seed based on the Julian time at execution, yielding a different set of random increases in $\ell_k$ each time reconstruction is performed.

The experimental data set, on the other hand, was used to determine the impact of NTVS in a realistic reconstruction scenario. The block-size was still chosen from the smaller end of an acceptable range of block-sizes because, although smaller blocks are more sensitive to noise, they also provide a greater material differentiation capability and an opportunity to assess the maximum potential benefits of NTVS for realistic experimental data sets.

Image reconstruction was performed within a $20\times 20\times 5 \textrm{ cm}^3$ volume with each voxel representing a volume of $1.0\times 1.0\times 2.5 \textrm{ mm}^3$, yielding $200\times 200$ image matrix for each slice.
\begin{figure}[h!]
    \centerline{\includegraphics[width=0.7\linewidth]{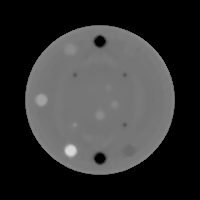}}
    \caption[justification = centering]{Representative reconstruction of the central slice of the \CTP phantom from simulated data.}
    \label{fig:sliceim}
\end{figure}
\begin{figure}[h!]
    \centerline{\includegraphics[width=0.95\linewidth]{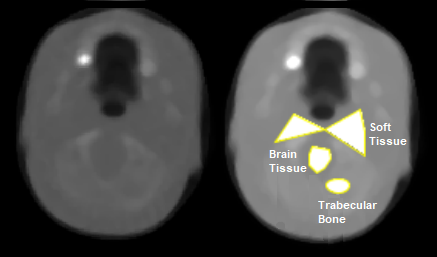}}
    \caption[justification = centering]{Representative reconstruction of the slice of the pediatric head phantom containing the analyzed regions of interest (left); the analyzed regions of interest are filled in white and labeled in the image on the right.}
    \label{fig:HN715}
\end{figure}
\subsection{Reconstruction Parameter Space}
The following describes the choices for reconstruction parameters that were systematically investigated in this work. Note that for the purpose of the investigations performed in this work, each of the following parameters of the parameter space remained constant for the duration of a particular reconstruction.
\subsubsection{Inclusion or exclusion of TV reduction requirement}
The primary structural change to the OTVS algorithm is the option to exclude the requirement that a perturbation reduces image TV, thereby eliminating the need to calculate and compare image TV before and after perturbations.  Hence, NTVS was investigated with and without this check.
\subsubsection{The number of TV perturbations per feasibility-seeking iteration}
After initial investigations with increasing $N$, results were found to degrade as $N$ increased beyond $N\approx 10$.  Therefore, in this work, the values of $N$ chosen were between 1 and 12, in increments of 1.
\subsubsection{The perturbation kernel coefficient}
Since the configuration of the OTVS algorithm effectively used $\alpha=0.5$ and the resulting perturbations did not negatively affect RSP accuracy~\cite{PSCR10}, this work only investigated with $\alpha\ge 0.5$ to determine how large it can be set without affecting RSP accuracy.  The values of $\alpha$ investigated in this work were $\alpha=0.5, 0.65, 0.75, 0.85, \textrm{ and } 0.95$.
\subsubsection{The choice of relaxation parameter in the feasibility-seeking algorithm}
In previous unpublished work $\lambda=0.0001$ yielded optimal results for a block-size containing 3200 proton histories; increasing $\lambda$ beyond this value results in increased standard deviations.  To investigate the interaction between TVS parameters and $\lambda$ and determine if NTVS is capable of reducing the increase in standard deviations, the values of $\lambda$ investigated in this work were $\lambda=0.0001, 0.00015,\textrm{ and }0.0002$.
\section{Computational Hardware and Performance Analysis}
Image reconstruction was executed on a single node of a compute cluster with input data read from a local solid state drive and the bulk of computation was performed in parallel on a single NVIDIA k40 GPU.  The parallel computational efficiency of the DROP algorithm increases as the number of histories per block increases since this permits better GPU utilization, but even with only 3200 histories per block, the total computation time from reading of input data from disk through the writing of reconstructed images to disk was, at most, about 6 minutes (for $k=12$ feasibility-seeking iterations and $N=12$ perturbation steps).

The central slices of the \CTP phantom containing the spherical inserts have the most complicated material composition and represent the greatest challenge to reconstruction.  Consequently, the data acquired for protons passing through these slices will have a greater variance in paths and WEPL values, which manifests in the corresponding slices of the reconstructed images as an increase in noise.  Hence, analysis of these slices provides a better basis for comparing the NTVS and OTVS algorithms.  Since the coplanar centers of the spherical inserts lie in the central slice, the comparative analyses performed in this work focused on this slice.  A representative reconstruction of this slice is shown in Figure~\ref{fig:sliceim}.

The image analysis program ImageJ2 1.51r~\cite{imagej2} was used to perform quantitative analyses of reconstructed image quality. The cylindrical material inserts of the \CTP phantom were analyzed by selecting a 7mm diameter circular region of interest (ROI) centered within the boundary of each insert and measuring the mean and standard deviation in reconstructed RSP of the voxels within the ROI selection. The polygon and ovular selection tools were used to measure the mean and standard deviation in reconstructed RSP within the more realistically complicated ROIs of the HN715 phantom. Although the finer structure of the HN715 phantom make it difficult to select an ROI of a single material, particularly for the brain tissue, ROIs were chosen from regions composed primarily of the material of interest and a minimal number of voxels of disparate material; the analyzed ROIs are shown shaded and labeled in Figure~\ref{fig:HN715}.

RSP error was calculated as the percentage difference between the mean measured RSP in an ROI and the RSP (a) defined for the material in the Geant4 simulation for the simulated data and (b) based on experimental material RSP investigations for the experimental data sets; the theoretical RSP used in the analyses of each material ROI of the (a) \CTP and (b) HN715 phantoms are listed in Tables~\ref{tab:CTP404-RSP} and~\ref{tab:HN715-RSP}, respectively. In accordance with~\cite{CCCNP10}, total variation was calculated as the sum of local variations over all voxels of the entire image for both the TV reduction requirement (when included) and the analysis of reconstructed images.
\section{Results}
In the following, we present results from an investigation of the multi-parameter space, including potential interactions between parameters, first for the preliminary investigation with the simulated \CTP data set and then for the definitive investigation with the experimental CTP404 and HN715 data sets. Note that each data point on the following plots represents a separate, complete reconstruction with the corresponding combination of reconstruction and superiorization parameters held fixed throughout the reconstruction.
\subsection{Simulated \CTP Data Set}\label{ssec:rSim}
\subsubsection{Number of TVS steps ($N$)}\label{sssec:rNSim}
\begin{figure}[h!]
\begin{minipage}[t]{.48\linewidth}
  \centering
    \centerline{\includegraphics[width=\linewidth]{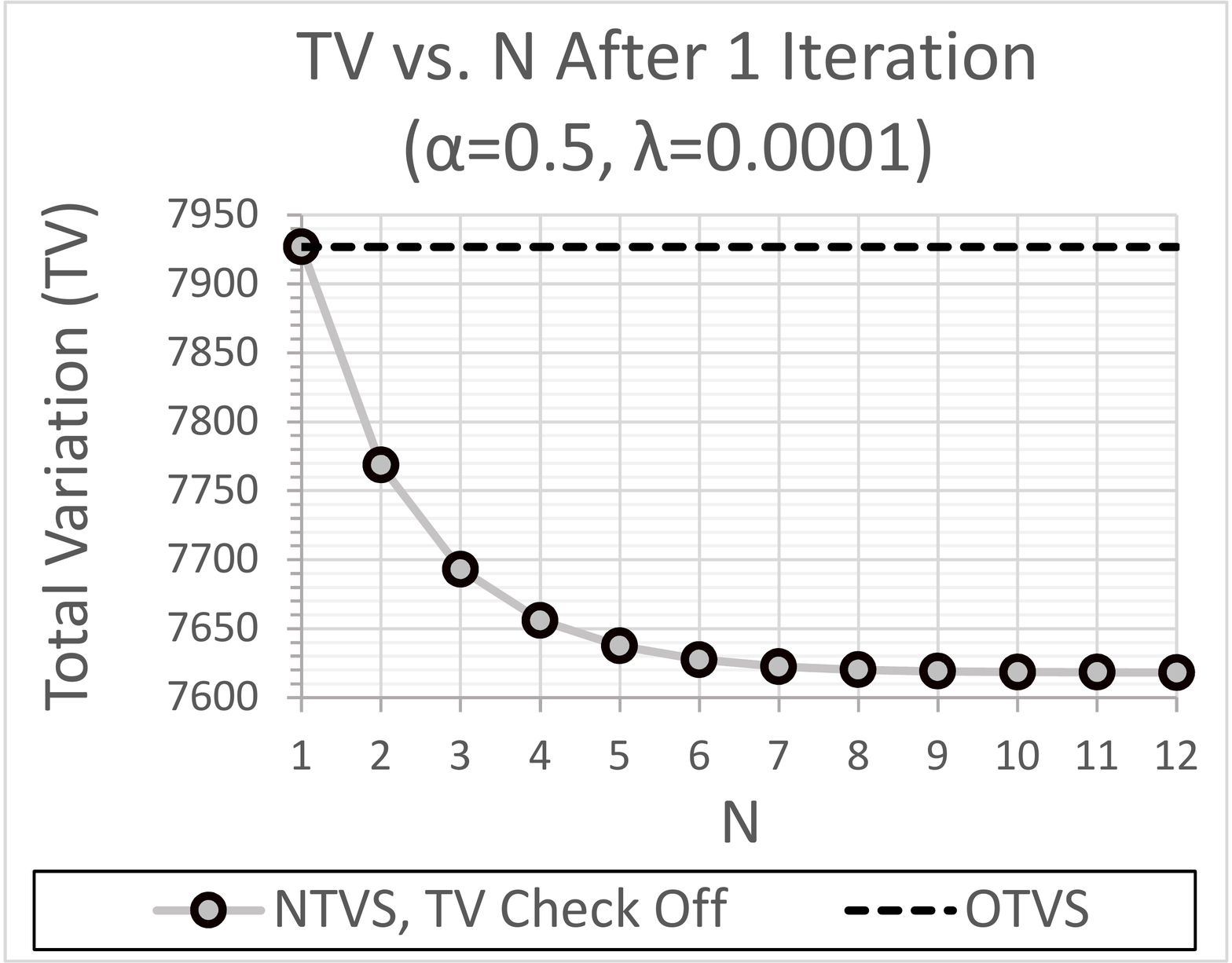}}
\end{minipage}
\hfill
\begin{minipage}[t]{.48\linewidth}
  \centering
     \centerline{\includegraphics[width=\linewidth]{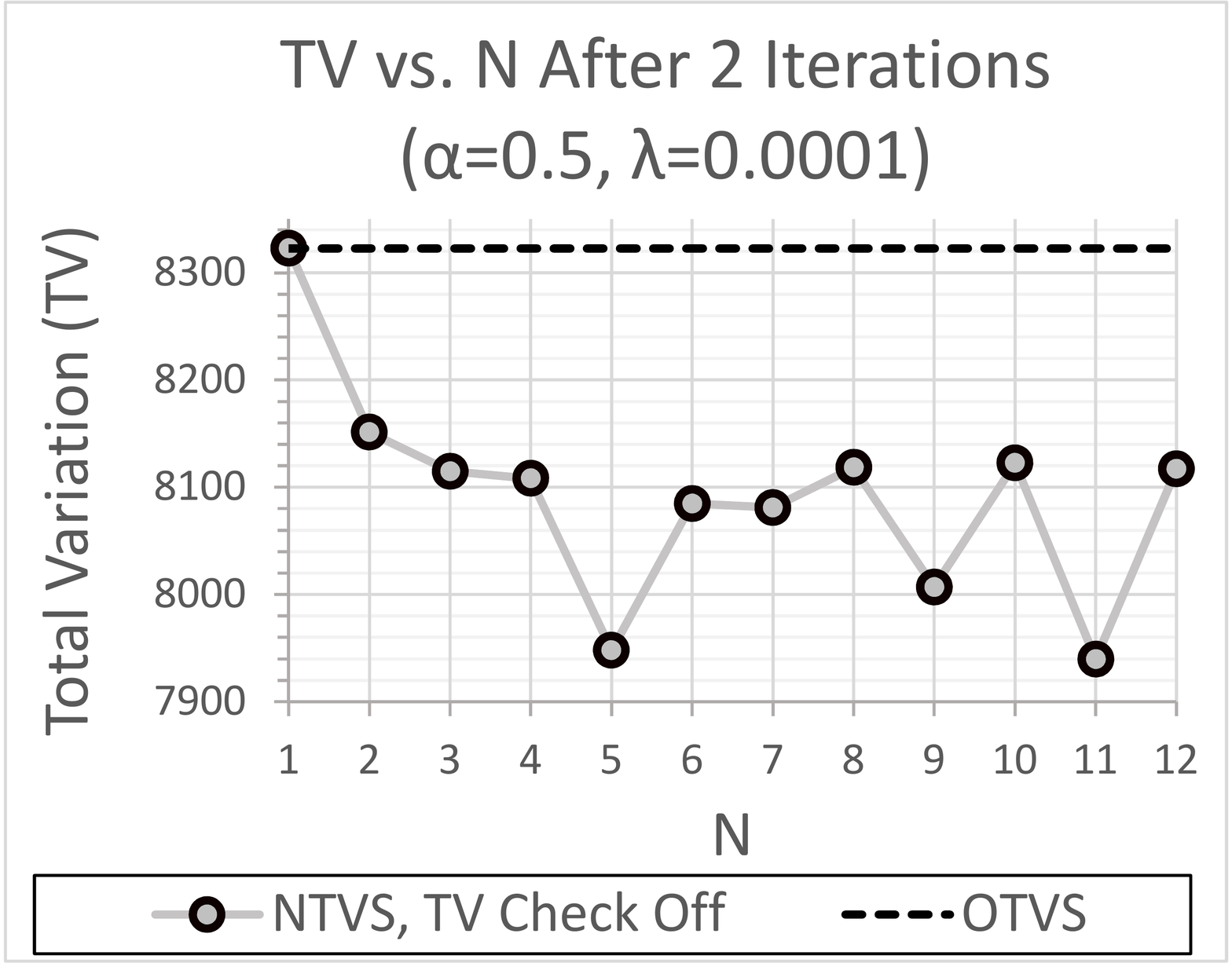}}
\end{minipage}
\vfill
\begin{minipage}[t]{.48\linewidth}
  \centering
    \centerline{\includegraphics[width=\linewidth]{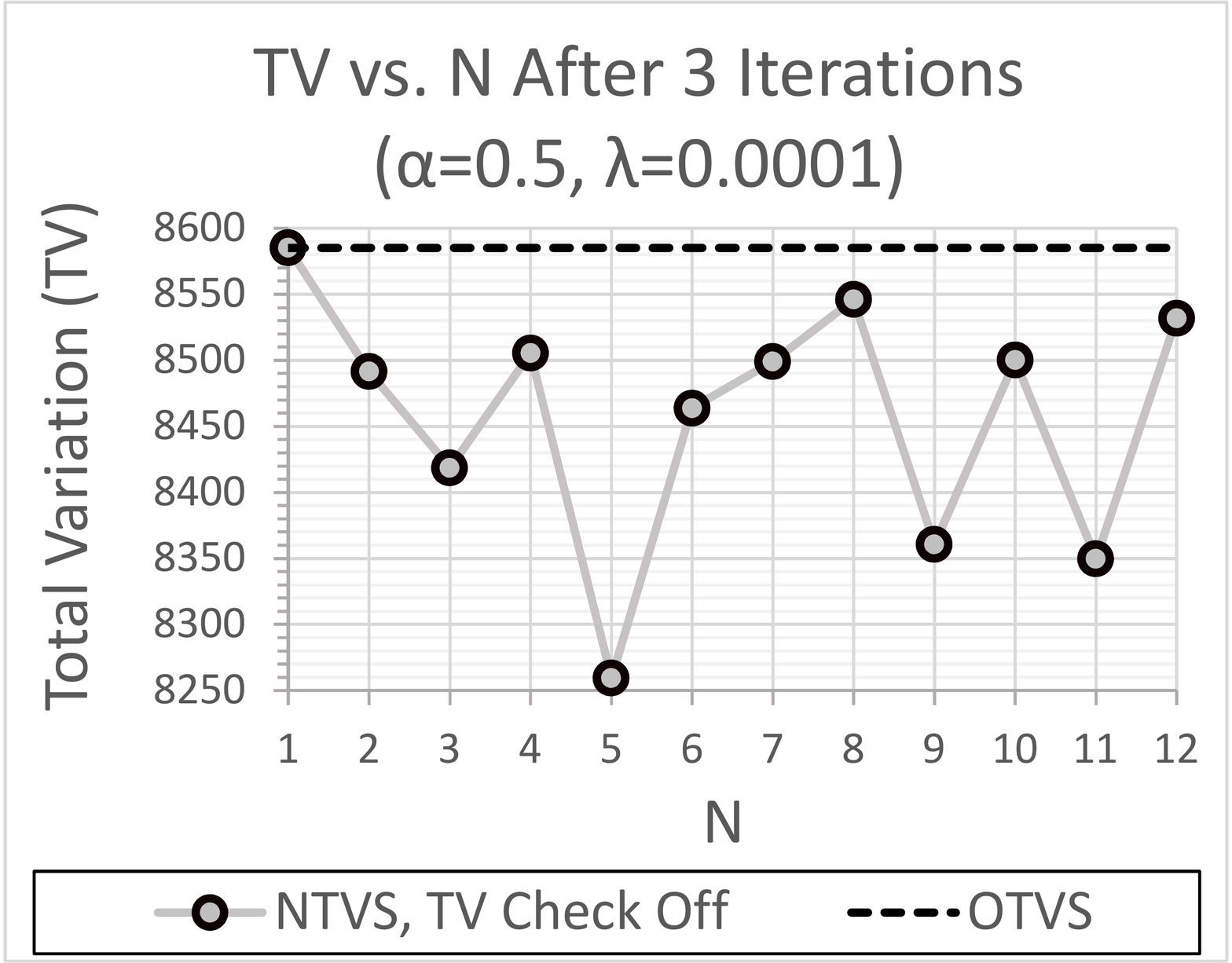}}
\end{minipage}
\hfill
\begin{minipage}[t]{.48\linewidth}
  \centering
     \centerline{\includegraphics[width=\linewidth]{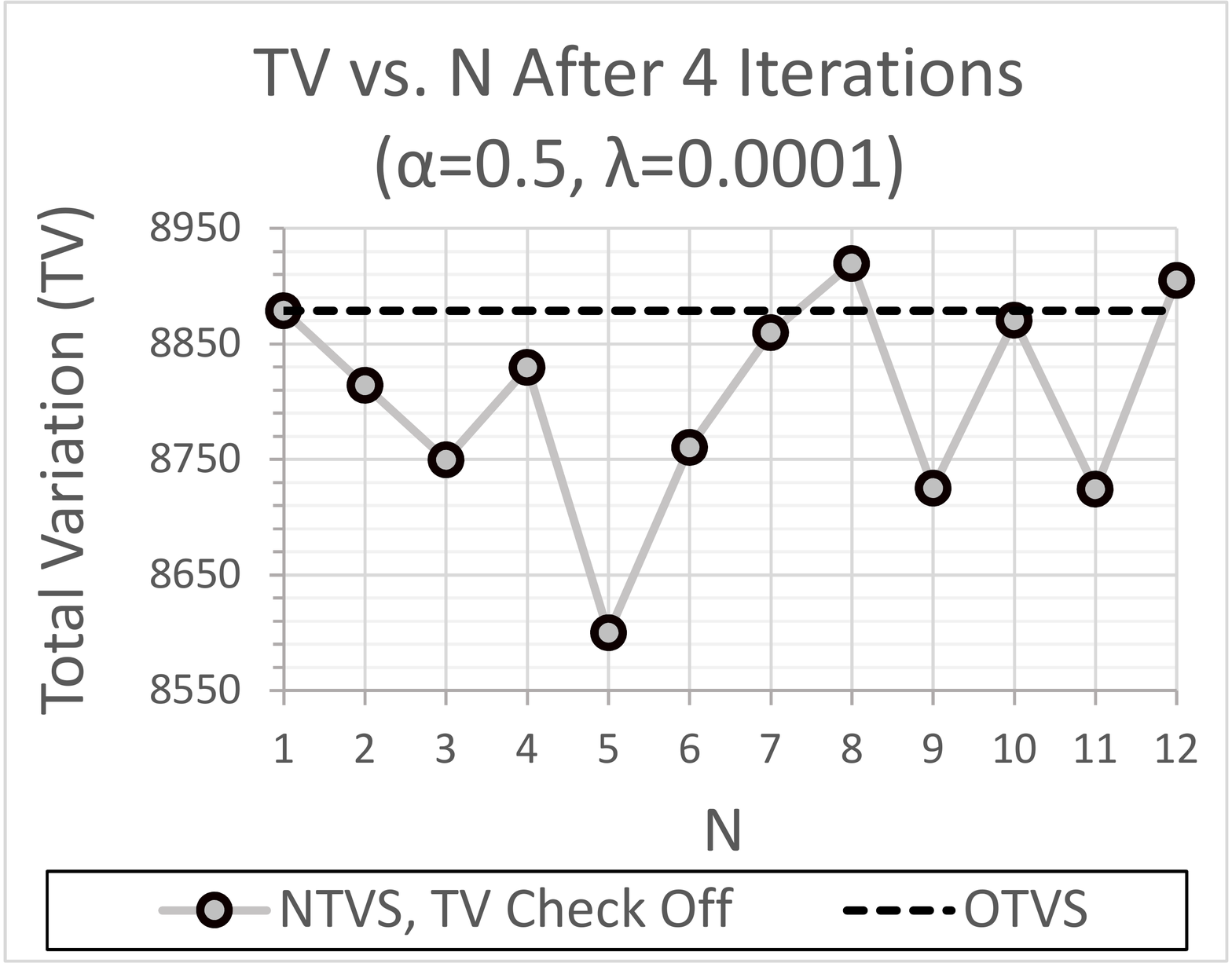}}
\end{minipage}
\caption{TV as a function of $N$ after each of the first 4 feasibility-seeking iterations for the simulated \CTP data set using OTVS and NTVS (TV reduction requirement excluded) with $\lambda=0.0001$ and $\alpha=0.5$.}
\label{fig:TVkSim}
\end{figure}

The number of TV perturbations per feasibility-seeking iteration, $N$, was varied between 1 and 12 in increments of 1.  Figure~\ref{fig:TVkSim} shows the dependence of TV as a function of $N$ for each of the first four feasibility-seeking iterations with the TV reduction requirement excluded. As will be shown later, a similar pattern was observed with the TV reduction requirement included.  The general effect of increasing $N$ was a reduction in TV that leveled off after $N\ge 5$ steps, as best seen in the $k=1$ iteration plot (top left of Figure~\ref{fig:TVkSim}).  An irregular oscillation in TV as a function of increasing $N$ appeared for $k\ge 2$ and increased in magnitude as the number of feasibility-seeking iterations, $k$, increased.

To determine whether the observed fluctuations were random, an analysis of 8 separate reconstructions with $N=5$, $\alpha=0.5$, and the TV reduction requirement excluded were performed for $k=12$ feasibility-seeking iterations. The random increase in $\ell$ between feasibility-seeking iterations was governed by a random number generator that was seeded with a value based on the Julian time at execution, resulting in an effectively random seeding of the random number generator. The standard deviation within the LDPE insert varied between reconstructions with a standard deviation of $\sigma_{\scriptscriptstyle LDPE}=0.00038$ (shown as an error bar on the point at $N=5$ in Figure~\ref{fig:TVSDvCheckSim}(b)); similar variations were also seen in the ROI of the other materials. Note that the standard deviation obtained within the LDPE insert at $N=5$ with the TV reduction requirement excluded was nearly $2\sigma_{\scriptscriptstyle LDPE}$ less than that obtained with the requirement included and just under $4\sigma_{\scriptscriptstyle LDPE}$ less than that obtained with OTVS. In addition, the standard deviation obtained with $N=5$ was at least $1.5\sigma_{\scriptscriptstyle LDPE}$ less than that obtained with any other value of $N$. These differences are large enough to conclude that the observed fluctuation in standard deviation as a function of $N$ was not random. 

For $3\le N\le 6$, there was a benefit from NTVS compared to OTVS, which persisted throughout all twelve feasibility-seeking iterations (see Figure~\ref{fig:TVSDvCheckSim}(a) and (b)).  However, for $N\ge 7$ the benefits of NTVS were increasingly lost as $N$ and $k$ increased.  This can be explained by the decreasing magnitude of TV reducing perturbations with increasing $N$ and the overall increase in TV from each feasibility-seeking iteration.  Although not shown here, a similar dependence on $N$ and $k$ was seen for regional standard deviations.  However, the benefit of NTVS in terms of standard deviation was consistently seen, including for $N\ge 7$, after twelve feasibility-seeking iterations (see, e.g., Figure~\ref{fig:TVSDvCheckSim}(b)).
\subsubsection{Inclusion/Exclusion of TV Reduction Requirement}\label{sssec:rTVSim}
\begin{figure}[h!]
\begin{minipage}{\linewidth}
  \centering
    \centerline{\includegraphics[width=0.7\linewidth]{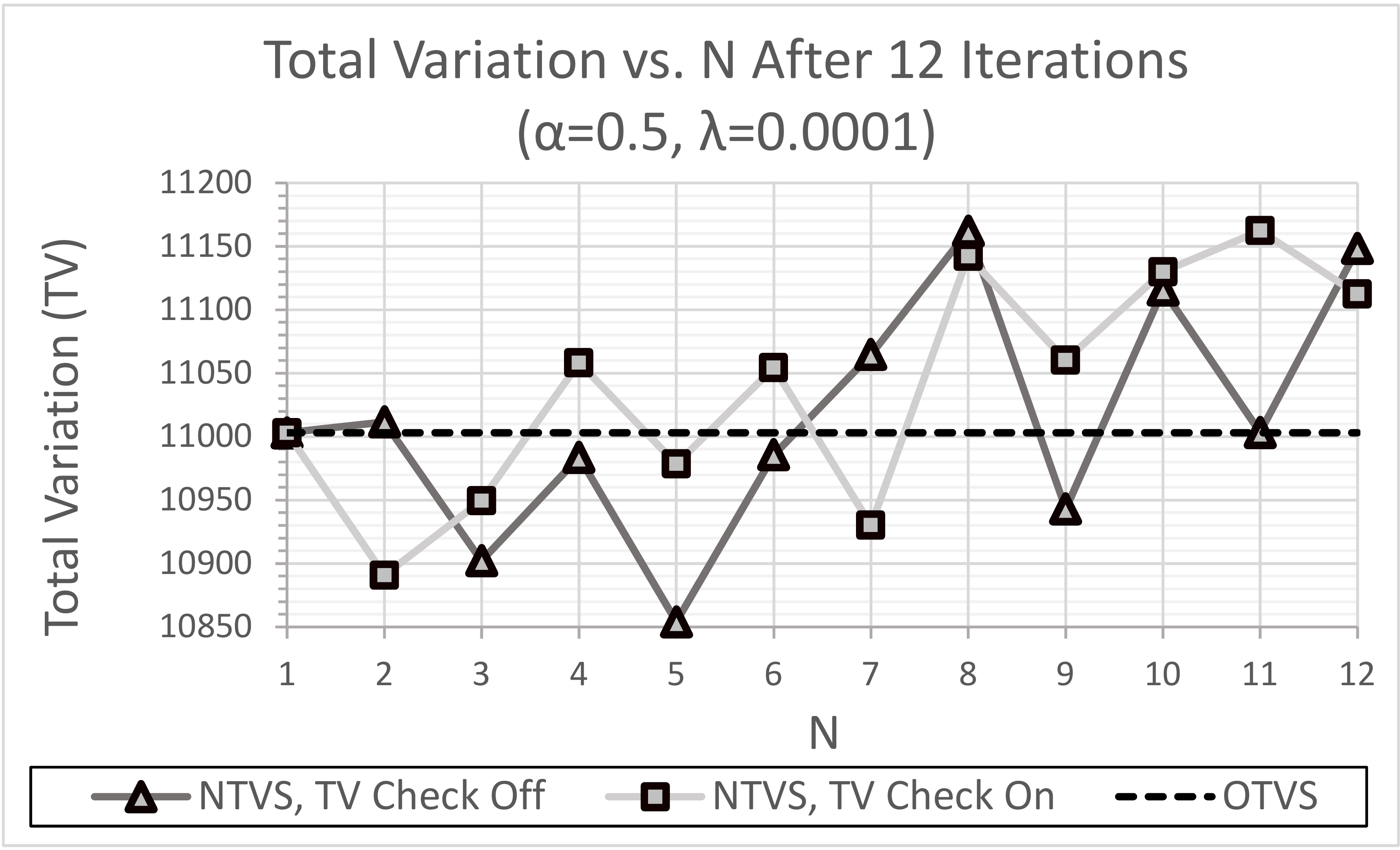}}
    \centerline{(a)}
\end{minipage}
\vfill
\vspace{2mm}
\begin{minipage}{\linewidth}
  \centering
  \centerline{\includegraphics[width=0.7\linewidth]{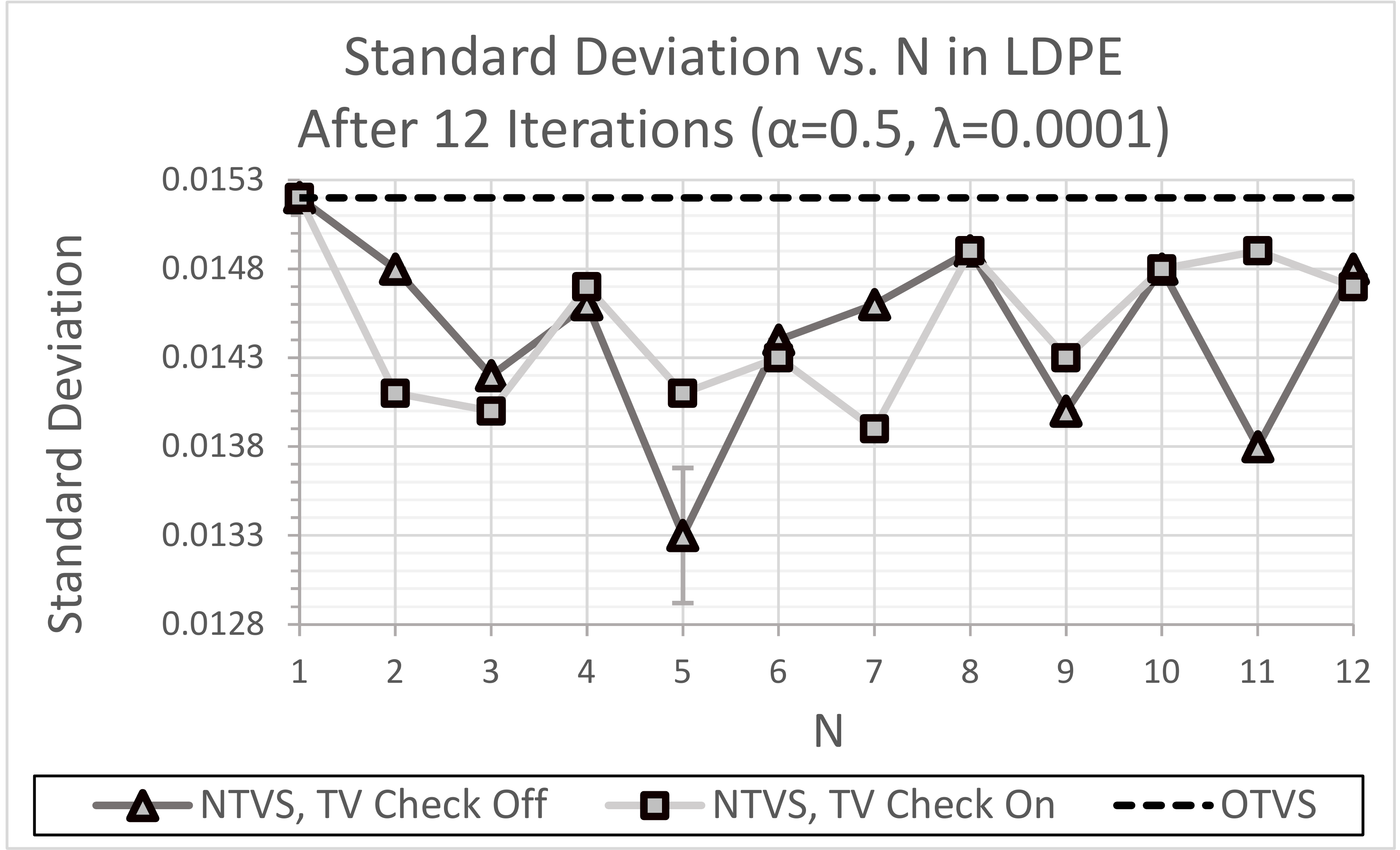}}
    \centerline{(b)}
\end{minipage}
\caption{(a) TV and (b) standard deviation (LDPE) as a function of $N$ after 12 feasibility-seeking iterations for the simulated \CTP data set using OTVS and NTVS including and excluding the TV reduction requirement with $\lambda=0.0001$ and $\alpha=0.5$. The error bar at $N=5$ denotes the variation in standard deviation ($\sigma=0.00038$) between 8 repetitions of reconstruction with $N=5$.}
    \label{fig:TVSDvCheckSim}
\end{figure}
To determine if the exclusion of the TV reduction requirement in the definition of the NTVS algorithm (Appendix~\ref{app:ntvs}) is an appropriate decision, reconstructions were also performed with a variation of the NTVS algorithm that included the TV reduction requirement; the definition of the algorithm used for these investigations is provided for reference at the end of Appendix~\ref{app:ntvs}.

Figures~\ref{fig:TVSDvCheckSim}(a)~and~\ref{fig:TVSDvCheckSim}(b) show the comparison of TV and standard deviation, respectively, for OTVS and NTVS with relaxation parameter $\lambda=0.0001$, median filter radius $r=2$ applied to the initial iterate~\cite{SKGPSS15}, and 12 feasibility-seeking iterations.  In each plot, the results for NTVS with and without inclusion of the TV reduction requirement are shown as a function of $N$.  The horizontal line corresponds to the result of OTVS ($N=1$, $\alpha=0.5$).

In the range of $3\le N \le 6$, including the TV reduction requirement had practically no benefit, whereas its removal yields up to a $5.7\%$ reduction in the standard deviation in RSP within the LDPE material insert and up to a $1.2\%$ reduction in overall TV.  Similar results were obtained for other values of $\alpha$, $\lambda$, and, in the case of standard deviation, for different materials. One can conclude that imposing the TV reduction requirement does not provide a consistent benefit in terms of TV and standard deviation. Therefore, for the remainder of the parameter space exploration, the TV reduction requirement was excluded.
\subsubsection{Perturbation Kernel ($\alpha$)}
\begin{figure}[h!]
\begin{minipage}[t]{\linewidth}
  \centering
     \centerline{\includegraphics[width=0.7\linewidth]{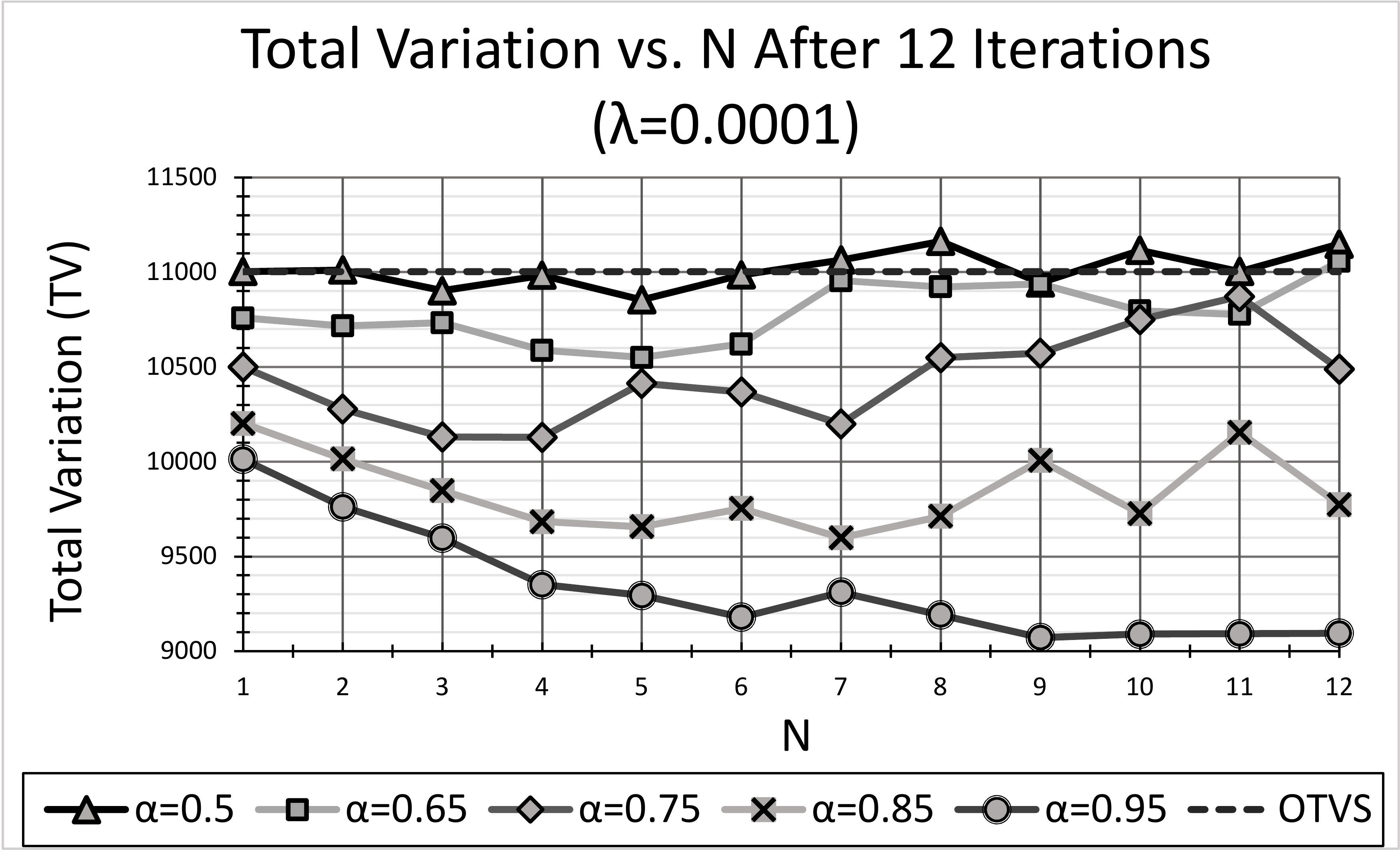}}
        \vspace{1mm}
\vfill\centerline{(a)}
\end{minipage}
\vfill
\vspace{2mm}
\begin{minipage}[t]{\linewidth}
  \centering
   \centerline{\includegraphics[width=0.7\linewidth]{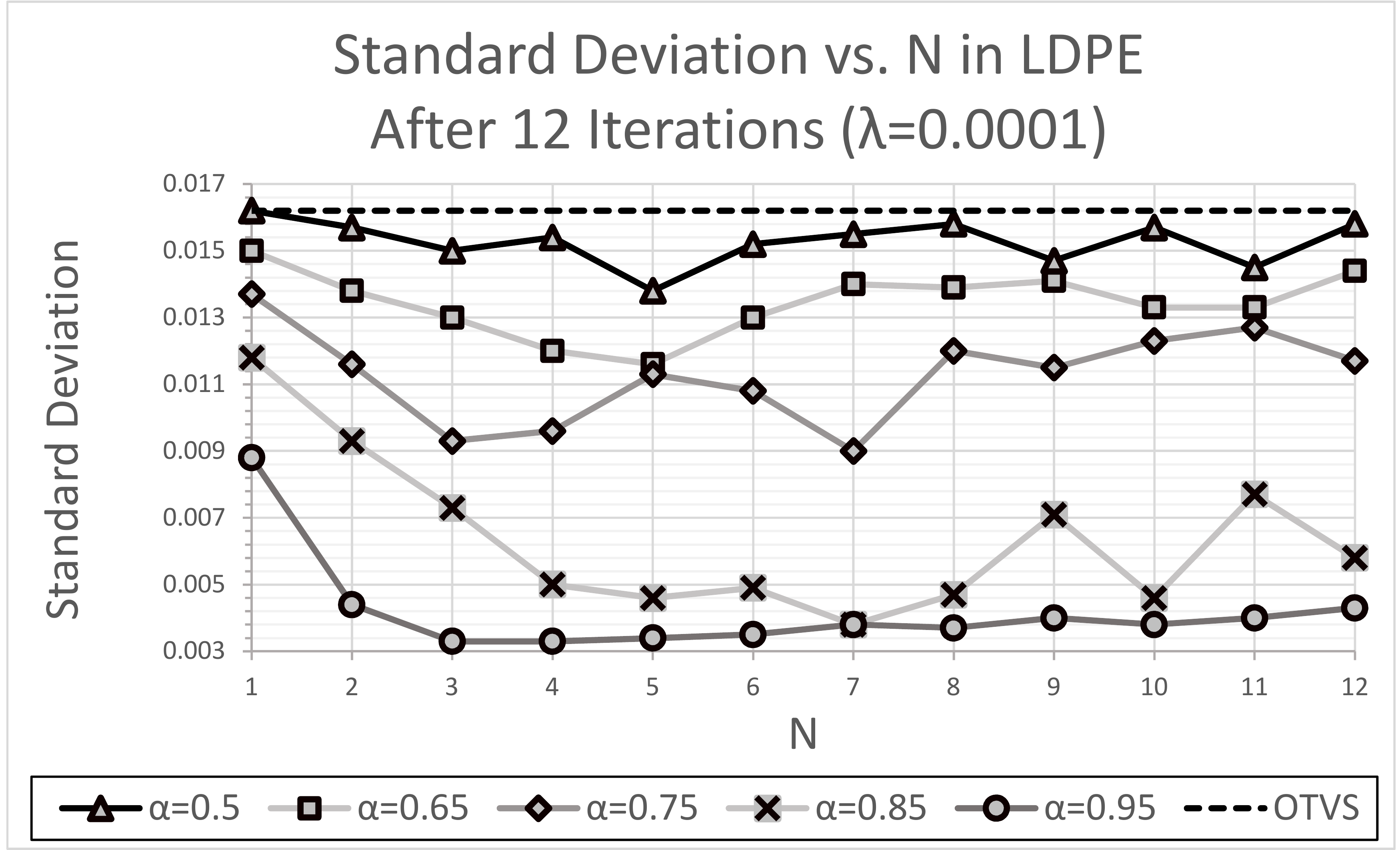}}
        \vspace{1mm}
\vfill\centerline{(b)}
\end{minipage}
\caption{(a) TV and (b) standard deviation (LDPE) as a function of $N$ after 12 feasibility-seeking iterations for the simulated \CTP data set using OTVS and NTVS (TV reduction requirement excluded) with $\lambda=0.0001$ and $\alpha=0.5$.}
\label{fig:TVSDvASim}
\end{figure}
Further investigations were performed to determine the effect of the perturbation kernel $\alpha$ (see step (\ref{alg:beta}) of the NTVS algorithm in Appendix~\ref{app:ntvs}) on TV and standard deviation for $0.5\le\alpha\le 0.95$ and $1\le N \le 12$.  Increasing $\alpha$ produces larger perturbations and results in the perturbation magnitude $\beta_k$ converging to zero more slowly.  Thus, one can expect a larger reduction of TV and standard deviation for larger values of $\alpha$.  Figures~\ref{fig:TVSDvASim}(a)~and~\ref{fig:TVSDvASim}(b) demonstrate this effect.
\begin{figure}[h!]
\begin{minipage}{\linewidth}
  \centering
     \centerline{\includegraphics[width=0.7\linewidth]{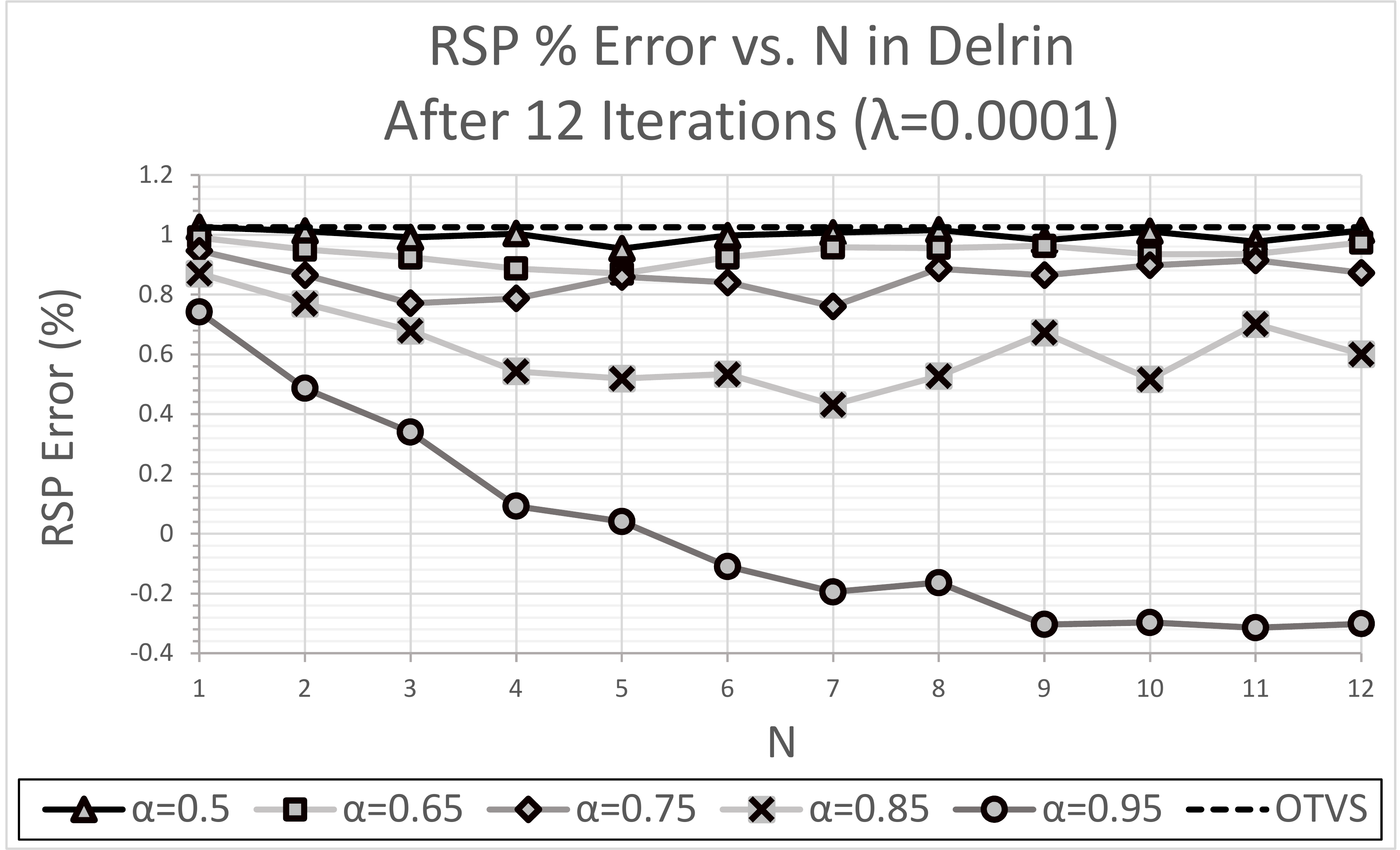}}
            \vspace{0.5mm}
\vfill\centerline{(a)}
\end{minipage}
\vfill
\vspace{2mm}
\begin{minipage}{\linewidth}
  \centering
   \centerline{\includegraphics[width=0.7\linewidth]{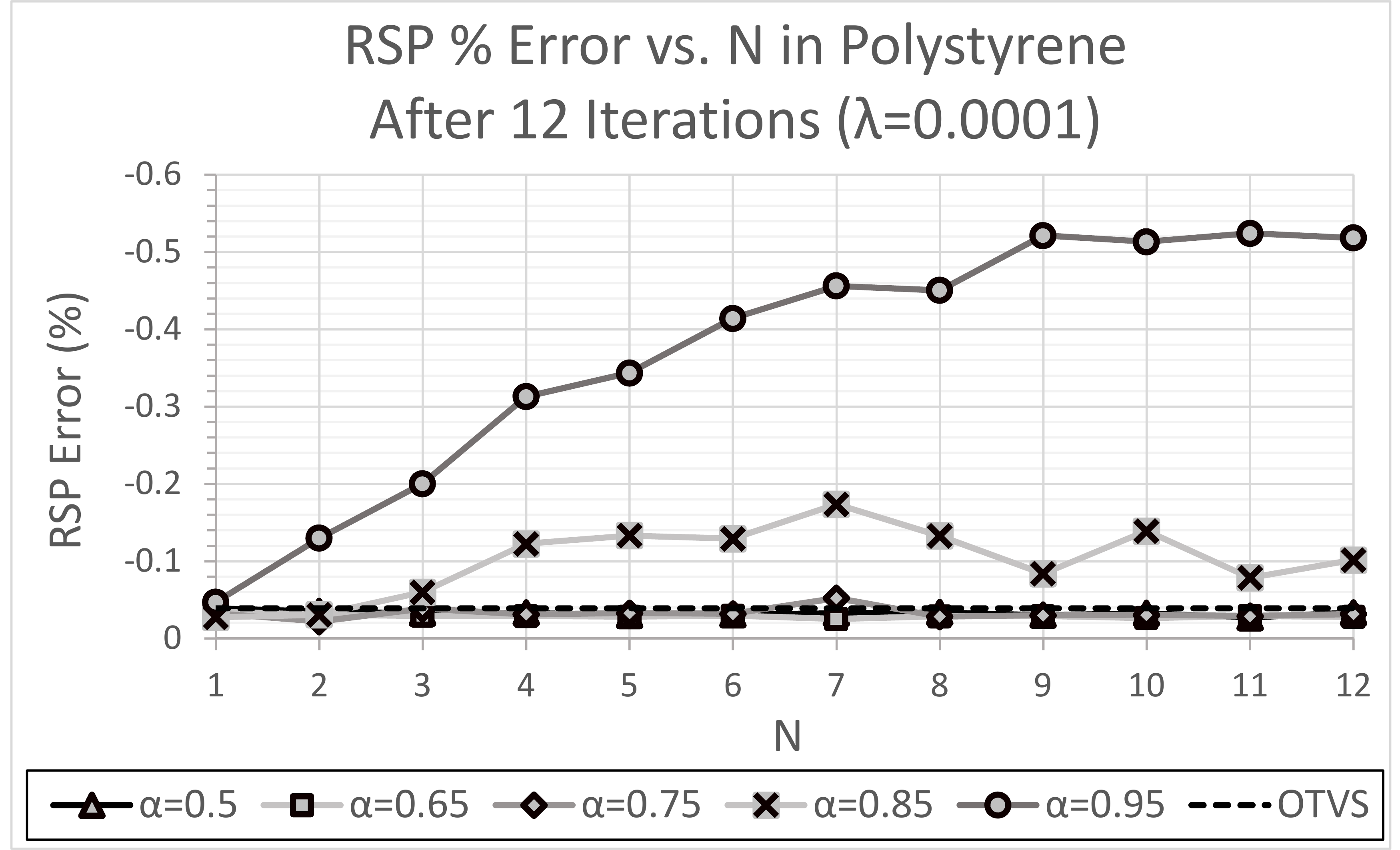}}
            \vspace{0.5mm}
\vfill\centerline{(b)}
\end{minipage}
\caption{RSP error in the (a) Delrin and (b) polystyrene ROIs as a function of $N$ after 12 feasibility-seeking iterations for the simulated \CTP data set using OTVS and NTVS (TV reduction requirement excluded) with $\lambda=0.0001$ and varying $\alpha$.}
\label{fig:RSPevASim}
\end{figure}

Figures~\ref{fig:RSPevASim}(a)~and~\ref{fig:RSPevASim}(b) show the effect of $\alpha$ on the accuracy of reconstructed RSP values in the Delrin and polystyrene inserts, respectively.  These two materials were chosen because they were most affected by the value of $\alpha$.  From these plots, one can see that for $\alpha > 0.75$, perturbations have a growing effect on RSP accuracy as $\alpha$ and $N$ increase.  This leads to changes in error greater than $1\%$ for Delrin and greater than $0.5\%$ for polystyrene.  Although increasing $\alpha$ to decrease TV and standard deviation is a worthwhile goal, one cannot do so without considering its effect on RSP error.  On the other hand, increasing $\alpha$ from $\alpha=0.5$ to $\alpha=0.75$ yielded up to a $39.3\%$ reduction in the standard deviation in RSP within the LDPE material insert and up to an $8.2\%$ reduction in overall TV without negatively impacting RSP error.
\subsubsection{Relaxation Parameter ($\lambda$)}
\begin{figure}[h!]
\begin{minipage}{\linewidth}
  \centering
     \centerline{\includegraphics[width=0.7\linewidth]{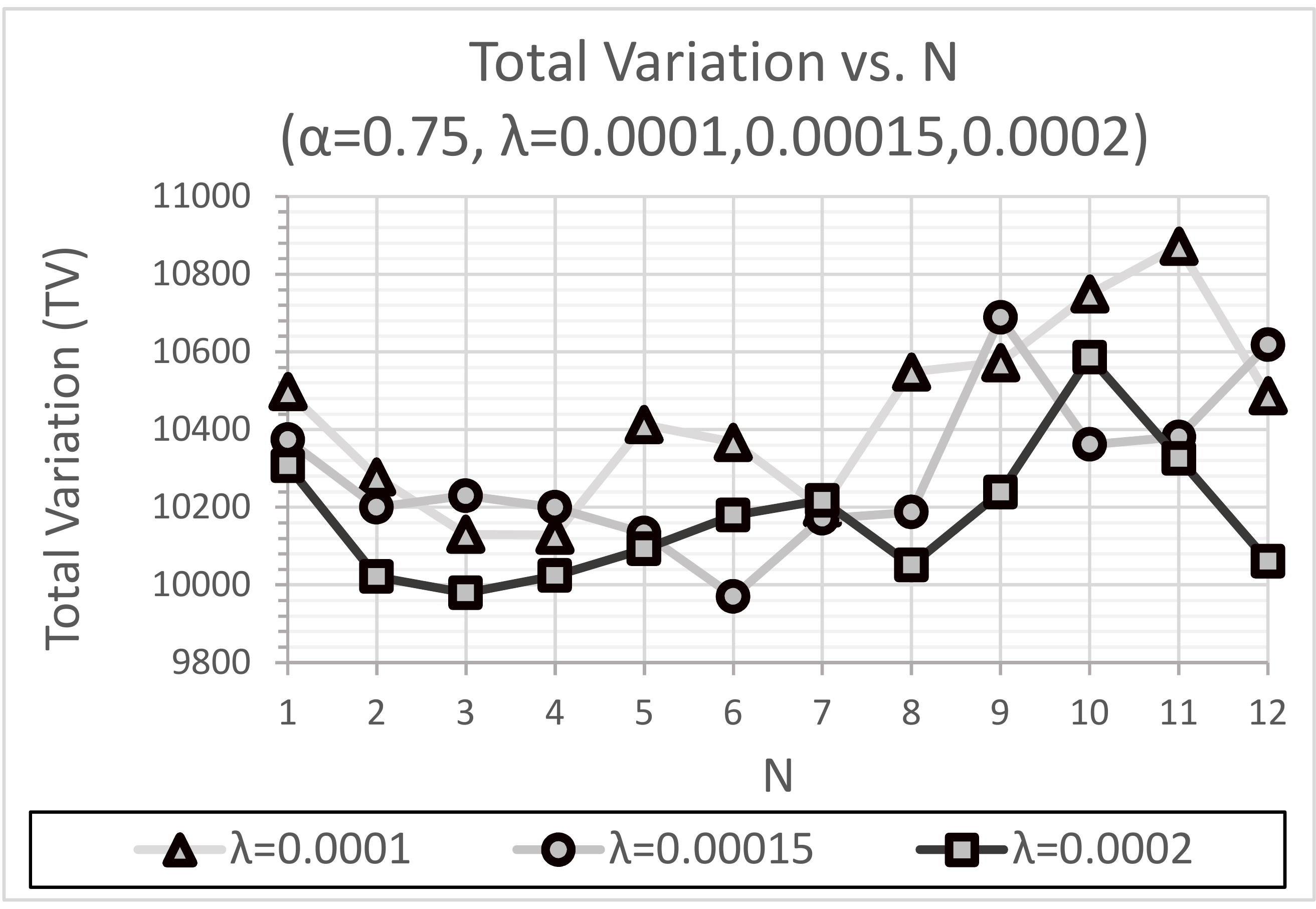}}
    \vspace{0.5mm}
\vfill\centerline{(a)}
\end{minipage}
\vfill
\vspace{2mm}
\begin{minipage}{\linewidth}
  \centering
    \centerline{\includegraphics[width=0.7\linewidth]{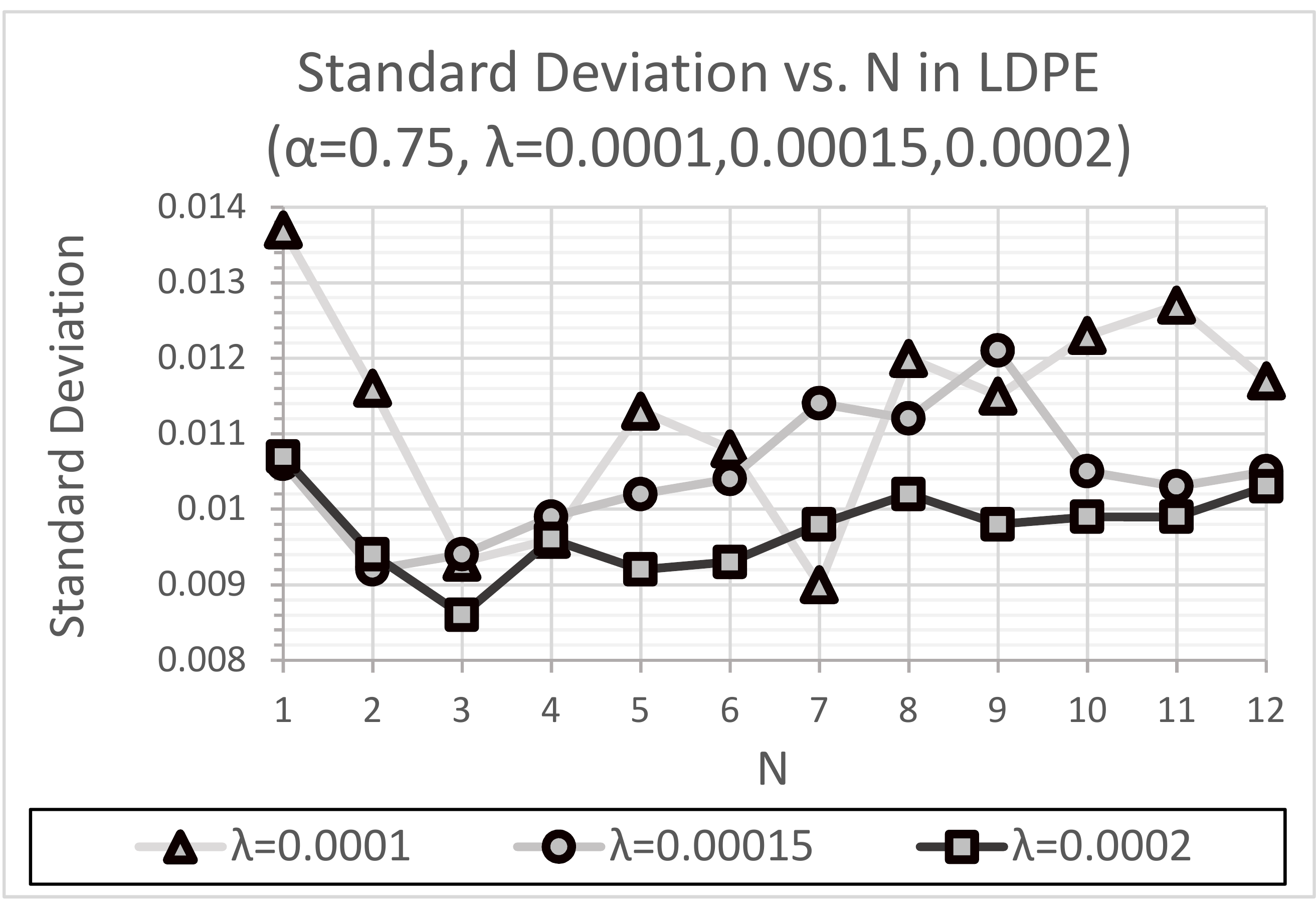}}
    \vspace{0.5mm}
\vfill\centerline{(b)}
\end{minipage}
\caption{(a) TV and (b) standard deviation (soft tissue) as a function of $N$ for $\lambda=0.0001$, $k=12$; $\lambda=0.00015$, $k=8$; and $\lambda=0.0002$, $k=6$ iterations, respectively, and $\alpha=0.75$ for the simulated \CTP data set.}
\label{fig:TVSDvLoptimalSim}
\end{figure}
Increasing the relaxation parameter accelerates the rate of convergence of the feasibility-seeking algorithm.  To investigate the impact of NTVS independent of convergence rate, the number of iterations was adjusted for $\lambda=0.00015$ and $\lambda=0.0002$ to obtain the same RSP accuracy as for $\lambda=0.0001$ and 12 iterations.  For this comparison, $\alpha=0.75$ was chosen.

Figures~\ref{fig:TVSDvLoptimalSim}(a)~and~\ref{fig:TVSDvLoptimalSim}(b) show TV and standard deviation within the LDPE ROI, respectively, for three combinations of $\lambda$ and $k$.  For most values of $N$, the relative improvements in TV and standard deviation increased as $\lambda$ increased.  Note that the trend for standard deviation was not as pronounced for other materials, but increasing $\lambda$ consistently produced comparable or larger reductions in TV and standard deviation in each material region.
\subsection{Experimental \CTP Data Set}\label{sssec:rExper}
The experimental \CTP data set was then used to reconstruct images for the same combination of parameter values from the reconstruction parameter space as for the simulated data set.
\subsubsection{Number of TVS steps ($N$)}\label{sssec:rNExper}
\begin{figure}[h!]
\begin{minipage}[t]{0.48\linewidth}
  \centering
    \centerline{\includegraphics[width=\linewidth]{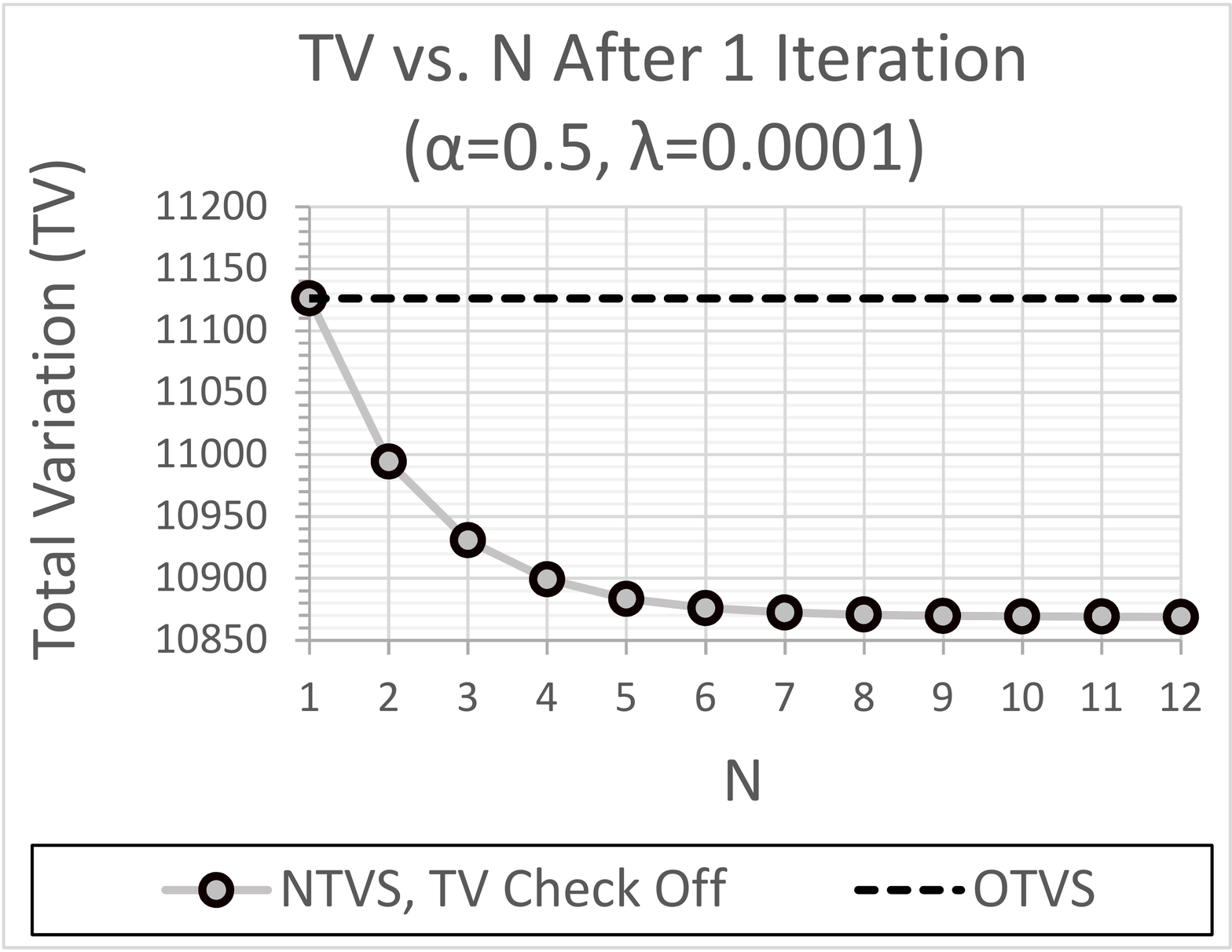}}
\end{minipage}
\hfill
\begin{minipage}[t]{0.48\linewidth}
  \centering
     \centerline{\includegraphics[width=\linewidth]{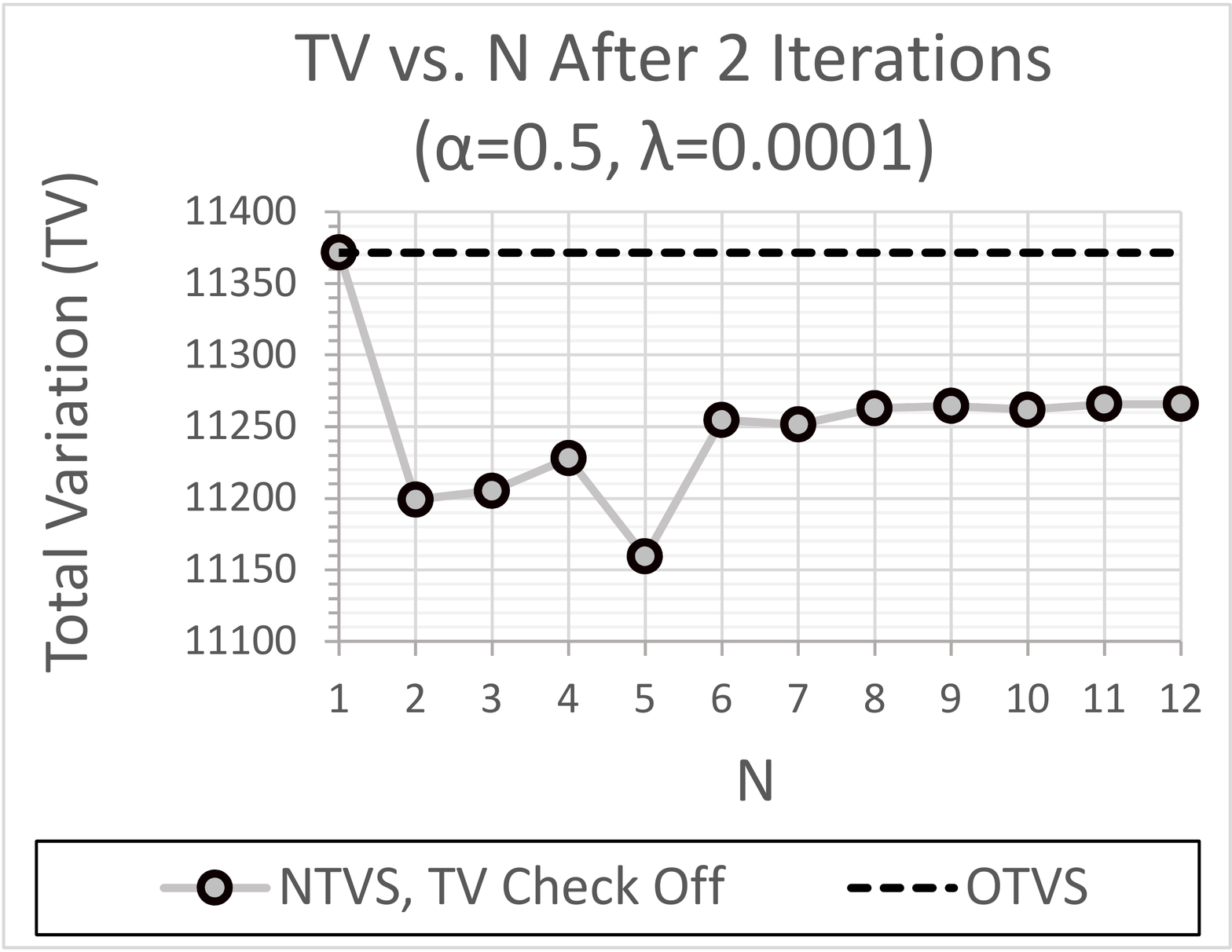}}
\end{minipage}
\vfill
\begin{minipage}[t]{0.48\linewidth}
  \centering
    \centerline{\includegraphics[width=\linewidth]{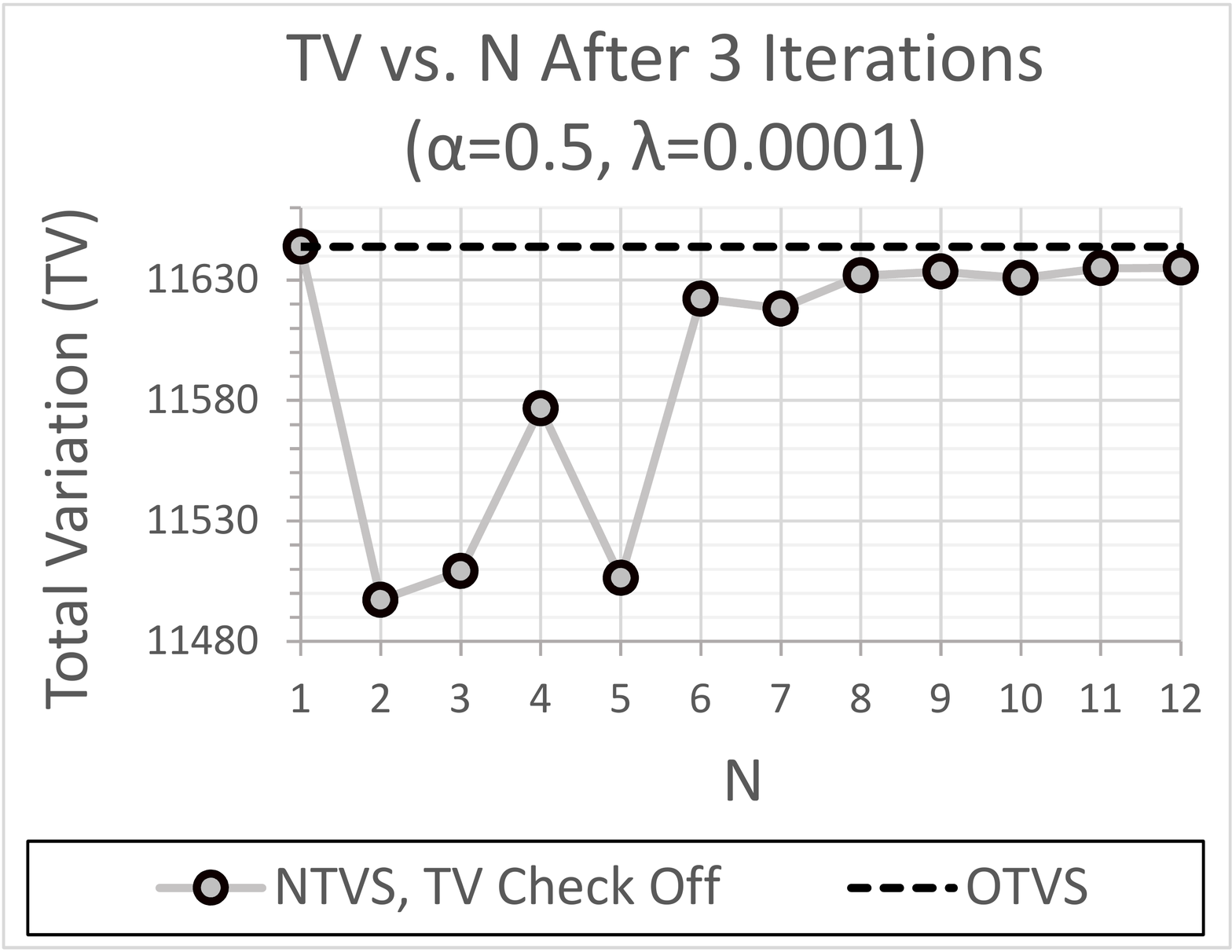}}
\end{minipage}
\hfill
\begin{minipage}[t]{0.48\linewidth}
  \centering
     \centerline{\includegraphics[width=\linewidth]{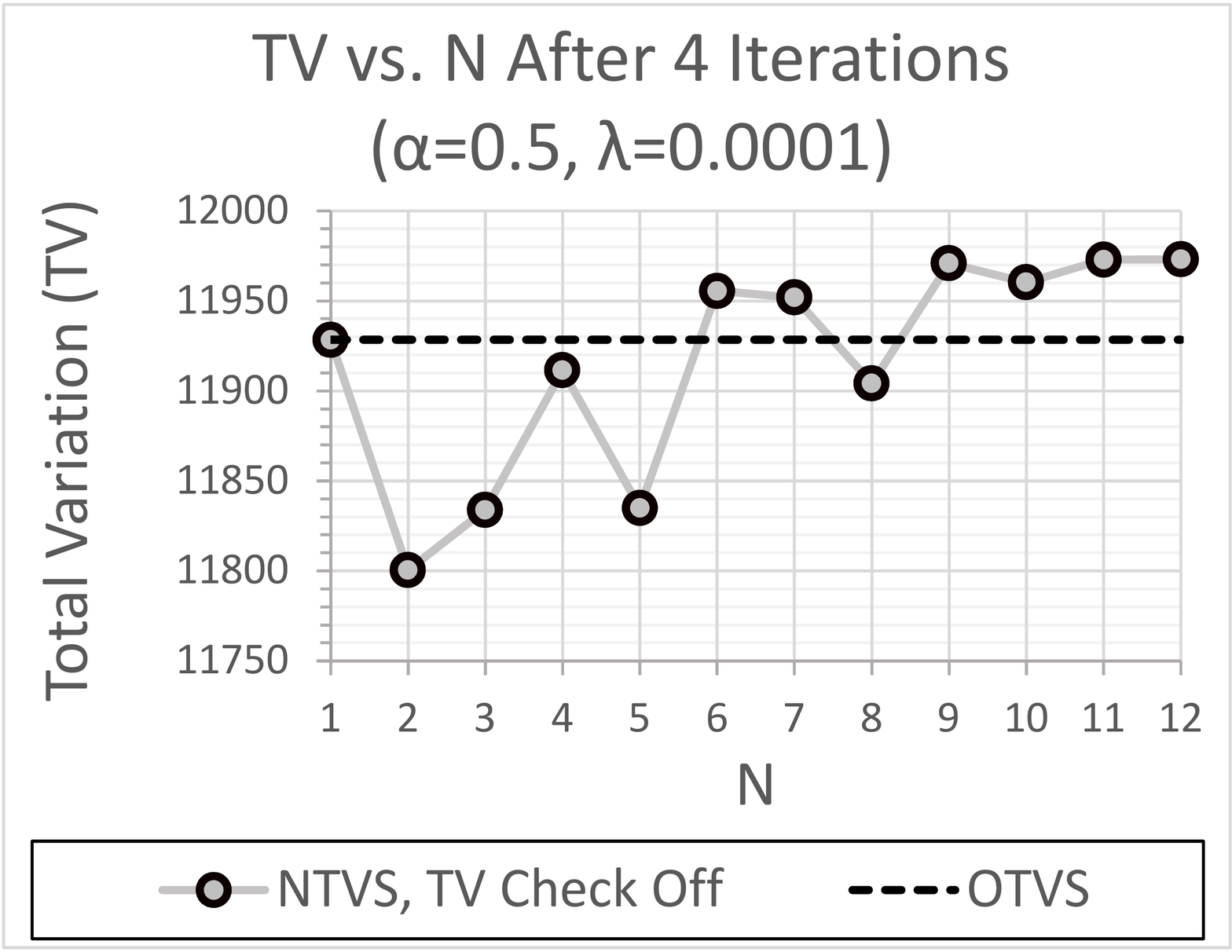}}
\end{minipage}
\caption{TV as a function of $N$ after each of the first 4 feasibility-seeking iterations for the experimental \CTP data set using OTVS and NTVS (TV reduction requirement excluded) with $\lambda=0.0001$ and $\alpha=0.5$ .}
\label{fig:TVkExper}
\end{figure}
The number of TV perturbations per feasibility-seeking iteration, $N$, was again varied sequentially between 1 and 12 for the experimental \CTP data set; Figure~\ref{fig:TVkExper} shows plots of TV as a function of $N$ for each of the first four feasibility-seeking iterations for the case where the TV reduction requirement is excluded. TV did not fluctuate as much as it did with the simulated data but the same general trend can be seen: increasing $N$ results in a monotonic reduction in TV for the first iteration, but as subsequent feasibility-seeking iterations are performed on the resulting image, the reductions in TV obtained by increasing $N$ reverse and eventually exceed the results obtained with OTVS. As with the simulated data, a consistent benefit was obtained by performing $3\le N\le 6$ TVS perturbations per feasibility-seeking iteration, but increasing beyond $N\ge 7$ results in an image whose perturbations place it in a less advantageous point in the solution space for feasibility-seeking.
\subsubsection{Inclusion/Exclusion of TV Reduction Requirement}\label{sssec:rTVExper}
\begin{figure}[h!]
\begin{minipage}{\linewidth}
  \centering
    \centerline{\includegraphics[width=0.65\linewidth]{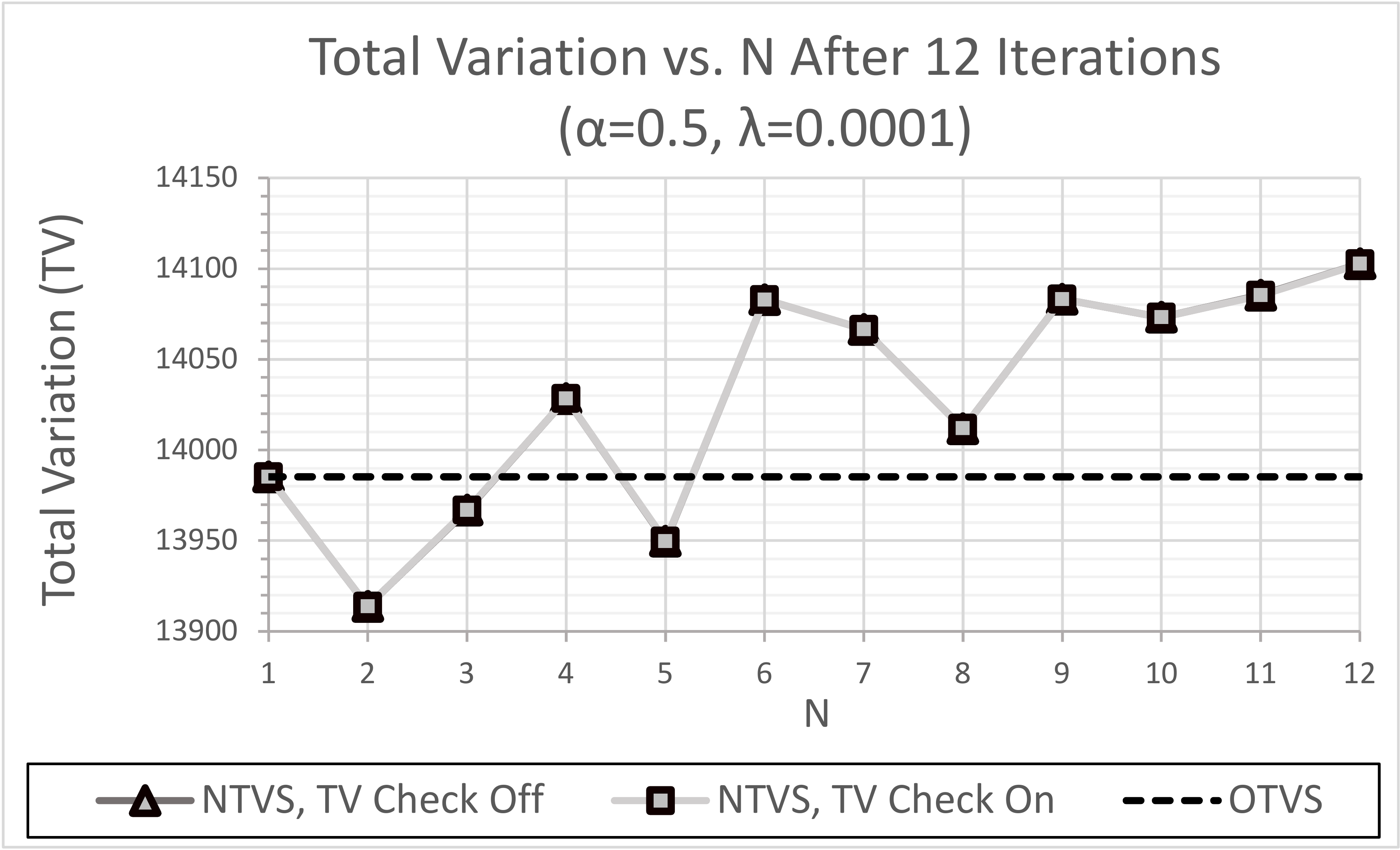}}
    \centerline{(a)}
\end{minipage}
\vfill
\vspace{2mm}
\begin{minipage}{\linewidth}
  \centering
  \centerline{\includegraphics[width=0.65\linewidth]{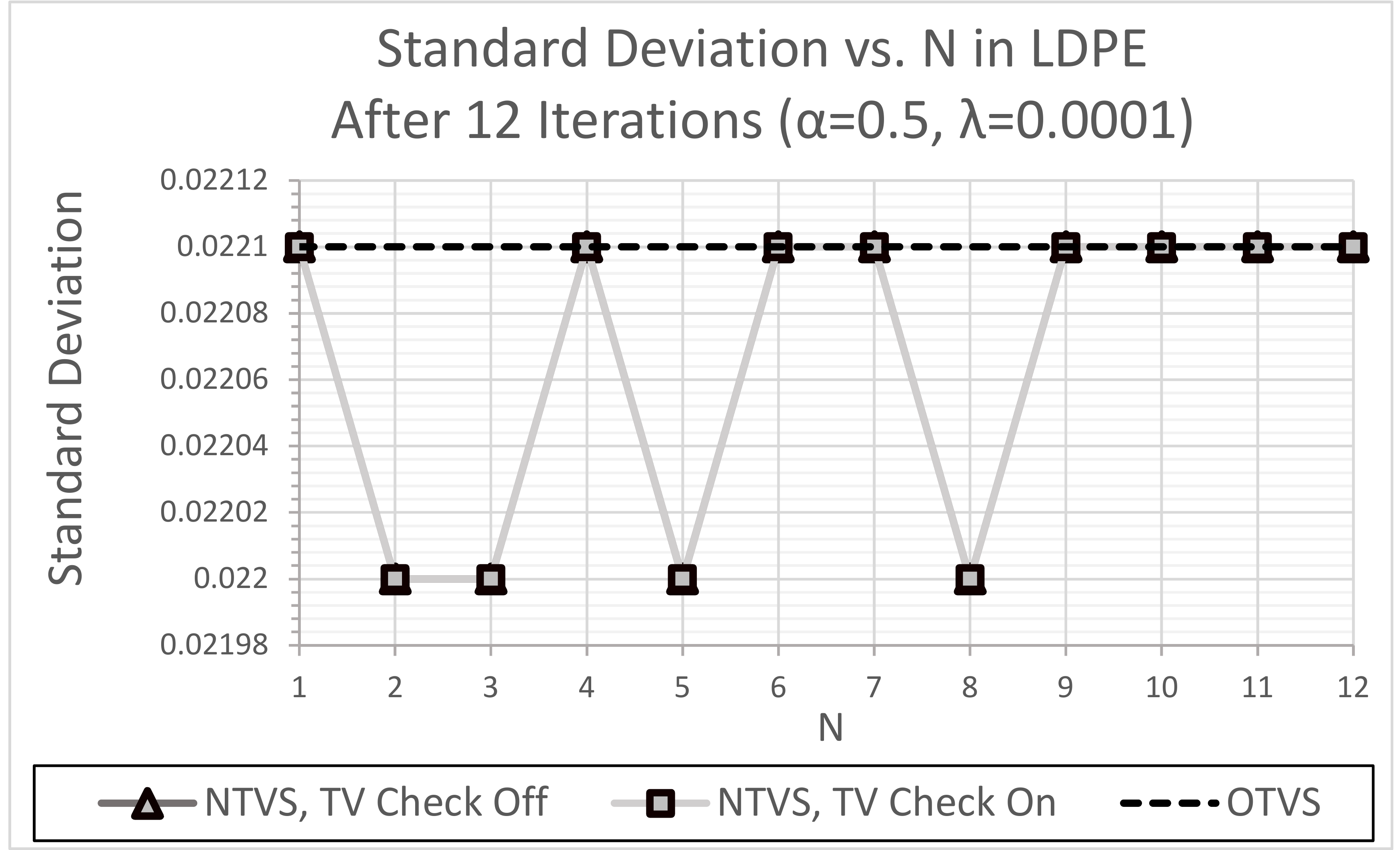}}
    \centerline{(b)}
\end{minipage}
\caption{(a) TV and (b) standard deviation as a function of $N$ after 12 feasibility-seeking iterations for the experimental \CTP data set using the OTVS algorithm and the NTVS algorithm including and excluding the TV reduction requirement with $\lambda=0.0001$ and $\alpha=0.5$ (note that the 2 NTVS curves overlap).}
    \label{fig:TVSDvCheckExper}
\end{figure}
\begin{figure}[h!]
\begin{minipage}{\linewidth}
  \centering
    \centerline{\includegraphics[width=0.65\linewidth]{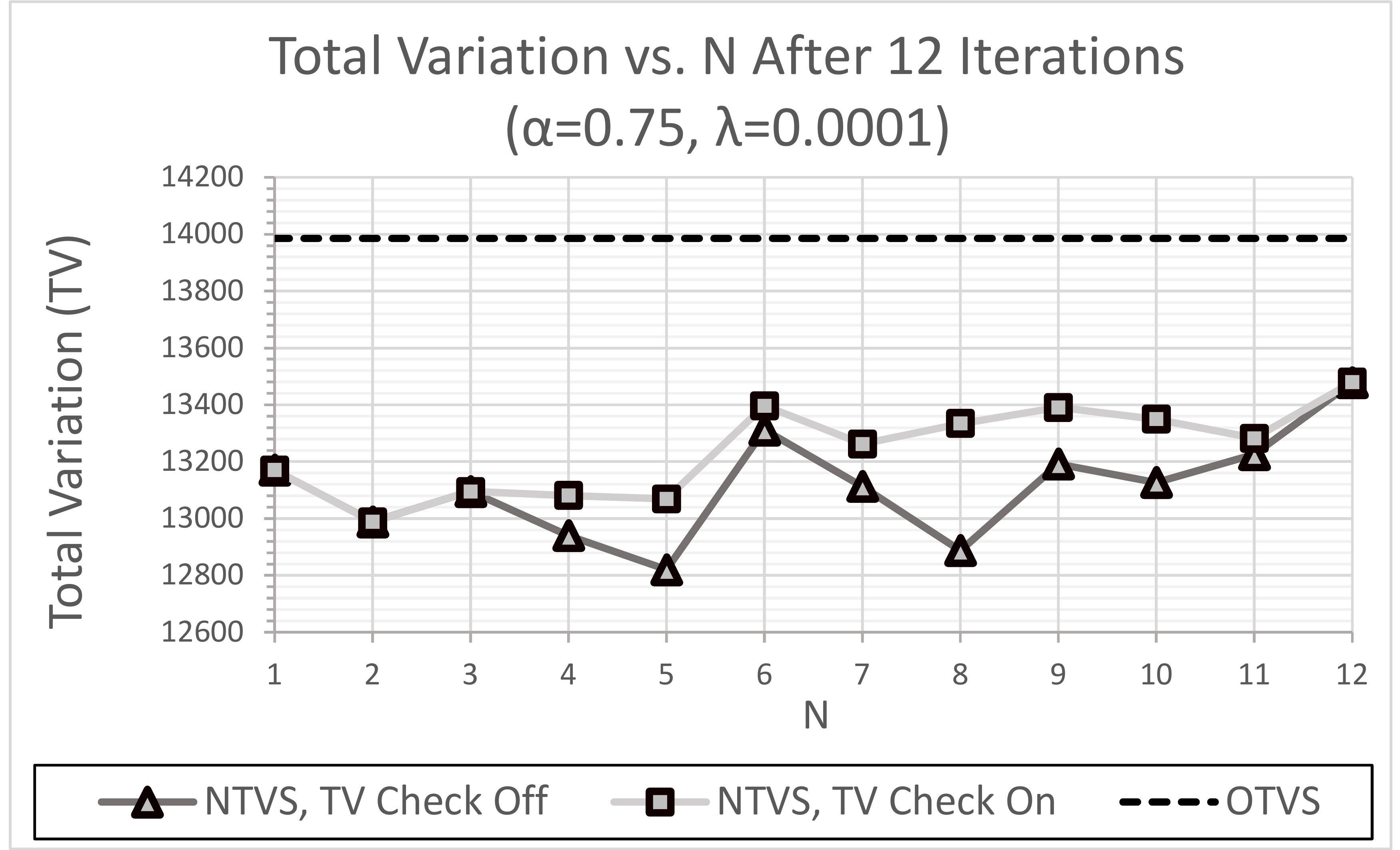}}
    \centerline{(a)}
\end{minipage}
\vfill
\vspace{2mm}
\begin{minipage}{\linewidth}
  \centering
  \centerline{\includegraphics[width=0.65\linewidth]{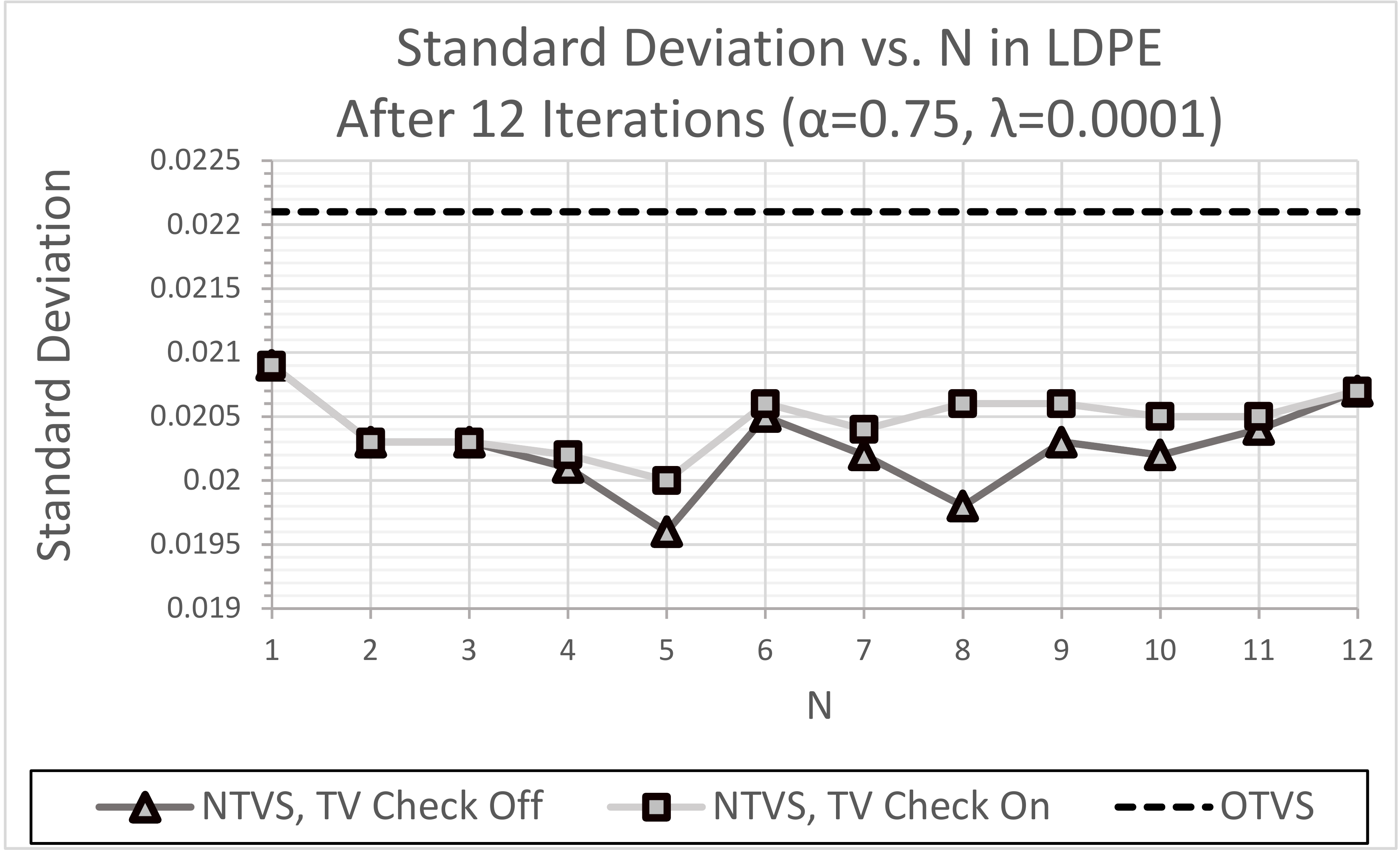}}
    \centerline{(b)}
\end{minipage}
\caption{(a) TV and (b) standard deviation as a function of $N$ after 12 feasibility-seeking iterations for the experimental \CTP data set using the OTVS algorithm and the NTVS algorithm including and excluding the TV reduction requirement with $\lambda=0.0001$ and $\alpha=0.75$.}
    \label{fig:TVSDvCheckExper2}
\end{figure}
An investigation that isolated the impact of the exclusion of the TV reduction requirement on NTVS results was also performed for the experimental \CTP data set. Figures~\ref{fig:TVSDvCheckExper}(a) and (b) show a comparison of TV and standard deviation, respectively, for inclusion and exclusion of the TV reduction requirement in reconstructions of the experimental data set with $\alpha=0.5$. In this case, the difference in TV and standard deviation is not large enough to discern between the lines representing inclusion and exclusion of the TV reduction requirement; this occurred for each value of $\lambda$ and, in the case of standard deviation, within the ROI of each material. However, for $\alpha=0.75$, the exclusion of the TV reduction requirement consistently resulted in smaller TV and standard deviation, as demonstrated by Figures~\ref{fig:TVSDvCheckExper2}(a) and (b); similar trends were also seen for other values of $\lambda$ and in the other cylindrical material inserts.

The scale of these plots also provides a better perspective on the reductions obtained with the larger $\alpha=0.75$ compared to those obtained with OTVS, indicating a consistent and sizeable reduction in TV and standard deviation for every value of $N$ investigated, including those with $N\ge 7$. These results demonstrate that the $50\%$ increase in $\alpha$ resulted in a reduction in TV and standard deviation with approximately the same magnitude as the largest difference in TV and standard deviation obtained with varying $N$. In particular, the difference between the standard deviation obtained with OTVS and $3\le N\le 6$ was more than twice as large as the maximum difference between results within this range of $N$, a trend that was also consistently seen within the ROIs of the other materials.
\subsubsection{Perturbation Kernel ($\alpha$)}\label{sssec:rAExper}
\begin{figure}[h!]
\begin{minipage}[t]{\linewidth}
  \centering
     \centerline{\includegraphics[width=0.7\linewidth]{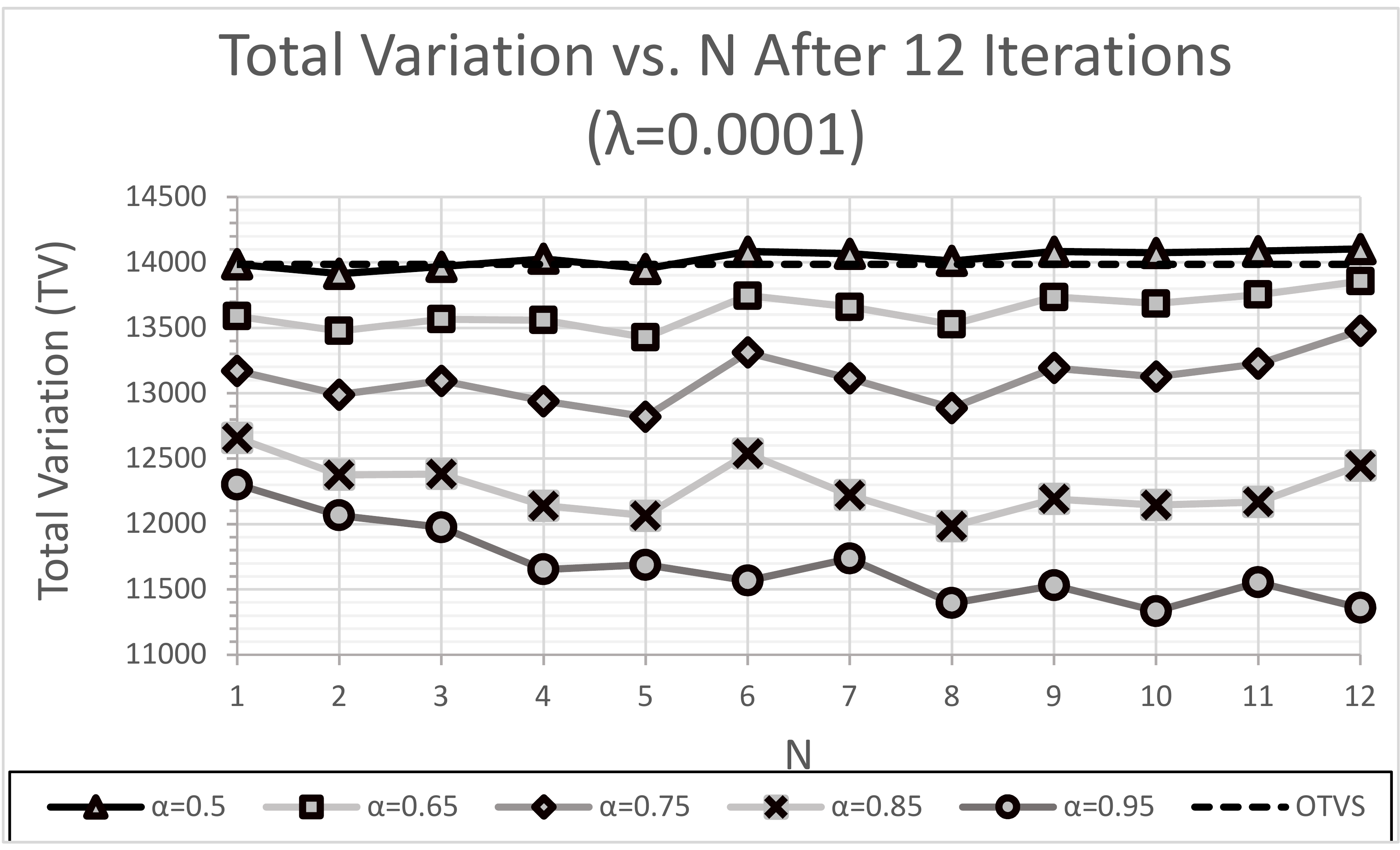}}
        \vspace{1mm}
\vfill\centerline{(a)}
\end{minipage}
\vfill
\vspace{2mm}
\begin{minipage}[t]{\linewidth}
  \centering
   \centerline{\includegraphics[width=0.7\linewidth]{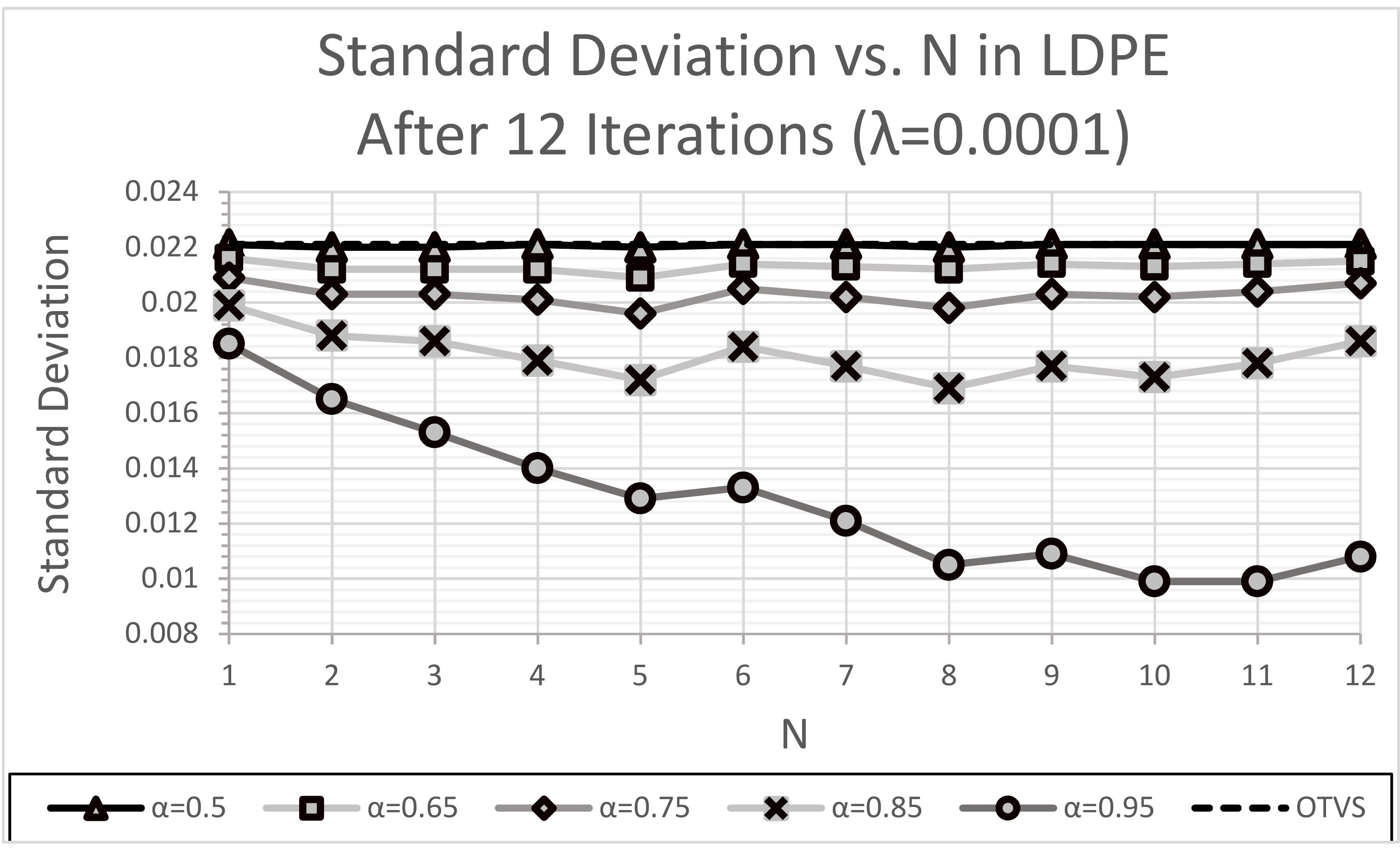}}
        \vspace{1mm}
\vfill\centerline{(b)}
\end{minipage}
\caption{(a) TV and (b) standard deviation (LDPE) as a function of $N$ after 12 feasibility-seeking iterations for the experimental \CTP data set using OTVS and NTVS (TV reduction requirement excluded) with $\lambda=0.0001$ and varying $\alpha$.}
\label{fig:TVSDvAExper}
\end{figure}
Increasing $\alpha$ produces larger perturbations and results in the perturbation magnitude $\beta_k$ converging to zero more slowly.  Thus, one can expect a larger reduction of TV and standard deviations for larger values of $\alpha$. Figures~\ref{fig:TVSDvAExper}(a) and (b), which show plots of TV and standard deviation as a function of $N$ for $\lambda=0.0001$ and with the TV reduction requirement excluded for reconstructions of the experimental data set, demonstrate this effect. Although these figures do not exhibit as strong of a dependence on $\alpha$ or $N$ as was observed with the simulated data, they do indicate the same general trend of decreasing TV and standard deviation as $\alpha$ increases, with the difference growing increasingly larger as $\alpha$ increases in steps of 0.1.
%
\begin{figure}[h!]
    \centering
    \centerline{\includegraphics[width=0.7\linewidth]{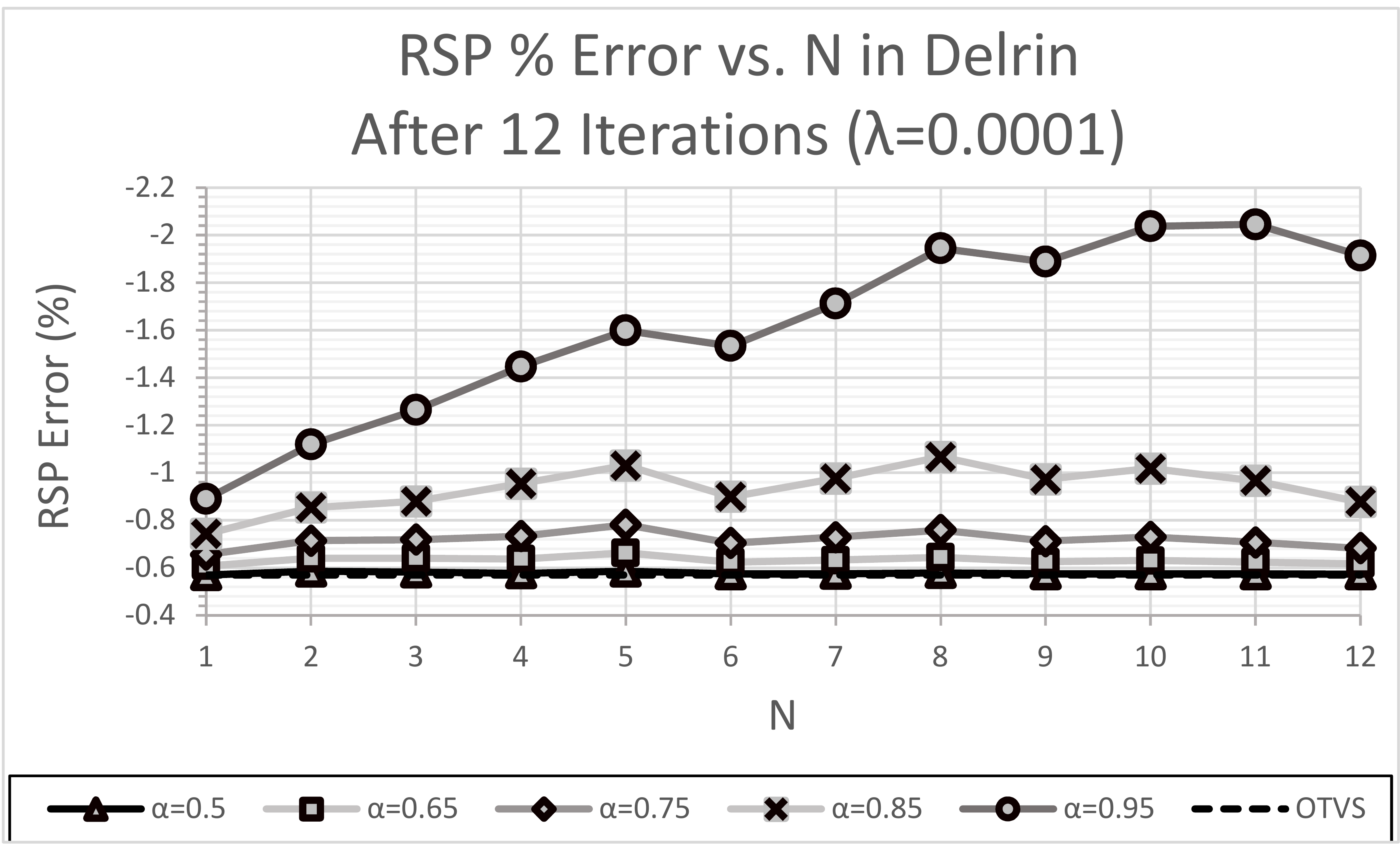}}
    \vspace{0.5mm}
    \caption{ RSP error (Delrin) for each value of $\alpha$ as a function of $N$ after 12 feasibility-seeking iterations for the experimental \CTP data set using OTVS and NTVS (TV reduction requirement excluded) with $\lambda=0.0001$.}
    \label{fig:RSPevAExper}
\end{figure}
Notice that, as was observed in Figures~\ref{fig:RSPevASim}(a)~and~(b) for the simulated data, increasing $\alpha$ beyond $\alpha\approx 0.75$ begins to have a significant impact on the reconstructed RSP and, hence, the RSP error. The direction in which the reconstructed RSP is driven (i.e. increases/decreases reconstructed RSP) is unpredictable, as demonstrated by the fact that an increasing $\alpha$ reduced the RSP error in the Delrin insert for the simulated data set, but Figure~\ref{fig:RSPevAExper} shows that an increasing $\alpha$ increased the RSP error in this insert in the case of the experimental data. In fact, for the experimental data set, increasing $\alpha>0.75$ resulted in an increase in RSP error within every cylindrical material insert.

\subsubsection{Relaxation Parameter ($\lambda$)}\label{sssec:rLExper}
\begin{figure}[h!]
\begin{minipage}{\linewidth}
  \centering
     \centerline{\includegraphics[width=0.65\linewidth]{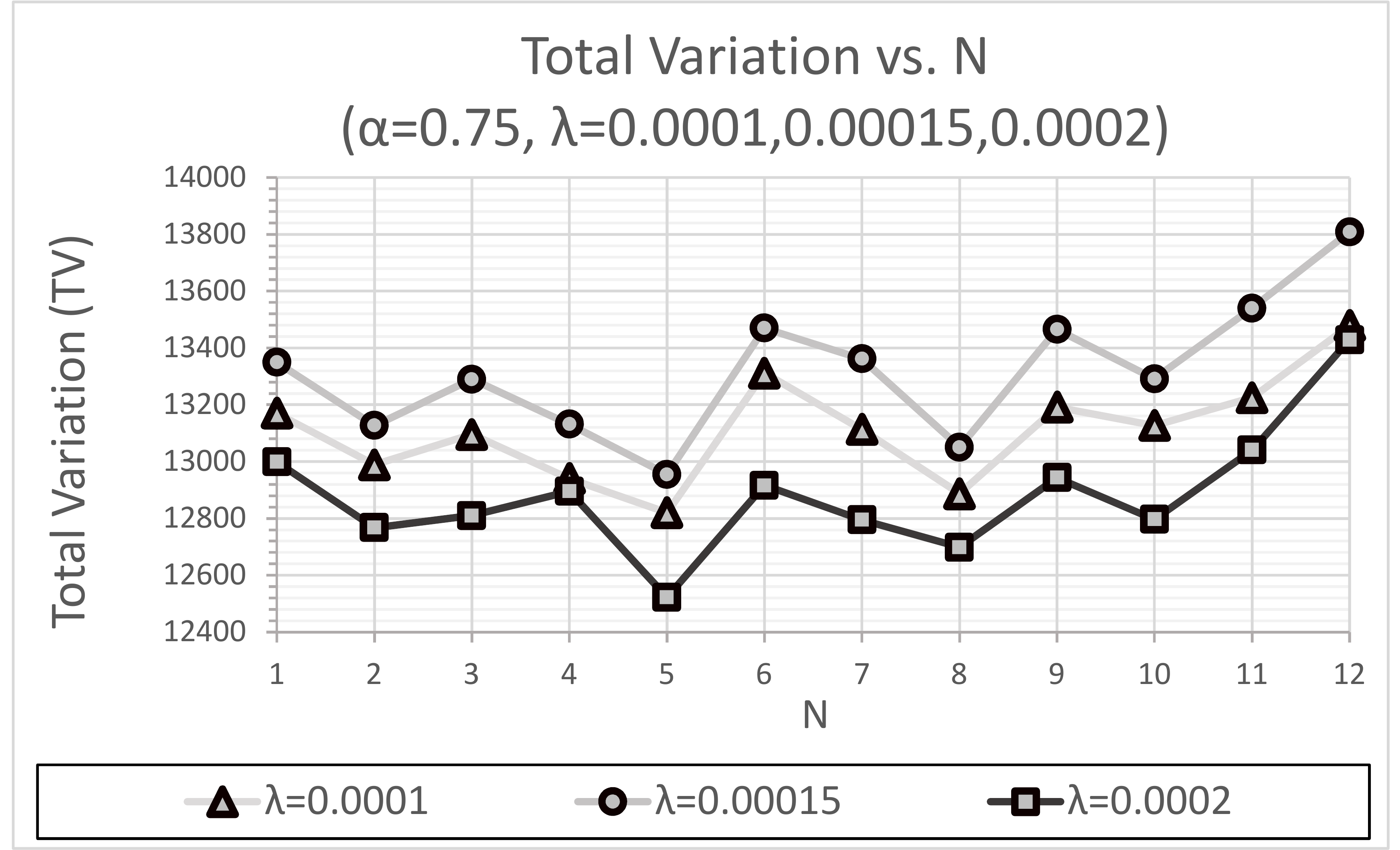}}
    \vspace{0.5mm}
\vfill\centerline{(a)}
\end{minipage}
\vfill
\vspace{2mm}
\begin{minipage}{\linewidth}
  \centering
    \centerline{\includegraphics[width=0.65\linewidth]{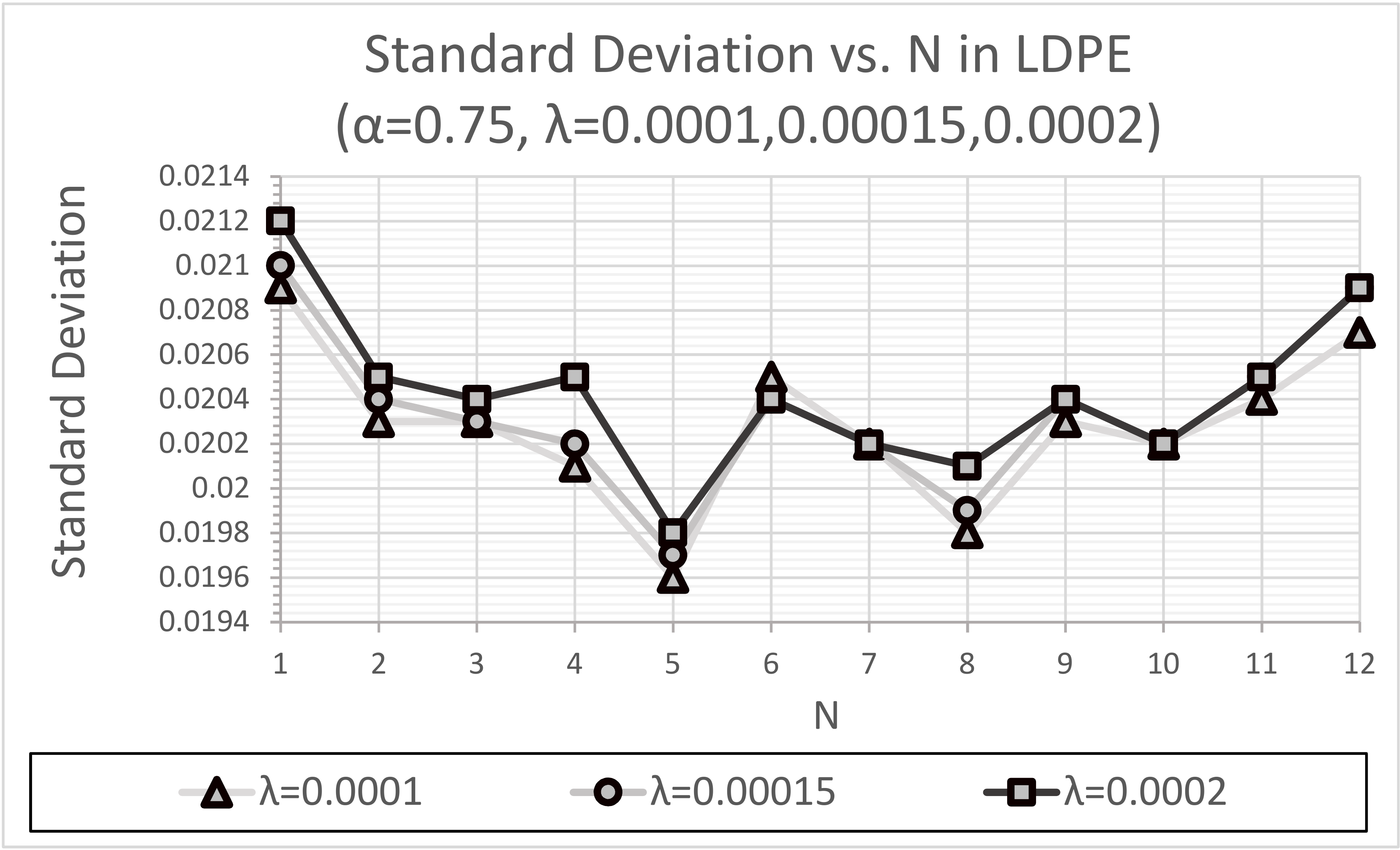}}
    \vspace{0.5mm}
\vfill\centerline{(b)}
\end{minipage}
\caption{(a) TV and (b) standard deviation (soft tissue) as a function of $N$ for $\lambda=0.0001$, $k=12$; $\lambda=0.00015$, $k=8$; and $\lambda=0.0002$, $k=6$ iterations, respectively, and $\alpha=0.75$ for the experimental \CTP data set.}
\label{fig:TVSvLoptimalExper}
\end{figure}
Comparisons of TV and standard deviation as a function of $N$ for varying relaxation parameter $\lambda$ are shown in Figures~\ref{fig:TVSvLoptimalExper}(a) and (b) for $\alpha=0.75$. As with the simulated data, the results for $\lambda=0.0001$ after $k=12$ feasibility-seeking iterations are shown and the number of iterations $k$ was chosen for $\lambda=0.00015,0.0002$ such that these reconstructions had converged to the same point (i.e., reached approximately the same RSP); this occurred at $k=8$ for $\lambda=0.00015$ and $k=6$ for $\lambda=0.0002$ for the experimental data as it did for the simulated data.

Figure~\ref{fig:TVSvLoptimalExper}(a) indicates that, for each value of $N$, increasing $\lambda$ results in a reduction in TV. This trend demonstrates the benefit of performing reconstruction with as few iterations as is necessary to obtain an acceptable level of convergence since a side effect of feasibility-seeking is a consistent increase in TV. On the other hand, the plot of standard deviation in Figure~\ref{fig:TVSvLoptimalExper}(b) indicates that, unlike with the simulated data, an increase in $\lambda$ results in a slight increase in standard deviation in the LDPE insert. For both TV and standard deviation, optimal results were obtained with $N=5$ for each value of $\lambda$ and within each material insert, with nearly identical behavior as a function of both $N$ and $\lambda$ seen in each insert.

\subsection{Experimental HN715 Pediatric Head Phantom Data Set}\label{sssec:rNHead}
The experimentally acquired data for the pediatric head phantom was reconstructed using the same set of parameter value combinations as those used to reconstruct the simulated and experimental \CTP phantom data sets. This phantom provides a considerably different material composition and internal structure to determine the impact these properties have on the behavior of the NTVS algorithm and the combination of parameter values that produce maximal benefit.
\subsubsection{Number of TVS steps ($N$)}\label{sssec:rNHead}
\begin{figure}[h!]
\begin{minipage}[t]{0.48\linewidth}
  \centering
    \centerline{\includegraphics[width=\linewidth]{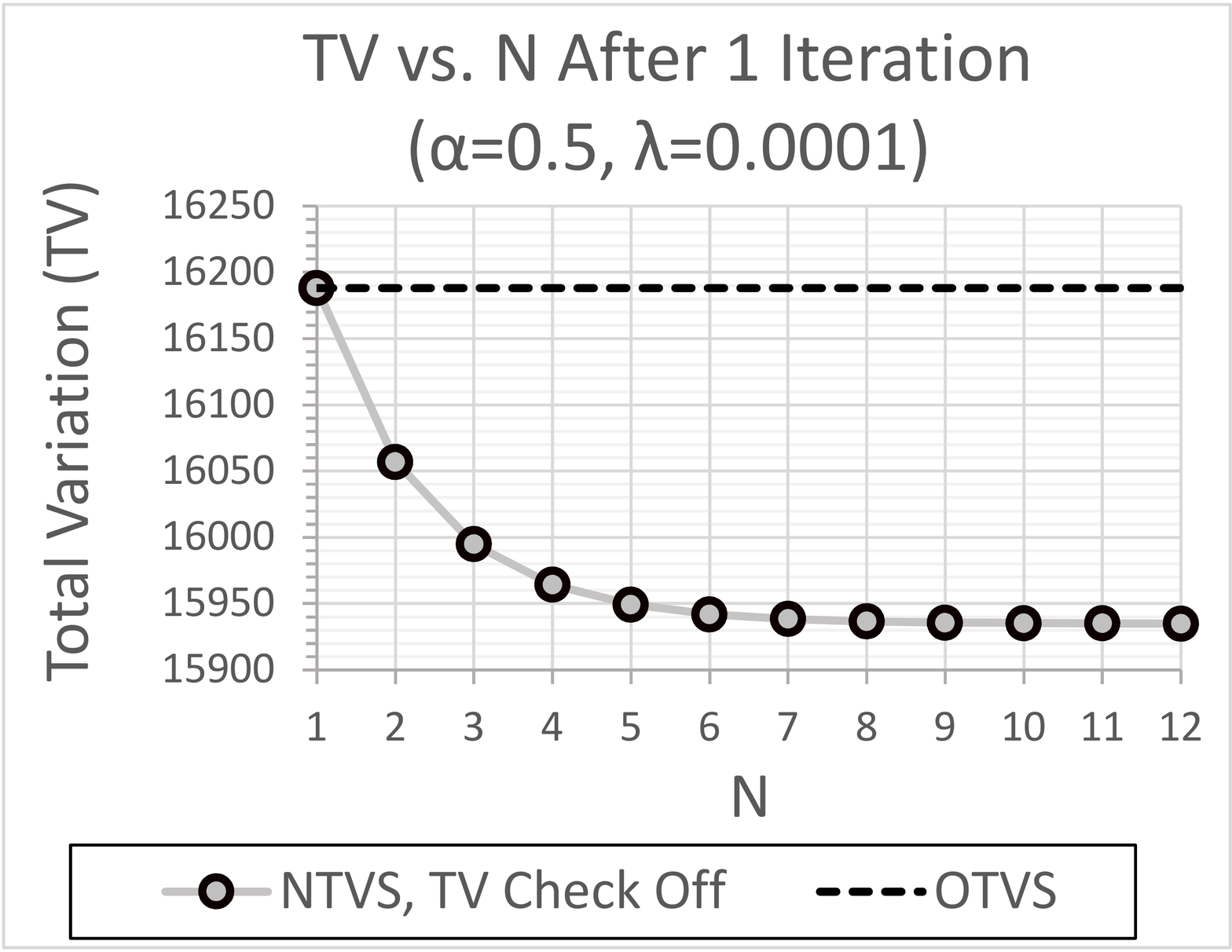}}
\end{minipage}
\hfill
\begin{minipage}[t]{0.48\linewidth}
  \centering
     \centerline{\includegraphics[width=\linewidth]{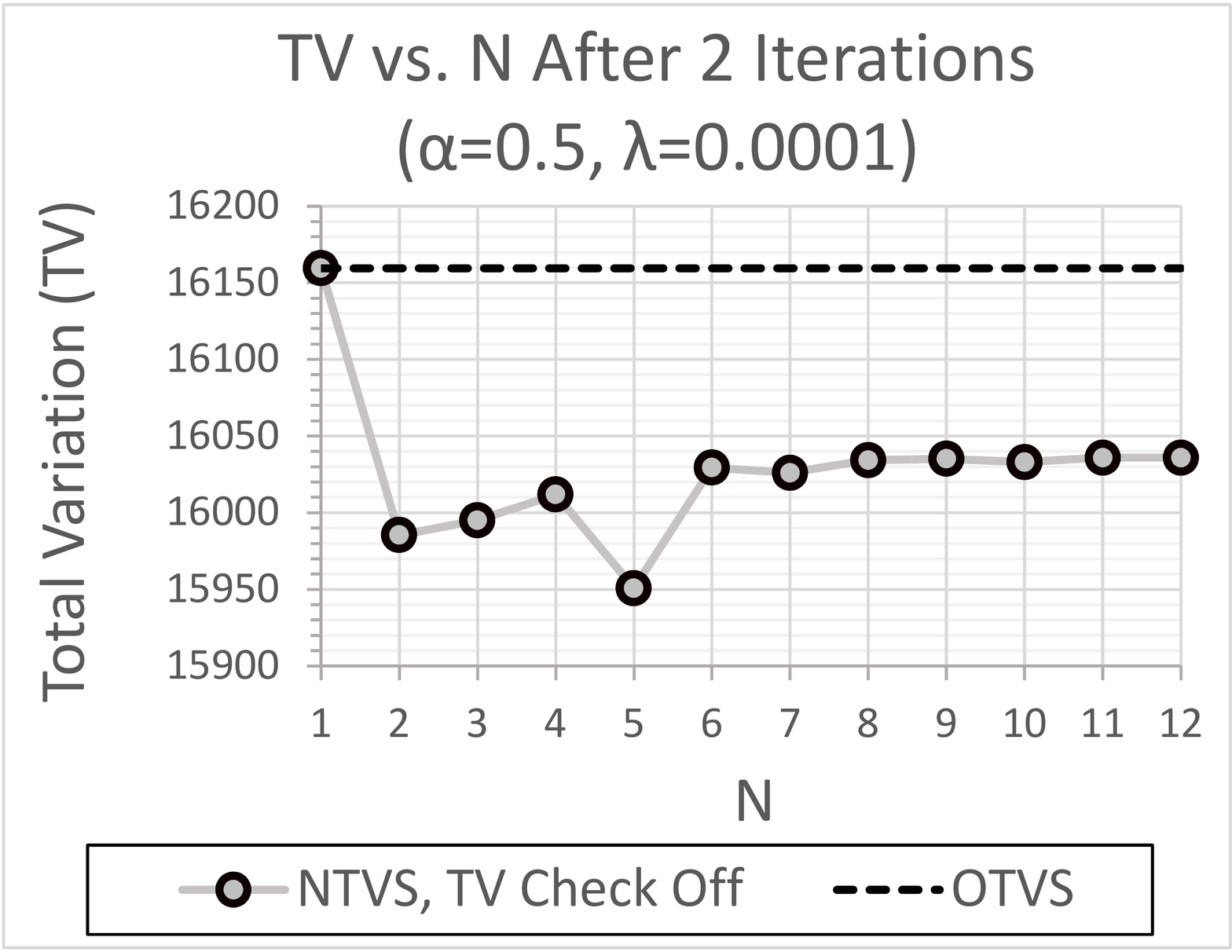}}
\end{minipage}
\vfill
\begin{minipage}[t]{0.48\linewidth}
  \centering
    \centerline{\includegraphics[width=\linewidth]{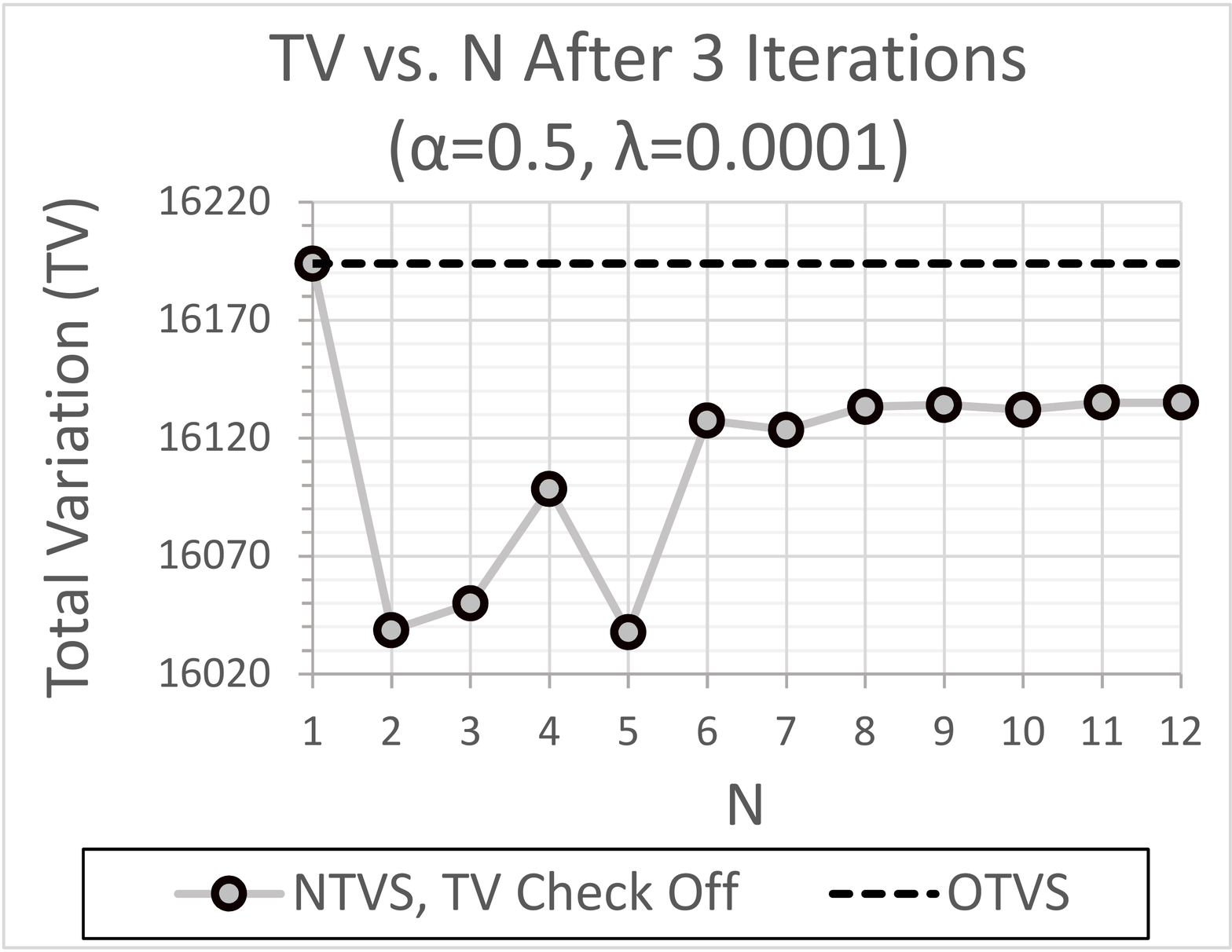}}
\end{minipage}
\hfill
\begin{minipage}[t]{0.48\linewidth}
  \centering
     \centerline{\includegraphics[width=\linewidth]{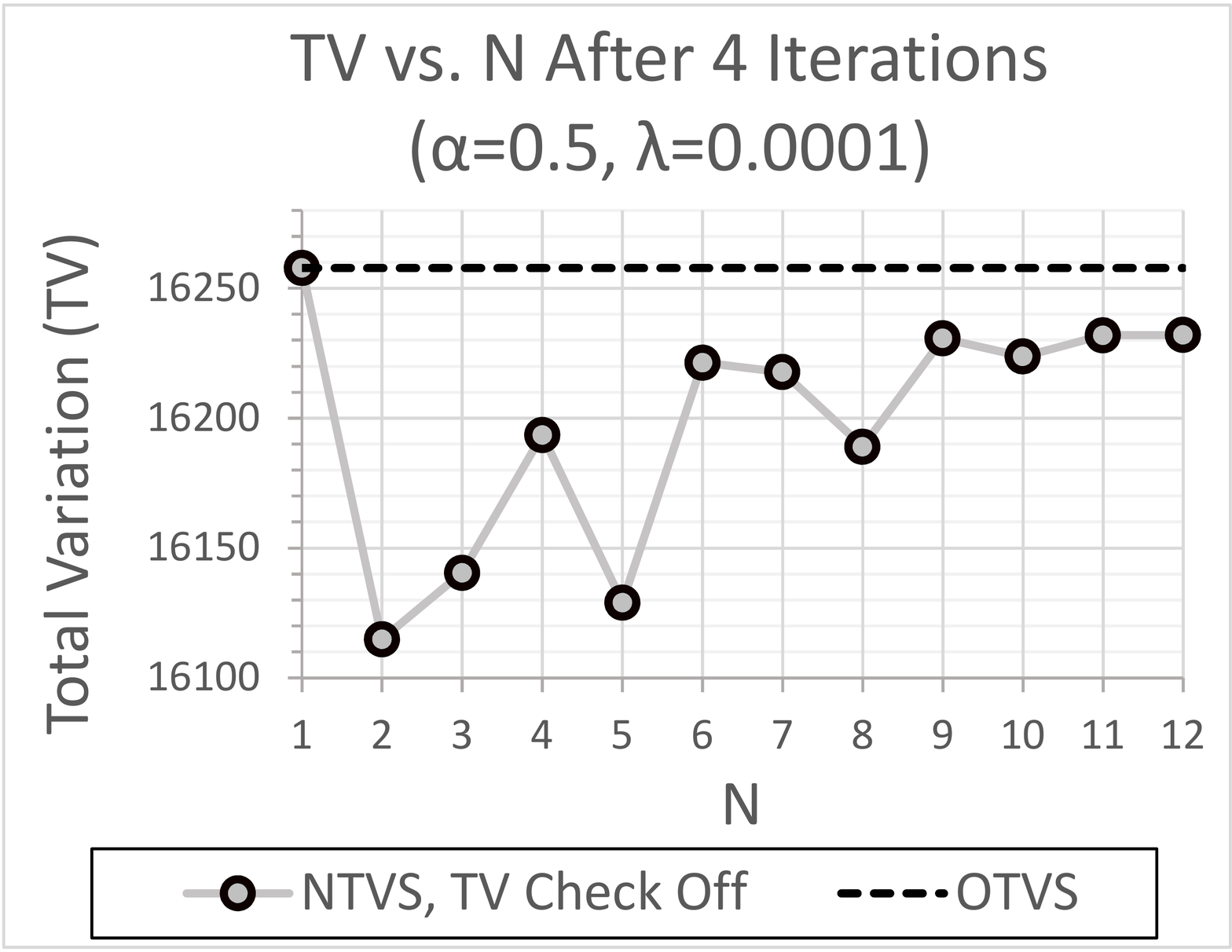}}
\end{minipage}
\caption{TV as a function of $N$ after each of the first 4 feasibility-seeking iterations for the experimental HN715 data set using OTVS and NTVS (TV reduction requirement excluded) with $\lambda=0.0001$ and $\alpha=0.5$ .}
\label{fig:TVkHead}
\end{figure}
Figure~\ref{fig:TVkHead} shows plots of TV as a function of $N$ for the first four feasibility-seeking iterations for the case where $\lambda=0.0001$ and the TV reduction requirement is excluded. These results are very similar to those of the experimental \CTP phantom, particularly for $N\le 6$, but unlike for both the simulated and experimental \CTP data sets, the benefits of NTVS do not degrade as quickly for $N\ge 7$ and continue to outperform OTVS for all values of $N$. However, the optimal values of $N$ after 4 feasibility-seeking iterations occur at $N=2$ and $N=5$ for all 3 data sets.

As previously noted, repeated reconstructions with the same value of $N$ yield variations in TV and standard deviation.
Again, the difference in TV and standard deviation as a function of $N$ is seen to be a property of the algorithm and its relationship with feasibility-seeking and not the result of the random variations arising from random increases in $\ell_k$. The objectives of feasibility-seeking and TVS are somewhat opposed; feasibility-seeking tends to amplify noise, thereby increasing TV, while each TVS perturbation may drive the solution to a more or less feasible solution. The resulting push back and forth begins to produce small differences in TV between successive values of $N$ after the first two feasibility-seeking iterations and these subsequently increase as each additional feasibility-seeking iteration amplifies the resulting differences. Simultaneously, TV perturbations and updates applied in feasibility-seeking both decrease in magnitude as $k$ increases, diminishing their ability to counteract the impact of a previous, less optimal solution. Hence, a solution that is less optimal after the first few iterations will rarely overcome its performance deficit and will far more often become increasingly suboptimal, particularly if parameter values are held fixed and not adapted based on performance as in the present case. Hence, values of $N$ that yield a larger reduction in TV early in reconstruction also experience a lesser amplification of noise at each feasibility-seeking iteration, resulting in a compounding effect that accounts for the relatively large differences in TV between similar values of $N$.

As can be seen in Figure~\ref{fig:TVSDvCheckHead}, showing the TV and standard deviation within the soft tissue ROI as a function of $N$ after all 12 feasibility-seeking iterations for $\lambda=0.0001$ and $\alpha=0.5$, NTVS including and excluding the TV reduction requirement both yield larger reductions in TV and standard deviation for every value of $N$ except for the slight increase in TV obtained with $N=12$. Repeating these reconstructions with $\alpha=0.75$ consistently yields images with significantly larger reductions in both TV and standard deviation for every value of $N$, with similar standard deviation results obtained for every material ROI. These results also demonstrate that the smallest reductions in TV and standard deviation obtained with $N=1$ and $N=12$ were approximately 50\% larger than the largest difference between varying values of $N$ and more than twice as large for $3\le N\le 6$.
\subsubsection{Inclusion/Exclusion of TV Reduction Requirement}\label{sssec:rTVHead}
\begin{figure}[h!]
\begin{minipage}{\linewidth}
  \centering
    \centerline{\includegraphics[width=0.65\linewidth]{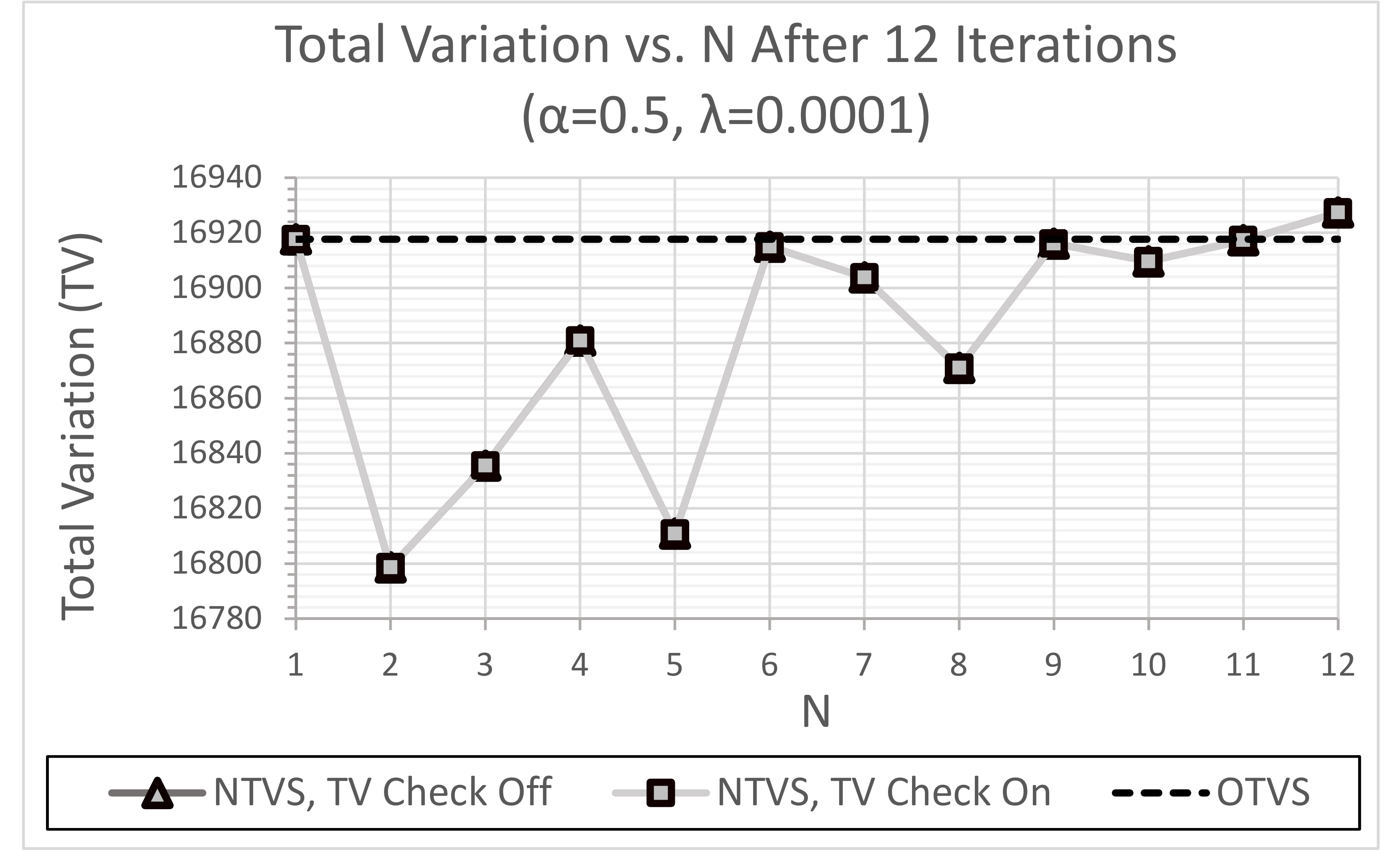}}
    \centerline{(a)}
\end{minipage}
\vfill
\vspace{2mm}
\begin{minipage}{\linewidth}
  \centering
  \centerline{\includegraphics[width=0.65\linewidth]{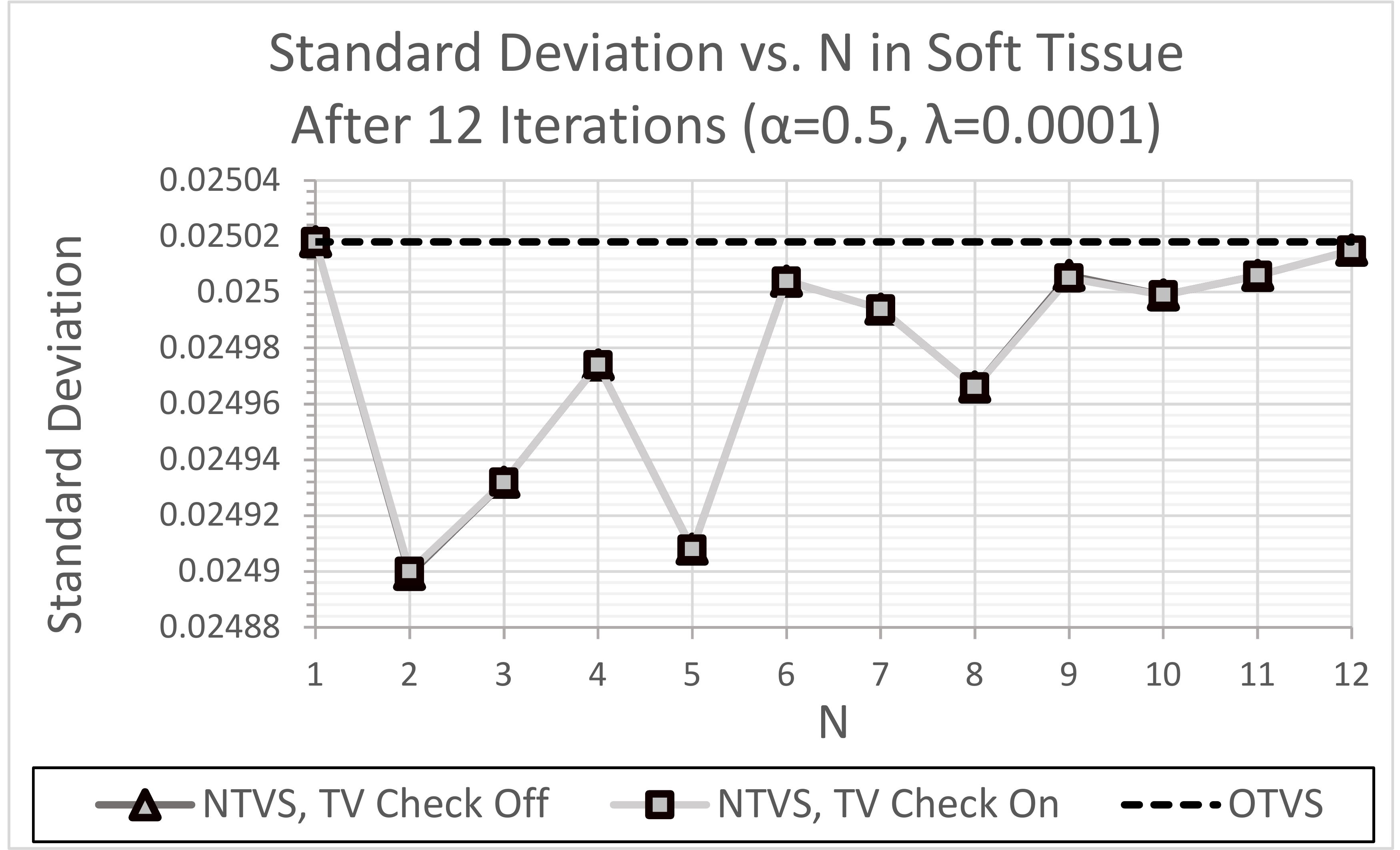}}
    \centerline{(b)}
\end{minipage}
\caption{(a) TV and (b) standard deviation as a function of $N$ after 12 feasibility-seeking iterations for the experimental HN715 data set using the OTVS algorithm and the NTVS algorithm including and excluding the TV reduction requirement with $\lambda=0.0001$ and $\alpha=0.5$ (note that the 2 NTVS curves overlap).}
\label{fig:TVSDvCheckHead}
\end{figure}
\begin{figure}[h!]
\begin{minipage}{\linewidth}
  \centering
    \centerline{\includegraphics[width=0.65\linewidth]{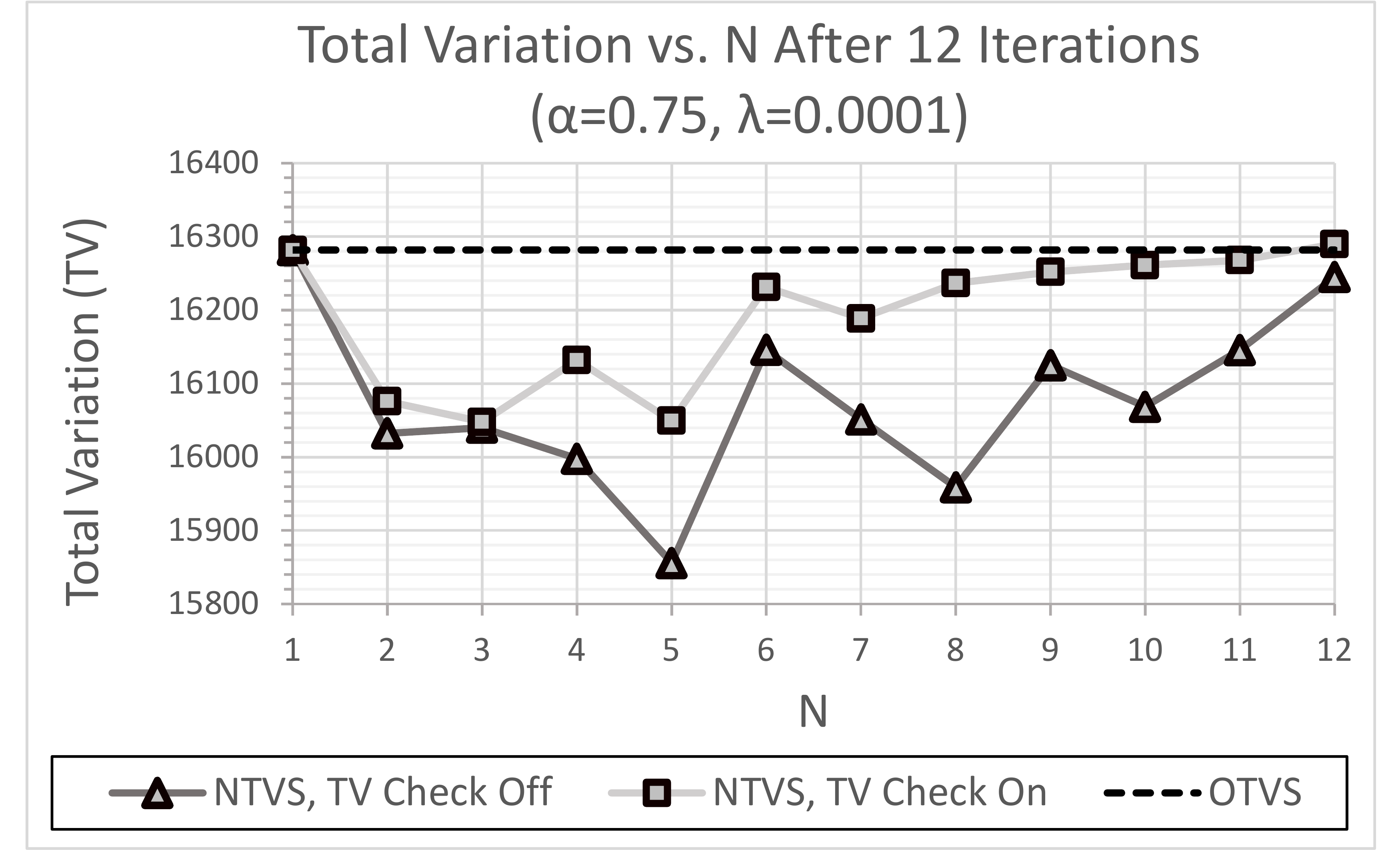}}
    \centerline{(a)}
\end{minipage}
\vfill
\vspace{2mm}
\begin{minipage}{\linewidth}
  \centering
  \centerline{\includegraphics[width=0.65\linewidth]{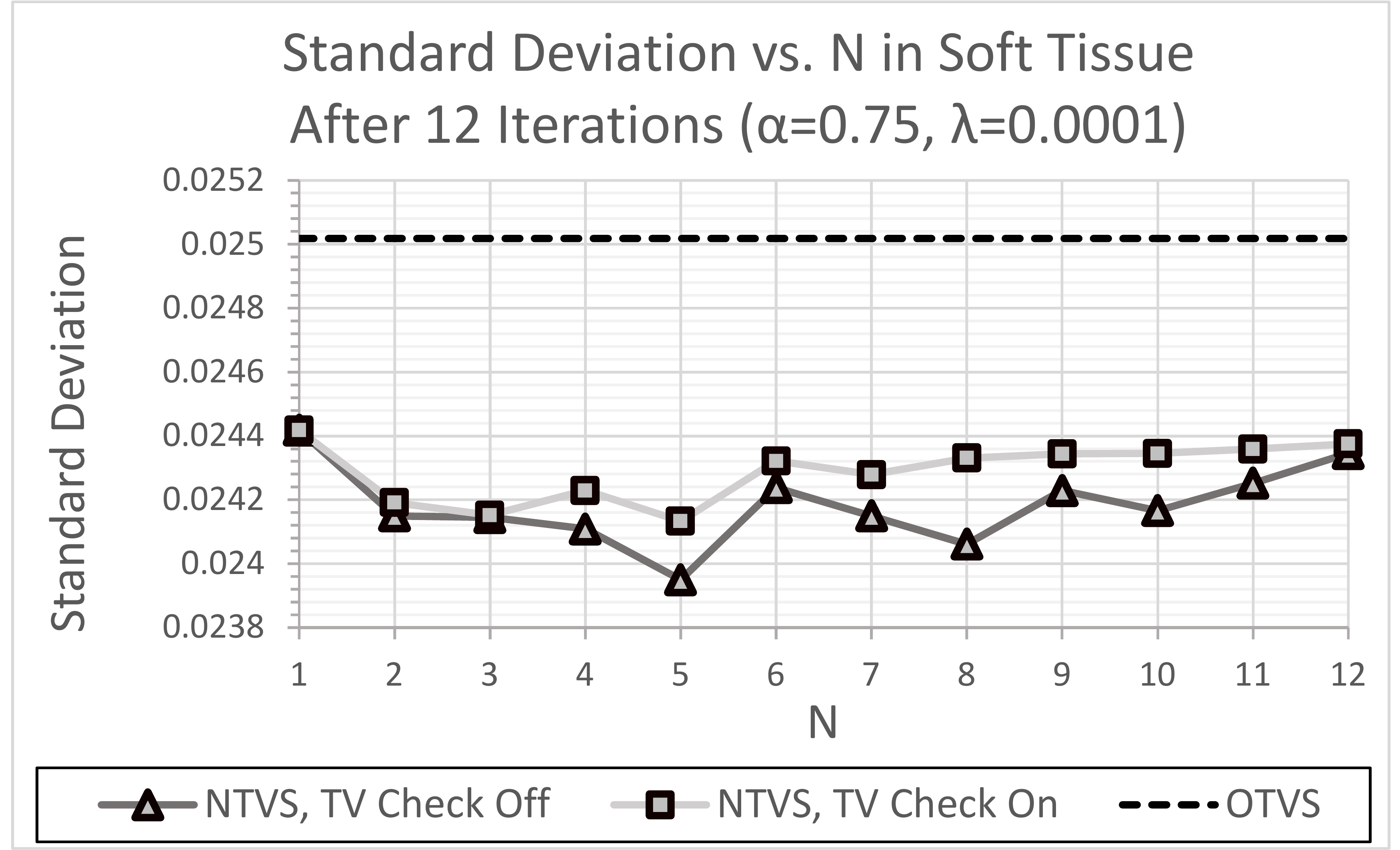}}
    \centerline{(b)}
\end{minipage}
\caption{(a) TV and (b) standard deviation as a function of $N$ after 12 feasibility-seeking iterations for the experimental HN715 data set using the OTVS algorithm and the NTVS algorithm including and excluding the TV reduction requirement with $\lambda=0.0001$ and $\alpha=0.75$.}
\label{fig:TVSDvCheckHead2}
\end{figure}
Comparisons of TV and standard deviation as a function of $N$ after 12 feasibility-seeking iterations are shown for OTVS and NTVS including and excluding the TV reduction requirement in Figures~\ref{fig:TVSDvCheckHead}(a) and (b), respectively. These are shown for $\lambda=0.0001$ and $\alpha=0.5$, which makes the reduction in $\beta_k$ with NTVS equivalent to that of OTVS. As with the experimental \CTP data set, the difference in TV and standard deviation between the results with and without the TV reduction requirement were not discernable for $\alpha=0.5$ and, again, independent of the value of $\lambda$ and material of the ROI. On the other hand, for $\alpha=0.75$, exclusion of the TV reduction requirement consistently yielded a larger reduction in TV and standard deviation for each value of $N$, as seen in Figures~\ref{fig:TVSDvCheckHead2}(a) and (b), respectively. Again, this was seen for all $\lambda$ and, in the case of the standard deviation, within the ROI of each material. As with the previous data sets, all subsequent analyses for this data set were performed using the NTVS algorithm with the TV reduction requirement excluded (as defined in Appendix~\ref{app:ntvs}).
\subsubsection{Perturbation Kernel ($\alpha$)}\label{sssec:rAHead}
\begin{figure}[h!]
\begin{minipage}[t]{\linewidth}
  \centering
     \centerline{\includegraphics[width=0.7\linewidth]{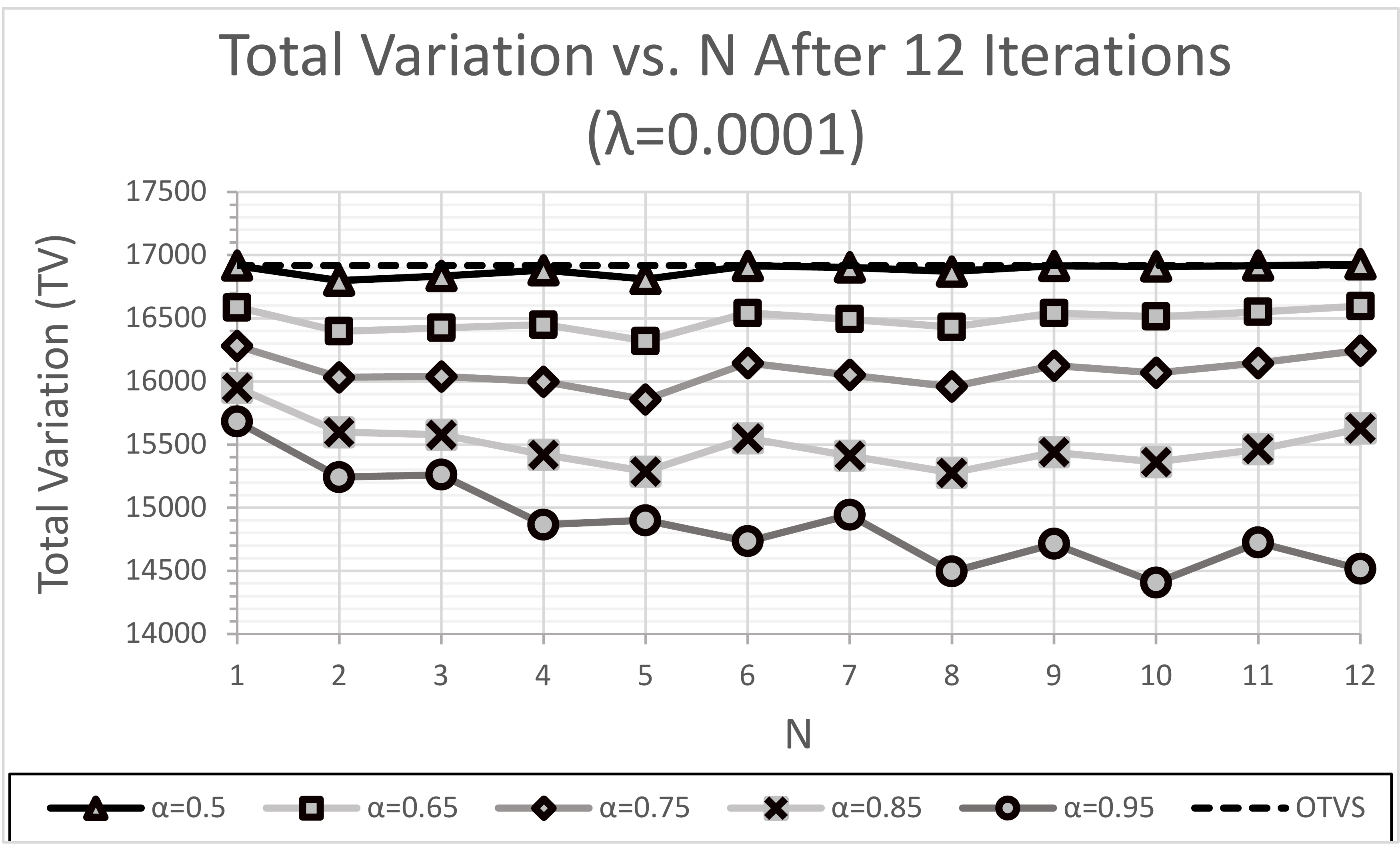}}
        \vspace{1mm}
\vfill\centerline{(a)}
\end{minipage}
\vfill
\vspace{2mm}
\begin{minipage}[t]{\linewidth}
  \centering
   \centerline{\includegraphics[width=0.7\linewidth]{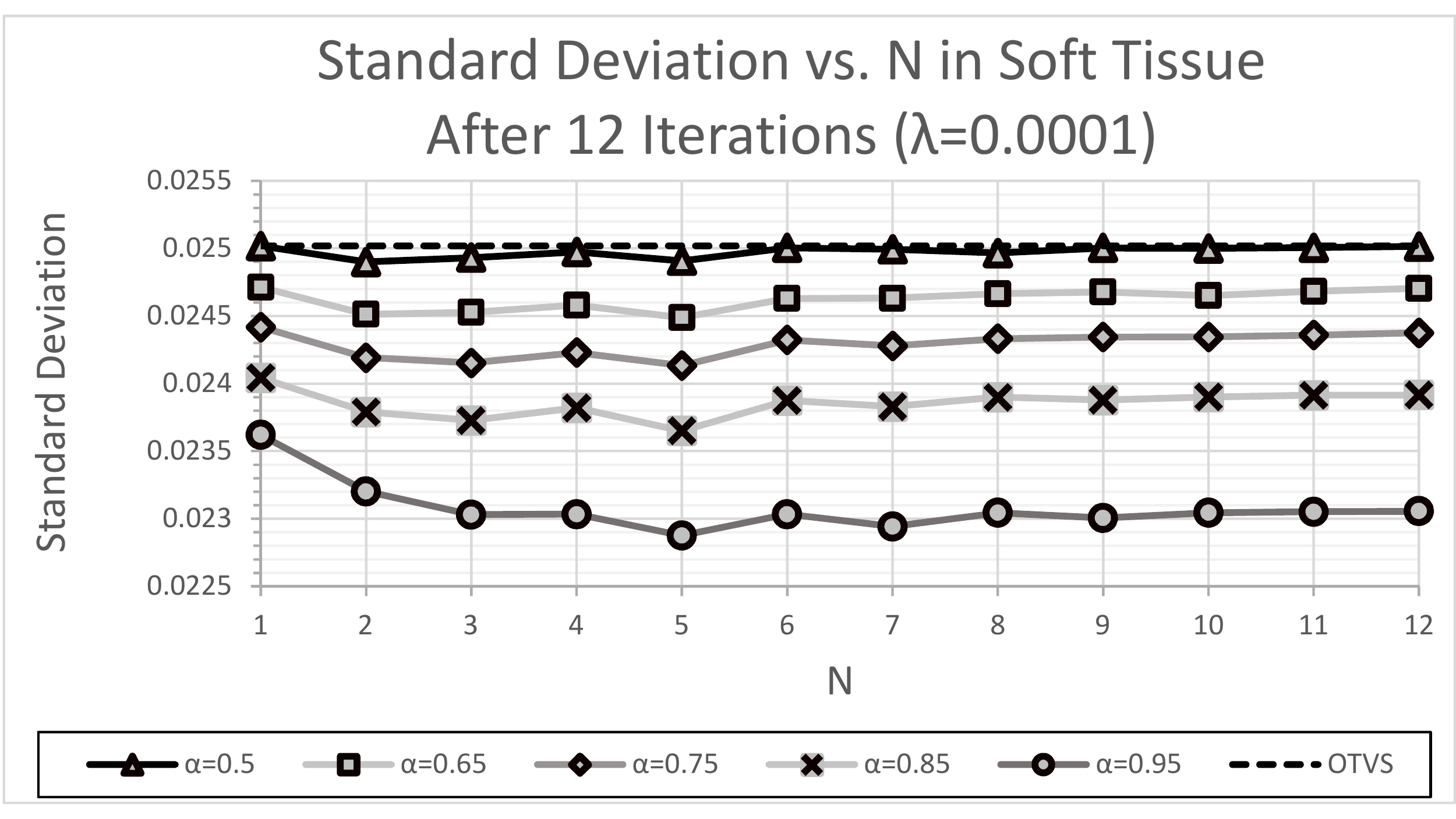}}
        \vspace{1mm}
\vfill\centerline{(b)}
        \vspace{1mm}
\end{minipage}
\caption{(a) TV and (b) standard deviation (soft tissue) as a function of $N$ after 12 feasibility-seeking iterations for the experimental HN715 data set using OTVS and NTVS (TV reduction requirement excluded) with $\lambda=0.0001$ and varying $\alpha$.}
\label{fig:TVSDvAHead}
\end{figure}
\begin{figure}[h!]
\begin{minipage}{\linewidth}
  \centering
   \centerline{\includegraphics[width=0.7\linewidth]{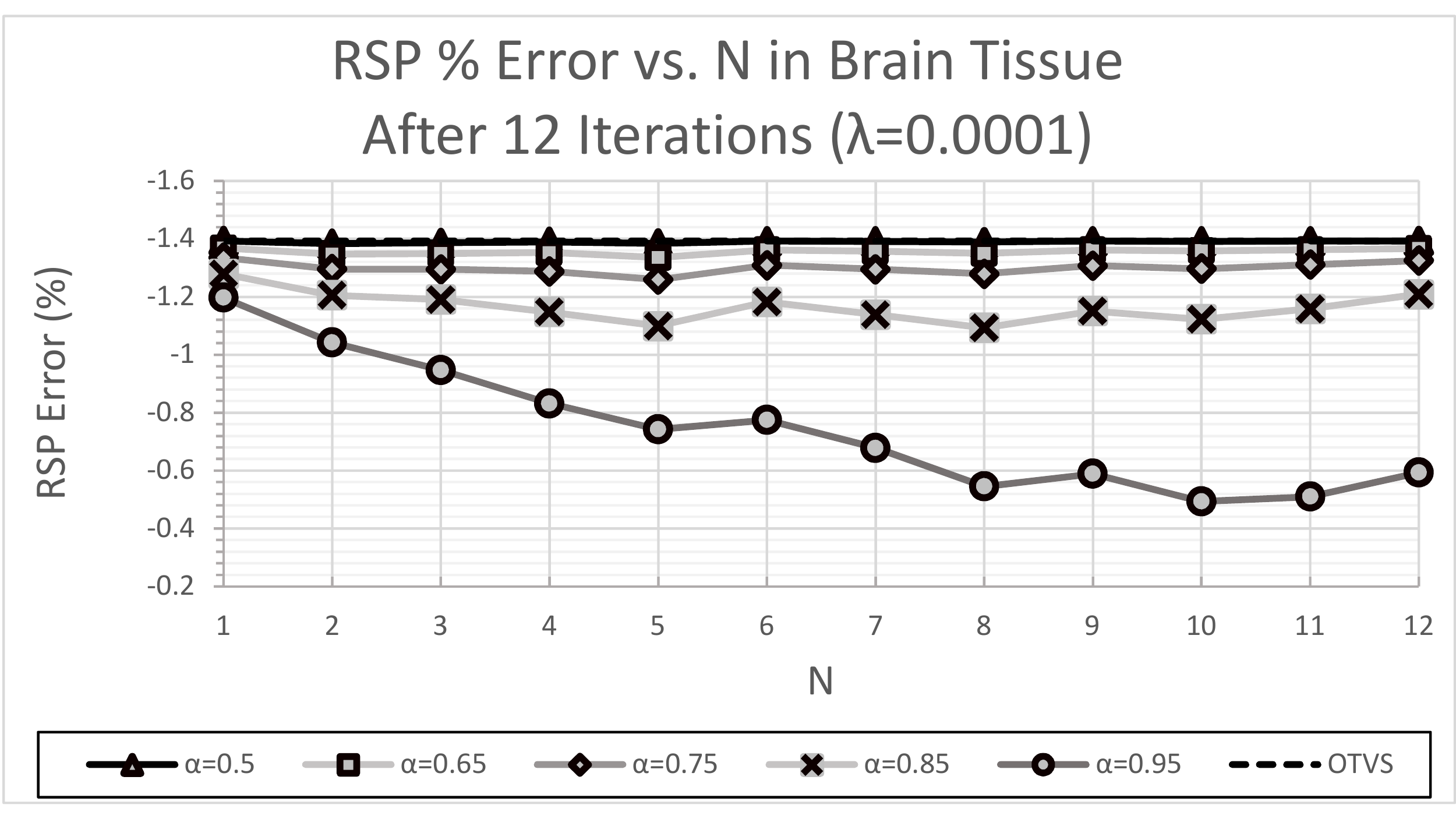}}
            \vspace{0.5mm}
\vfill\centerline{(b)}
\end{minipage}
\vfill
\vspace{2mm}
\begin{minipage}{\linewidth}
  \centering
   \centerline{\includegraphics[width=0.7\linewidth]{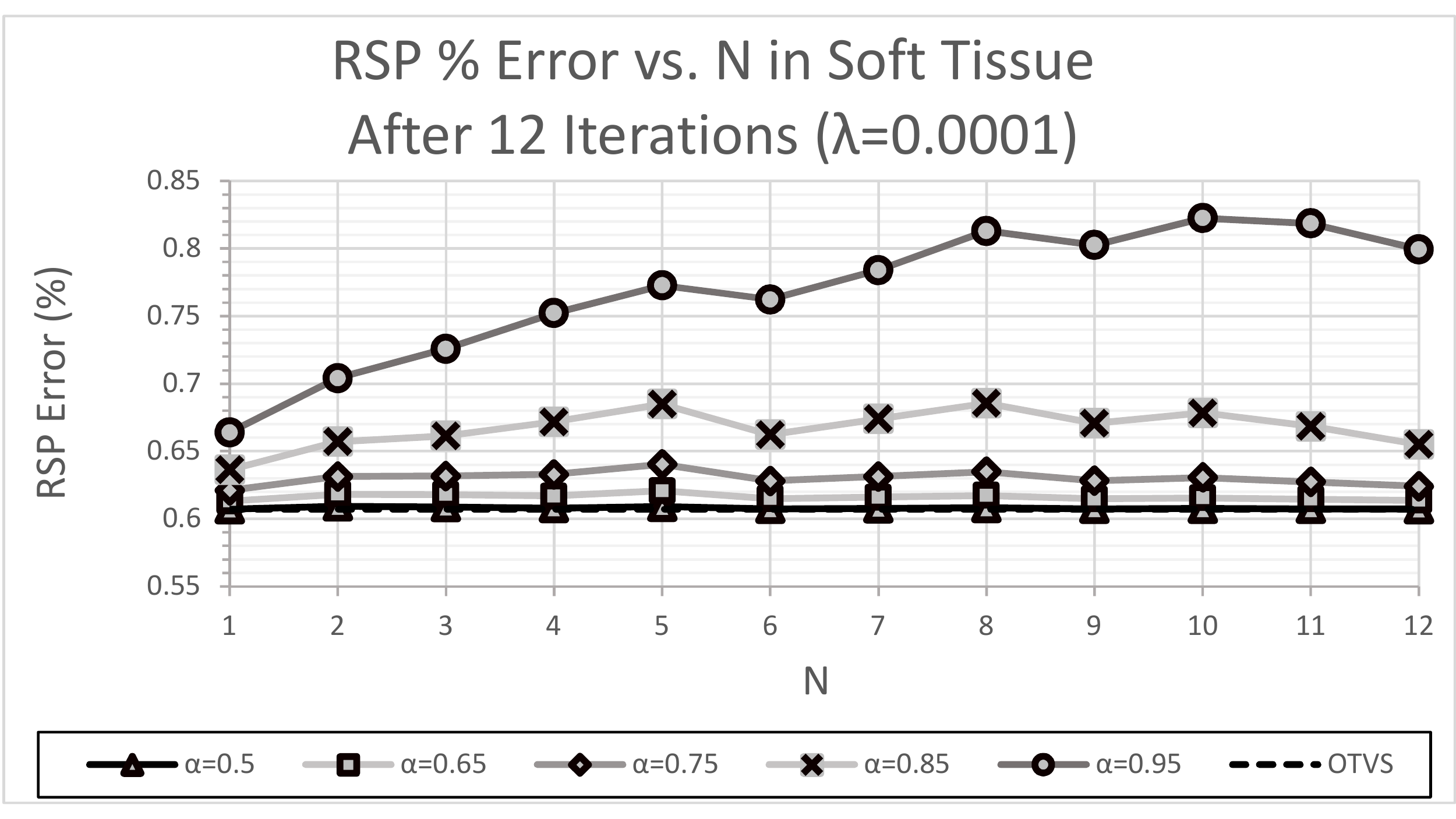}}
            \vspace{0.5mm}
\vfill\centerline{(b)}
\end{minipage}
\caption{RSP error in the (a) brain tissue and (b) soft tissue ROIs as a function of $N$ after 12 feasibility-seeking iterations for the experimental HN715 data set using OTVS and NTVS (TV reduction requirement excluded) with $\lambda=0.0001$ and varying $\alpha$.}
\label{fig:RSPevAHead}
\end{figure}
Plots of TV and standard deviation in the ROI of soft tissue ROI a function of $N$ for $\lambda=0.0001$ and with the TV reduction requirement excluded are shown for each value of $\alpha$ in Figure~\ref{fig:TVSDvAHead} for the HN715 data set; as with the other data sets, the standard deviation results for the ROIs of the other materials showed a similar trend as a function of $\alpha$. As with the simulated and experimental \CTP data sets, TV and standard deviation decreased as $\alpha$ increased, but the standard deviation was less sensitive to the value of $N$ than observed with the \CTP data sets.

Figure~\ref{fig:RSPevAHead} once again demonstrates the impact that values of $\alpha>0.75$ had on reconstructed RSP error within the different materials inserts. Unlike the RSP reconstructed from the experimental \CTP data set, the RSP reconstructed from the experimental HN715 data set was driven in unpredictable directions depending on the particular material insert (as it was with the simulated \CTP data set), improving accuracy within some material inserts while decreasing accuracy in others.
\subsubsection{Relaxation Parameter ($\lambda$)}\label{sssec:rLHead}
\begin{figure}[h!]
\begin{minipage}{\linewidth}
  \centering
     \centerline{\includegraphics[width=0.65\linewidth]{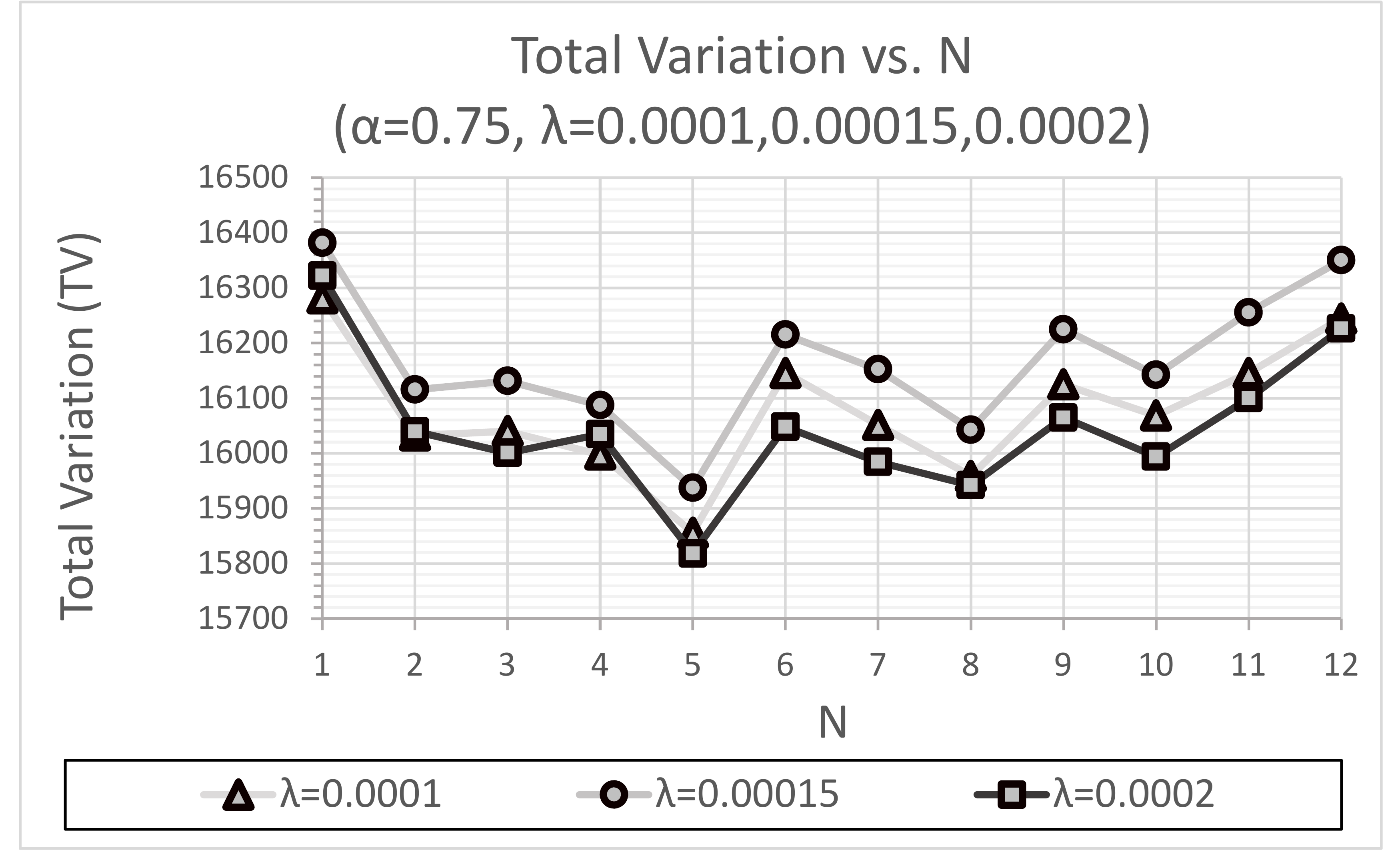}}
    \vspace{0.5mm}
\vfill\centerline{(a)}
\end{minipage}
\vfill
\vspace{2mm}
\begin{minipage}{\linewidth}
  \centering
    \centerline{\includegraphics[width=0.65\linewidth]{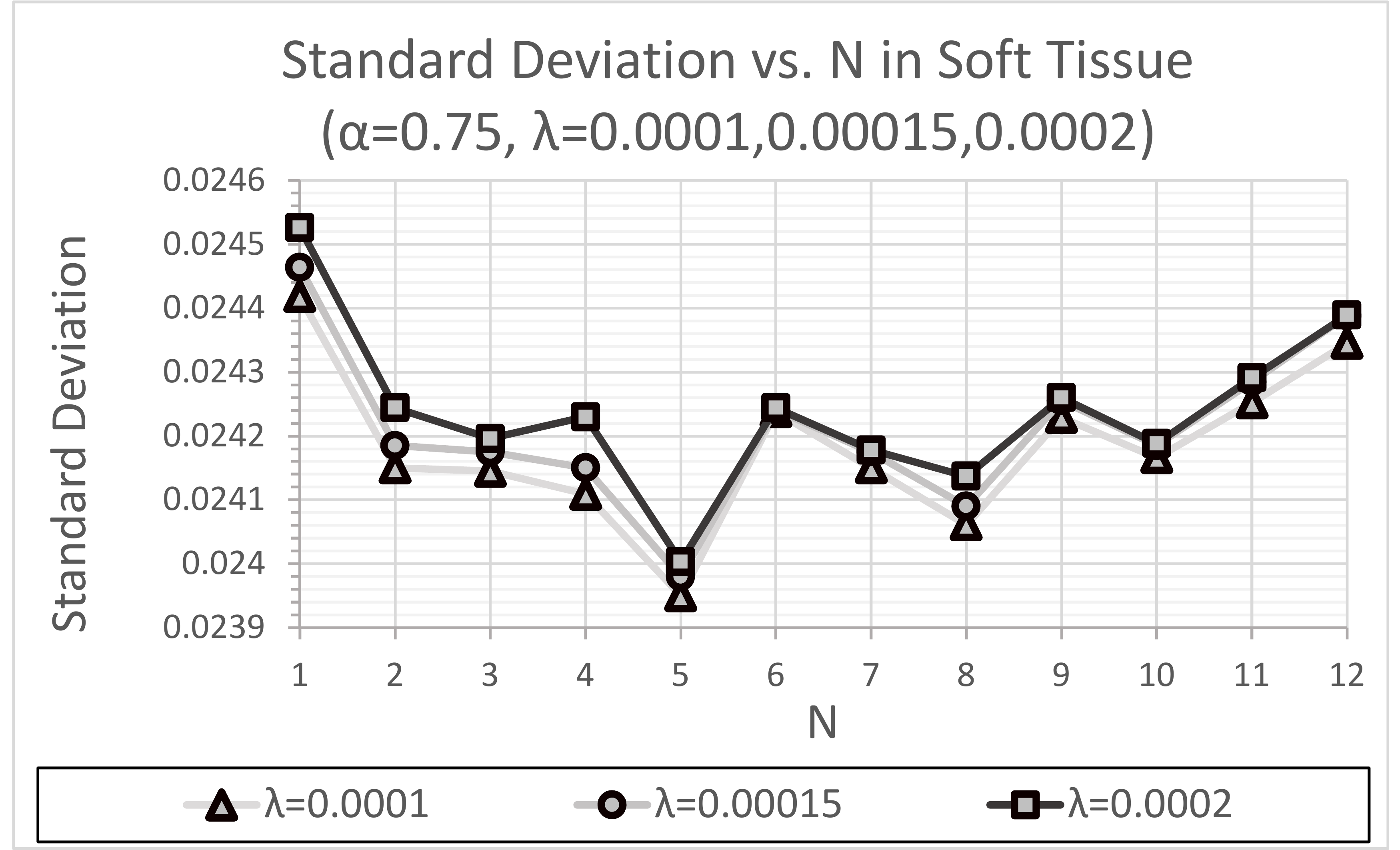}}
    \vspace{0.5mm}
\vfill\centerline{(b)}
\end{minipage}
\caption{(a) TV and (b) standard deviation (soft tissue) as a function of $N$ for $\lambda=0.0001$, $k=12$; $\lambda=0.00015$, $k=8$; and $\lambda=0.0002$, $k=6$ iterations, respectively, and for $\alpha=0.75$ for the experimental HN715 data set.}
\label{fig:TVSvLoptimalHead}
\end{figure}
Figures~\ref{fig:TVSvLoptimalHead}(a) and (b) shows comparisons of TV and standard deviation within the ROI of soft tissue, respectively, as a function of $N$ for varying relaxation parameter $\lambda$ with $\alpha=0.75$ and excluding the TV reduction requirement. Plots of standard deviation for the ROI of other materials displayed the same dependence on $N$ and $\lambda$. The number of feasibility-seeking iterations $k$ for $\lambda=0.00015$ and $\lambda=0.0002$ were again chosen such that these reconstructions reached the same point in convergence as the $\lambda=0.0001$ reconstructions yielded after 12 feasibility-seeking iterations; this resulted in the same feasibility-seeking iteration numbers $k=8$ and $k=6$ for $\lambda=0.00015$ and $\lambda=0.0002$, respectively, as for the simulated and experimental \CTP data sets.

As was seen with the \CTP data sets, increasing $\lambda$ consistently yielded larger reductions in TV for each value of $N$. The standard deviation obtained within the ROI of soft tissue was similar to those obtained with the experimental \CTP data set, demonstrating a slight increase in standard deviation as $\lambda$ increases for each value of $N$. However, the results for $3\le N\le 6$ are consistently better than those obtained with $N=1$ for NTVS and with OTVS for the less noise sensitive $\lambda=0.0001$ (see Figure~\ref{fig:TVSDvCheckExper}).
\begin{figure}[h!]
    \centering
    \centerline{\includegraphics[width=0.65\linewidth]{{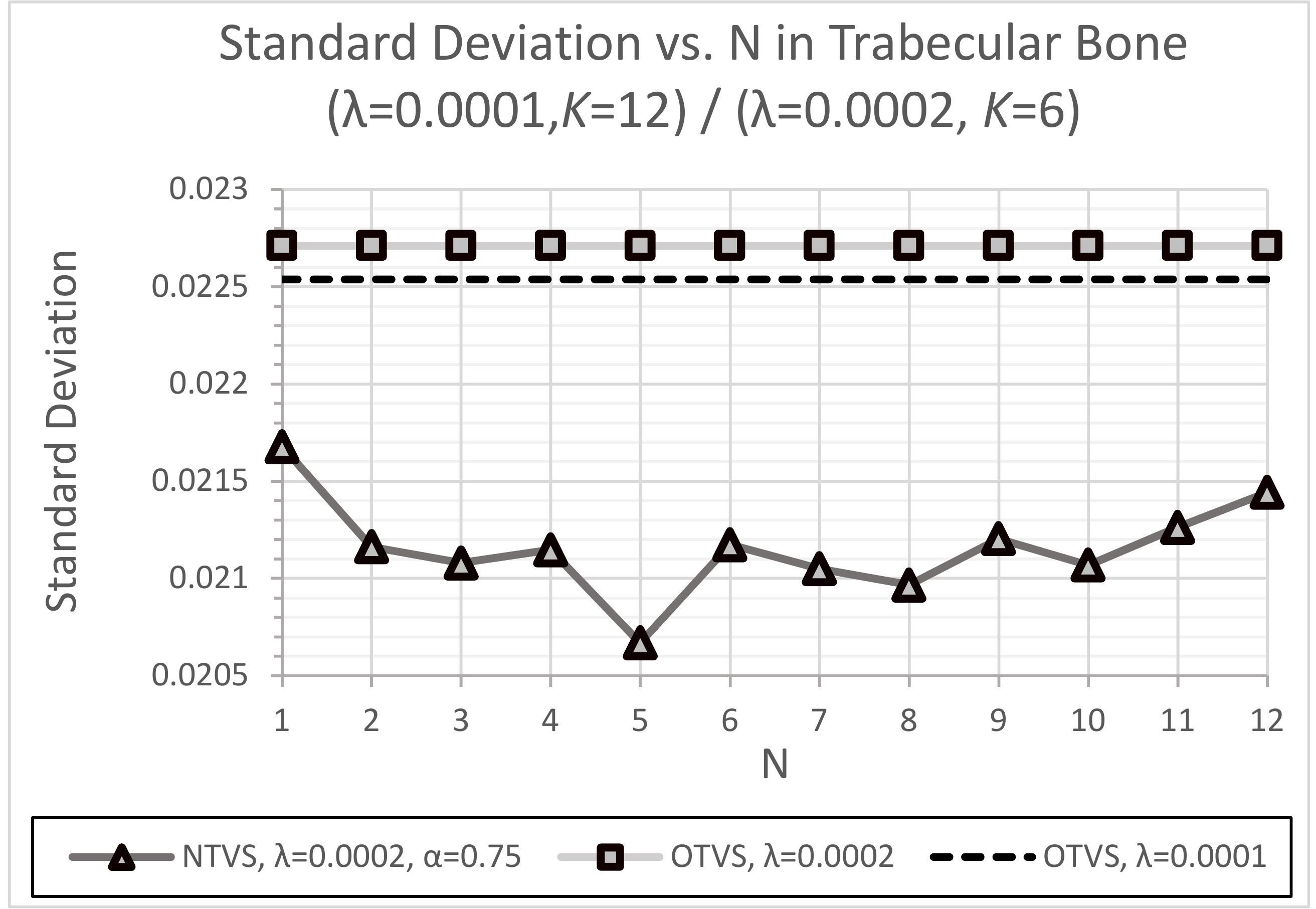}}}
\caption{Standard deviation in the ROI of trabecular bone as a function of $N$ after 6 feasibility-seeking iterations for $\lambda=0.0002$ as compared to OTVS with $\lambda=0.0001$ and $\lambda=0.0002$.}
    \label{fig:TVSDvLcomp}
\end{figure}

The benefit of NTVS in permitting reconstruction with larger $\lambda$ can be seen by considering the plots of standard deviation within the ROI of the trabecular bone for $\lambda=0.0002$ as compared to OTVS for $\lambda=0.0001$ and $\lambda=0.0002$, as shown in Figure~\ref{fig:TVSDvLcomp}. The $\lambda=0.0001$ solution with OTVS yields a noticeably smaller standard deviation for the same level of RSP accuracy, a trend that has previously been encountered for different phantom data sets, leading to the choice of $\lambda=0.0001$ for reconstructions for block sizes up to $250,000$ histories. However, using NTVS with $\alpha=0.75$, the $\lambda=0.0002$ solution yields smaller standard deviations than those obtained with OTVS and $\lambda=0.0001$ for all values of $N$ and with negligible differences in standard deviation obtained with NTVS and $\lambda=0.0001$ for many values of $N$ (particularly at the often optimal $N=5$; see Figure~\ref{fig:TVSvLoptimalHead}(b)). The larger reduction in TV also obtained with $\lambda=0.0002$ leads to the conclusion that reconstruction with $\lambda=0.0002$ is now an appropriate choice made possible by introduction of the NTVS algorithm.
\section{Discussion}
In this work, we have investigated the impact of the innovative changes made to the original version of the DROP-TVS algorithm (OTVS) on both a cylindrical phantom with material inserts (simulated and experimental data sets) and an anthropomorphic head phantom closely resembling a human head (experimental data set). Whereas the changes in noise parameters (TV and standard deviation) introduced by modifications of the OTVS algorithm, leading to the NTVS algorithm are admittedly small, less than 5\% improvements for most parameter variations investigated, and for some instances no improvement was seen, we feel one can learn from these small improvements and they are expected to be proportionally larger with noisier data sets such as those introduced by very low fluences or fluence-modulated pCT methods~\cite{DDRHB17,DJPDP18}. We thus feel that it is worthwhile sharing the experiences made with the innovative algorithmic structures introduced into NTVS and the fact that a repeated TV check is actually not required or even leads to inferior results.

The reconstruction parameter space investigated in this work contained 360 parameter value combinations (2 \{TV check on/off\} $\cdot$ 5 \{alpha\} $\cdot$ 3 \{lambda\} $\cdot$ 12 \{N\}), requiring 360 separate reconstructions for each of the investigated data sets. The benefit of NTVS was dependent on the number of perturbations per feasibility-seeking iteration, $N$, with the largest benefit consistently attained for $3\le N\le 6$, typically optimal with $N=5$.  For $N\ge 7$, these benefits decreased as the number of feasibility-seeking iterations, $k$, increased except for $\alpha\ge 0.85$, but these benefits were negated by the fact that the RSP error with these $\alpha$ is affected in an unpredictable and often counterproductive direction.  This can be understood as the effect that larger $N$ have on the magnitude of perturbations $\beta_k$ as $k$ increases.  With $\ell_k$ increasing by 1 after each of the $N$ perturbations, increasing $N$ results in $\beta_k=\alpha^{\ell_k}$ decreasing more quickly as $k$ increases.  Hence, for larger $N$, meaningful perturbations persist for a smaller number of feasibility-seeking iterations unless $\alpha$ is close to 1.0, in which case perturbations can decrease too slowly and result in inappropriately large perturbations as reconstruction nears convergence.

A remarkable finding of our investigations was a fluctuating reduction of TV and standard deviation as a function of $N$ for all three data sets, which persisted for each of the 30 different parameter value combinations for each value of $N$. It was important to determine if these fluctuations were an inherent and reproducible characteristic of NTVS or if the random decrease in $\ell_k$ or some other aspect of TVS or feasibility-seeking accounted for this observation. Hence, reconstructions were performed repeatedly (8 times) for $N=5$ and the same parameter value combination from the reconstruction parameter space to determine the variation in reconstructed images between independent reconstructions. This analysis demonstrated that the bounded randomness of $\ell$ did not produce large enough variations to account for the observed fluctuations. In general terms, the fluctuations are the result of the opposing objectives and resulting effects on TV of the alternating applications of TVS and feasibility-seeking.



The inclusion of the TV reduction requirement results in the image being perturbed with perturbations of successively smaller magnitude due to the resulting increment of $\ell_k$ each time a perturbation fails to decrease the image TV. Since a failure of this requirement almost always happens during the early feasibility-seeking iterations while it is far from convergence, the resulting increase in $\ell_k$ after each failure results in all subsequent perturbations having a smaller perturbation magnitude $\beta_{k}=\alpha_k^{\ell_k}$ for each of the $N$ perturbation steps of the remaining $K-k$ iterations. By excluding the TV reduction requirement, the magnitude of subsequent perturbations is preserved throughout the remainder of reconstruction. Hence, although a perturbation applied early in feasibility-seeking may temporarily increase TV slightly, the cumulative effect of larger perturbations throughout the remainder of reconstruction usually results in a larger overall reduction in TV and, consequently, standard deviation. For the range $3\le N\le 6$, the removal of the TV reduction requirement produced at least comparable and often superior TV and standard deviation results for both the simulated and experimental data sets, particularly with $\alpha=0.75$ as seen most clearly with the experimental data sets.

Removing the TV reduction requirement also improves computation time by eliminating the conditional branch that prevents full parallelization of the superiorization algorithm and eliminating repeated perturbations until an improved TV is achieved.  There are global calculations within the TVS algorithm, such as the $\ell_2$ (discrete-space) norm used to normalize perturbation vectors, which act as a bottleneck in an explicit and direct implementation. However, such data dependencies can be eliminated by performing these calculations in each thread rather than communicating these from a central location.  Hence, there are no real data dependencies and the parallelization made possible by removing the conditional branch reduces NTVS computation time by up to 30\% (estimated based on a count of the reduced number of sequential computational operations).

An appealing aspect of NTVS is the added ability to control the perturbation kernel $\alpha$, which was previously held constant in OTVS at a value of $\alpha=0.5$.  Increasing the perturbation kernel $\alpha$ yields larger reductions in TV and standard deviations. However, it was found that as $\alpha$ increased beyond $\alpha\approx0.75$, perturbations began to affect reconstructed RSP values in an unpredictable and region-dependent manner. The direction that the RSP was driven was shown not to be an inherent property of the phantom geometry and/or composition, as demonstrated by the observation that the RSP within the Delrin insert of the \CTP phantom was driven in opposite directions for the simulated and experimental data sets, respectively. Thus, we suggest using $\alpha=0.75$ as this maximized the benefits that an increasing $\alpha$ have on TV and standard deviation while avoiding the unpredictable and potentially negative impact of larger $\alpha$ values.


Another benefit of NTVS is that it allows feasibility-seeking to be performed with a larger relaxation parameter $\lambda$ than was appropriate with OTVS ($\lambda=0.0001$, $k=12$).  It was found that with NTVS, the same RSP error can be obtained with $\lambda=0.0002$ after performing $k=6$ feasibility-seeking iterations without producing larger standard deviations, as previously experienced in practice with OTVS, which lead to the choice of $\lambda=0.0001$ in previously published work with the simulated data set~\cite{SWCSS15,SKGPSS15}.  Arriving at an acceptable solution in $k=6$ feasibility-seeking iterations also offers substantial computational benefit. As mentioned previously, feasibility-seeking increases TV at each feasibility-seeking iteration $k$. Since reconstruction with a larger $\lambda$ reaches the same point in convergence at an earlier iteration $k$ while perturbations are still larger, an image with a smaller TV can be obtained by performing fewer iterations.


In the work presented here, each of the TVS parameter values was held fixed throughout a reconstruction.  One possible direction to explore in future work is investigating how parameter values can be varied \textit{during} reconstruction to produce greater benefits with NTVS.

Another interesting question to explore is if the diminishing benefits for $N\ge 7$ are due to an excessive use of TVS per feasibility-seeking iteration or if this is simply a consequence of $\beta_k$ decreasing too quickly as a function of $k$, perhaps resulting in an under-utilization of TVS at larger values of $k$.  Note that the value of the perturbation kernel $\alpha$ determines not only the initial perturbation magnitude $\beta_k(k=1)$, but also the rate at which $\beta_k$ decreases after each perturbation.

One would like the ability to control the initial perturbation magnitude $\beta_k(N=1)$ as a function of $k$ while independently determining the rate at which $\beta_k$ decreases between each of the $N$ perturbations per feasibility-seeking iteration.  This is not possible with $\beta_k=\alpha^\ell$ since the value of $\ell$ implicitly depends on both $n$ and $k$. Hence, an interesting direction to explore is the introduction of another parameter $\gamma$ that independently controls the rate at which perturbation magnitude decreases as a function of $k$; the parameter $\alpha$ would then control only the rate at which $\beta_{k,n}$ decreases between each of the $n=1,2,\ldots,N$ perturbation steps. This makes $\beta_{k,n}$ an explicit rather than an implicit function of $n$ and $k$, eliminating the need to randomly increase $\ell$ between feasibility-seeking iterations to prevent perturbation magnitude decreasing too quickly as a function of $k$. By reformulating the perturbation magnitude as $\beta_{k,n}=\alpha^n\gamma^{f(k)}$, with $0< \alpha,\gamma< 1$ and $f(k)$ chosen such that $\lim_{k\to\infty}f(k)=\infty$ (e.g., $f(k)=k$), the rate at which $\beta_{k,n}$ decreases as a function of $n$ and $k$ can then be controlled independently while preserving the superiorization requirement that $\lim_{k\to\infty}\beta_{k,n}=0$.
\section{Conclusions}
The investigations performed in this work demonstrate that the modifications implemented by the NTVS algorithm provide clear advantages over the OTVS algorithm in terms of both quality and computational cost.  Future work should include investigating whether varying parameters during reconstruction or controlling the decrease of the perturbation magnitude independently during iterations and repeated perturbation steps can further increase the advantages of the NTVS algorithm.
\appendices
\section{Definition of Terms}\label{app:defterms}
The list below defines the terms and mathematical notation used in describing the OTVS and NTVS algorithms:
\begin{itemize}
\item $k$ : overall cycle \#, i.e., $k$-th iteration of feasibility-seeking and TV perturbations.
\item $K$ : total \# of cycles, i.e., total \# of iterations of feasibility-seeking and TV perturbations.
\item $n$ : TV perturbation step \#, $1\le n\le N$.
\item $x^k$ : image vector $x$ at cycle $k$.
\item $\bar{x}$ : initial iterate $x^0$ of image reconstruction.
\item $N$ : \# of TV perturbation steps per feasibility-seeking iteration.
\item $\alpha$ : perturbation kernel, $0< \alpha< 1$.
\item $\ell_{k}$ : perturbation kernel exponent.
\item $\beta_{k,n}$: perturbation coefficient $\beta_{k,n}=\alpha^{\ell_{k}}$ at TV perturbation step $n$ and feasibility-seeking iteration $k$.
\item $\phi$ : the target function to which superiorization is applied; here, $\phi=\textrm{TV}$, the total variation of the image vector.
\item $\phi(x^{k,n})$ :  TV of image vector $x^{k,n}$ at TV perturbation step $n$ and feasibility-seeking iteration $k$.
\item $v^{k,n}$ :  normalized non-ascending perturbation vector for $\phi$ at $x^{k,n}$, i.e., $$v^{k,n}=-\frac{\nabla\phi(x^{k,n})}{\left\vert\left\vert\nabla\phi(x^{k,n})\right\vert\right\vert}=\phi'(x^{k,n})$$.
\item $P_T$ : projection operator representative of an iterative feasibility-seeking algorithm.
\end{itemize}
\clearpage
\section{OTVS Algorithm}\label{app:otvs}
The pseudocode definition of the OTVS algorithm is written as follows:
    \begin{algorithmic}[1]
        \State set $k = 0$
        \State set $\ell = 0$
        \State set $\beta =1$
        \State set $x^k = \bar{x}$
        \While{$k<K$}
            \State set $v^{k}=\phi'(x^{k})$
            \State set $loop = true$
            \While{\textit{loop}}
                \State set $z = x^{k} + \beta v^{k}$
                \If{$\phi(z) \le \phi(x^k)$}\label{alg:TVcheck}
                    \State set $x^{k} = z$
                    \State set $loop = false$
                \EndIf
                \State set $\ell = \ell + 1$
                \State set $\beta = (\frac{1}{2})^{\ell}$ \qquad(originally $\beta \leftarrow\beta /2$)
            \EndWhile
        \State set $x^{k+1} = P_T(x^{k})$
        \State set $k = k + 1$
        \EndWhile
    \end{algorithmic}\label{alg:OTVS}
{%
\section{NTVS Algorithm}\label{app:ntvs}
A pseudocode definition of the NTVS algorithm is written as follows:
    \begin{algorithmic}[1]
    \State set $k = 0$
    \State set $\ell_{-1} = 0$
    \State set $x^k = \bar{x}$
    \While{$k<K$}
    \State set $n = 0$
    \State set $\ell_{k} = \textrm{rand}(k,\ell_{k-1})$\label{alg:lk}
    \State set $x^{k,n} = x^k$
    \While{$\boldsymbol{n < N}$}\label{alg:N}
            \State set $v^{k,n}=\phi'(x^{k,n})$
            \State set $\beta_{k,n} = \alpha^{\ell_{k}}$\label{alg:beta}
            \State set $x^{k,n+1} = x^{k,n} + \beta_{k,n}v^{k,n}$
            \State set $n = n + 1$
            \State set $\ell_{k} = \ell_{k} + 1$
        \EndWhile
        \State set $x^{k+1} = P_T(x^{k,N})$
        \State set $k = k + 1$
    \EndWhile
    \end{algorithmic}\label{alg:NTVS}
}
\newpage%
{%
\vspace{1cm}
*\textbf{FOR REFERENCE ONLY}: A pseudocode definition of the NTVS algorithm with the TV reduction requirement included:
    \begin{algorithmic}[1]
    \State set $k = 0$
    \State set $\ell_{-1} = 0$
    \State set $x^k = \bar{x}$
    \While{$k<K$}
    \State set $n = 0$
    \State set $\ell_{k} = \textrm{rand}(k,\ell_{k-1})$\label{alg:lk}
    \State set $x^{k,n} = x^k$
    \While{$\boldsymbol{n < N}$}\label{alg:N}
            \State set $v^{k,n}=\phi'(x^{k,n})$
            \State set $\beta_{k,n} = \alpha^{\ell_{k}}$\label{alg:beta}
            \State set $loop = true$
            \While{\textit{loop}}
                \State set $z^{k,n} = x^{k,n} + \beta_{k,n}v^{k,n}$
                \If{$\phi(z^{k,n}) \le \phi(x^{k,n})$}\label{alg:TVcheckvar}
                    \State set $x^{k,n} = z^{k,n}$
                    \State set $loop = false$
                \EndIf
                \State set $\ell_{k} = \ell_{k} + 1$
            \EndWhile
            \State set $n = n + 1$
        \EndWhile
        \State set $x^{k+1} = P_T(x^{k,N})$
        \State set $k = k + 1$
    \EndWhile
    \end{algorithmic}\label{alg:NTVSvar}
}
\section*{Acknowledgment}
We greatly appreciate the constructive comments of three anonymous reviewers which helped us to significantly improve this paper. The research in proton CT was supported by the National Institute of Biomedical Imaging and Bioengineering (NIBIB) of the National Institute of Health (NIH) and the National Science Foundation (NSF) award number R01EB013118, and the United States - Israel Binational Science Foundation (BSF) grant no. 2009012, and is currently supported by BSF grant no. 2013003. The content of this paper is solely the responsibility of the authors and does not necessarily represent the official views of NBIB or NIH.  The support of UT Southwestern and State of Texas through a Seed Grants in Particle Therapy award is gratefully acknowledged.
\ifCLASSOPTIONcaptionsoff
  \newpage
\fi
\bibliography{pct}

\begin{thebibliography}{10}
\providecommand{\url}[1]{#1}
\csname url@samestyle\endcsname
\providecommand{\newblock}{\relax}
\providecommand{\bibinfo}[2]{#2}
\providecommand{\BIBentrySTDinterwordspacing}{\spaceskip=0pt\relax}
\providecommand{\BIBentryALTinterwordstretchfactor}{4}
\providecommand{\BIBentryALTinterwordspacing}{\spaceskip=\fontdimen2\font plus
\BIBentryALTinterwordstretchfactor\fontdimen3\font minus
  \fontdimen4\font\relax}
\providecommand{\BIBforeignlanguage}[2]{{%
\expandafter\ifx\csname l@#1\endcsname\relax
\typeout{** WARNING: IEEEtran.bst: No hyphenation pattern has been}%
\typeout{** loaded for the language `#1'. Using the pattern for}%
\typeout{** the default language instead.}%
\else
\language=\csname l@#1\endcsname
\fi
#2}}
\providecommand{\BIBdecl}{\relax}
\BIBdecl

\bibitem{Cormack76}
A.~Cormack and A.~Koehler, ``Quantitative proton tomography: preliminary
  experiments,'' \emph{Physics in Medicine \& Biology}, vol.~21, no.~4, pp.
  560--569, 1976.

\bibitem{Hanson78}
K.~M. Hanson, J.~N. Bradbury, T.~M. Cannon, R.~L. Hutson, D.~B. Laubacher,
  R.~Macek, M.~A. Paciotti, and C.~A. Taylor, ``The application of protons to
  computed tomography,'' \emph{IEEE Transactions on Nuclear Science}, vol.~25,
  no.~1, pp. 657--660, Feb 1978.

\bibitem{Hanson79}
K.~M. Hanson, ``Proton computed tomography,'' \emph{IEEE Transactions on
  Nuclear Science}, vol.~26, no.~1, pp. 1635--1640, Feb 1979.

\bibitem{Hanson81}
K.~M. Hanson, J.~N. Bradbury, T.~M. Cannon, R.~L. Hutson, D.~B. Laubacher,
  R.~J. Macek, M.~A. Paciotti, and C.~A. Taylor, ``Computed tomography using
  proton energy loss,'' \emph{Physics in Medicine \& Biology}, vol.~26, no.~6,
  pp. 965--983, 1981.

\bibitem{JBCGK16}
\BIBentryALTinterwordspacing
R.~P. Johnson, V.~A. Bashkirov, G.~Coutrakon, V.~Giacometti, P.~Karbasi, N.~T.
  Karonis, C.~Ordo{\~n}ez, M.~Pankuch, H.~F.-W. Sadrozinski, K.~E. Schubert,
  and R.~W. Schulte, ``Results from a prototype proton-{CT} head scanner,''
  \emph{Conference on the Application of Accelerators in Research and Industry,
  CAARI 2016, 30 October - 4 November 2016, Ft. Worth, TX, USA}, Jul 2017.
  [Online]. Available: \url{https://arxiv.org/pdf/1707.01580}
\BIBentrySTDinterwordspacing

\bibitem{BJSS16}
V.~A. Bashkirov, R.~P. Johnson, H.~F.-W. Sadrozinski, and R.~W. Schulte,
  ``Development of proton computed tomography detectors for applications in
  hadron therapy,'' \emph{Nuclear Instruments \& Methods in Physics Research
  Section A: Accelerators, Spectrometers, Detectors and Associated Equipment},
  vol. 809, pp. 120--129, 2016.

\bibitem{BCSBB17}
\BIBentryALTinterwordspacing
M.~Bruzzi, C.~Civinini, M.~Scaringella, D.~Bonanno, M.~Brianzi, M.~Carpinelli,
  G.~Cirrone, G.~Cuttone, D.~L. Presti, G.~Maccioni, S.~Pallotta, N.~Randazzo,
  F.~Romano, V.~Sipala, C.~Talamonti, and E.~Vanzi, ``Proton computed
  tomography images with algebraic reconstruction,'' \emph{Nuclear Instruments
  \& Methods in Physics Research Section A: Accelerators, Spectrometers,
  Detectors and Associated Equipment}, vol. 845, pp. 652--655, May 2017,
  {Proceedings of the Vienna Conference on Instrumentation 2016}. [Online].
  Available:
  \url{http://www.sciencedirect.com/science/article/pii/S0168900216304454}
\BIBentrySTDinterwordspacing

\bibitem{Johnson18}
\BIBentryALTinterwordspacing
R.~P. Johnson, ``Review of medical radiography and tomography with proton
  beams,'' \emph{Reports on Progress in Physics}, vol.~81, no.~1, p. 016701,
  2018. [Online]. Available:
  \url{http://stacks.iop.org/0034-4885/81/i=1/a=016701}
\BIBentrySTDinterwordspacing

\bibitem{Paganetti12}
H.~Paganetti, ``Range uncertainties in proton therapy and the role of {M}onte
  {C}arlo simulations,'' \emph{Physics in Medicine \& Biology}, vol.~57,
  no.~11, pp. R99–--R117, May 2012.

\bibitem{SPTS08}
R.~W. Schulte, S.~N. Penfold, J.~Tafas, and K.~E. Schubert, ``A maximum
  likelihood proton path formalism for application in proton computed
  tomography,'' \emph{Medical Physics}, vol.~35, pp. 4849--4856, Nov 2008.

\bibitem{FDDBS15}
C.-A.~C. Fekete, P.~Doolan, M.~F. Dias, L.~Beaulieu, and J.~Seco, ``Developing
  a phenomenological model of the proton trajectory within a heterogeneous
  medium required for proton imaging,'' \emph{Physics in Medicine \& Biology},
  vol.~60, no.~13, pp. 5071--5082, 2015.

\bibitem{PSCBM10}
S.~N. Penfold, R.~W. Schulte, Y.~Censor, V.~A. Bashkirov, S.~A. McAllister,
  K.~E. Schubert, and A.~B. Rosenfeld, ``Block-iterative and string-averaging
  projection algorithms in proton computed tomography image reconstruction,''
  in \emph{Biomedical Mathematics: Promising Directions in Imaging, Therapy
  Planning and Inverse Problems}, Y.~Censor, M.~Jiang, and G.~Wang, Eds., The
  Huangguoshu International Interdisciplinary Conference.\hskip 1em plus 0.5em
  minus 0.4em\relax Medical Physics, 2010, pp. 347--367.

\bibitem{HGDC12}
\BIBentryALTinterwordspacing
G.~T. Herman, E.~Gardu{\~n}o, R.~Davidi, and Y.~Censor, ``Superiorization: An
  optimization heuristic for medical physics,'' \emph{Medical Physics},
  vol.~39, no.~9, pp. 5532--5546, 2012. [Online]. Available:
  \url{http://dx.doi.org/10.1118/1.4745566}
\BIBentrySTDinterwordspacing

\bibitem{PSCR10}
S.~N. Penfold, R.~W. Schulte, Y.~Censor, and A.~B. Rosenfeld, ``Total variation
  superiorization schemes in proton computed tomography image reconstruction,''
  \emph{Medical Physics}, vol.~37, pp. 5887--5895, 2010.

\bibitem{CDH10}
Y.~Censor, R.~Davidi, and G.~T. Herman, ``Perturbation resilience and
  superiorization of iterative algorithms,'' \emph{Inverse problems}, vol.~26,
  p. 65008, Jun 2010.

\bibitem{Censor14}
Y.~Censor, ``Weak and strong superiorization: Between feasibility-seeking and
  minimization,'' \emph{Analele Stiintifice ale Universitatii Ovidius
  Constanta, Seria Matematica}, vol.~23, pp. 41--54, Oct 2014.

\bibitem{CDHST14}
\BIBentryALTinterwordspacing
Y.~Censor, R.~Davidi, G.~T. Herman, R.~W. Schulte, and L.~Tetruashvili,
  ``Projected subgradient minimization versus superiorization,'' \emph{Journal
  of Optimization Theory and Applications}, vol. 160, no.~3, pp. 730--747, Mar
  2014. [Online]. Available: \url{https://doi.org/10.1007/s10957-013-0408-3}
\BIBentrySTDinterwordspacing

\bibitem{website:Censor}
Y.~Censor, ``Superiorization and perturbation resilience of algorithms: A
  bibliography compiled and continuously updated,''
  http://math.haifa.ac.il/yair/bib-superiorization-censor.html.

\bibitem{CHJ17}
\BIBentryALTinterwordspacing
Y.~Censor, G.~T. Herman, and M.~Jiang, ``Superiorization: Theory and
  applications,'' \emph{{\textup{Special Issue of}} Inverse Problems}, vol.~33,
  no.~4, p. 040301, 2017. [Online]. Available:
  \url{http://stacks.iop.org/0266-5611/33/i=4/a=040301}
\BIBentrySTDinterwordspacing

\bibitem{HWF17}
\BIBentryALTinterwordspacing
T.~Humphries, J.~Winn, and A.~Faridani, ``Superiorized algorithm for
  reconstruction of {CT} images from sparse-view and limited-angle
  polyenergetic data,'' \emph{Physics in Medicine \& Biology}, vol.~62, no.~16,
  pp. 6762--6783, 2017. [Online]. Available:
  \url{http://stacks.iop.org/0031-9155/62/i=16/a=6762}
\BIBentrySTDinterwordspacing

\bibitem{HZM17}
E.~Helou, M.~Zibetti, and E.~Miqueles, ``Superiorization of incremental
  optimization algorithms for statistical tomographic image reconstruction,''
  \emph{Inverse Problems}, vol.~33, no.~4, p. 044010, 2017.

\bibitem{GH17}
\BIBentryALTinterwordspacing
E.~Gardu{\~n}o and G.~T. Herman, ``Computerized tomography with total variation
  and with shearlets,'' \emph{Inverse Problems}, vol.~33, no.~4, p. 044011,
  2017. [Online]. Available:
  \url{http://stacks.iop.org/0266-5611/33/i=4/a=044011}
\BIBentrySTDinterwordspacing

\bibitem{YCW17}
\BIBentryALTinterwordspacing
Q.~Yang, W.~Cong, and G.~Wang, ``Superiorization-based multi-energy {CT} image
  reconstruction,'' \emph{Inverse Problems}, vol.~33, no.~4, p. 044014, 2017.
  [Online]. Available: \url{http://stacks.iop.org/0266-5611/33/i=4/a=044014}
\BIBentrySTDinterwordspacing

\bibitem{PC15}
S.~N. Penfold and Y.~Censor, ``Techniques in iterative proton {CT} image
  reconstruction,'' \emph{Sensing and Imaging}, vol.~16, no.~1, Oct 2015.

\bibitem{BR67}
R.~Bracewell and A.~Riddle, ``Inversion of fan beam scawns in radio
  astronomy,'' \emph{Astrophysics Journal}, vol. 150, pp. 427--434, 1967.

\bibitem{RL71}
G.~Ramanchandran and A.~Lakshminarayanan, ``Three dimensional reconstructions
  from radiographs and electron micrographs: Application of convolution instead
  of {F}ourier transforms,'' \emph{Proceedings of the National Academy of
  Sciences, USA}, vol.~68, pp. 2236--2240, 1971.

\bibitem{CCCNP10}
V.~Caselles, A.~Chambolle, D.~Cremers, M.~Novaga, and T.~Pock, ``An
  introduction to total variation for image analysis,'' in \emph{Theoretical
  Foundations and Numerical Methods for Sparse Recovery, De Gruyter}, 2010.

\bibitem{Censor17}
\BIBentryALTinterwordspacing
Y.~Censor, ``Can linear superiorization be useful for linear optimization
  problems?'' \emph{Inverse Problems}, vol.~33, no.~4, p. 044006, 2017.
  [Online]. Available:
  \url{http://iopscience.iop.org/article/10.1088/1361-6420/33/4/044006}
\BIBentrySTDinterwordspacing

\bibitem{HD08}
\BIBentryALTinterwordspacing
G.~T. Herman and R.~Davidi, ``Image reconstruction from a small number of
  projections,'' \emph{Inverse Problems}, vol.~24, no.~4, p. 045011, 2008.
  [Online]. Available: \url{http://stacks.iop.org/0266-5611/24/i=4/a=045011}
\BIBentrySTDinterwordspacing

\bibitem{DHC09}
\BIBentryALTinterwordspacing
R.~Davidi, G.~T. Herman, and Y.~Censor, ``Perturbation-resilient
  block-iterative projection methods with application to image reconstruction
  from projections,'' \emph{International Transactions in Operational
  Research}, vol.~16, no.~4, pp. 505--524, 2009. [Online]. Available:
  \url{https://onlinelibrary.wiley.com/doi/abs/10.1111/j.1475-3995.2009.00695.x}
\BIBentrySTDinterwordspacing

\bibitem{BDHK07}
D.~Butnariu, R.~Davidi, G.~T. Herman, and I.~G. Kazantsev, ``Stable convergence
  behavior under summable perturbations of a class of projection methods for
  convex feasibility and optimization problems,'' \emph{IEEE Journal of
  Selected Topics in Signal Processing}, vol.~1, no.~4, pp. 540--547, Dec 2007.

\bibitem{Langthaler14}
O.~Langthaler, ``Incorporation of the superiorization methodology into
  biomedical imaging software,'' Salzburg University of Applied Sciences,
  Salzburg, Austria, and the Graduate Center of the City University of New
  York, NY, USA, {Marshall Plan Scholarship Report}, September 2014, 76 pages.

\bibitem{Prommegger14}
B.~Prommegger, ``Verification and evaluation of superiorized algorithms used
  in. biomedical imaging: Comparison of iterative algorithms with and without
  superiorization for image reconstruction from projections,'' Salzburg
  University of Applied Sciences, Salzburg, Austria, and the Graduate Center of
  the City University of New York, NY, USA, {Marshall Plan Scholarship Report},
  October 2014, 84 pages.

\bibitem{Havas16}
C.~Havas, ``Revised implementation and empirical study of maximum likelihood
  expectation maximization algorithms with and without superiorization in image
  reconstruction,'' Salzburg University of Applied Sciences, Salzburg, Austria,
  and the Graduate Center of the City University of New York, NY, USA,
  {Marshall Plan Scholarship Report}, October 2016, 49 pages.

\bibitem{geant4}
\BIBentryALTinterwordspacing
S.~Agostinelli, J.~Allison, K.~Amako \emph{et~al.}, ``Geant4 - a simulation
  toolkit,'' \emph{Nuclear Instruments \& Methods in Physics Research Section
  A: Accelerators, Spectrometers, Detectors and Associated Equipment}, vol.
  506, no.~3, pp. 250--303, 2003. [Online]. Available:
  \url{http://www.sciencedirect.com/science/article/pii/S0168900203013688}
\BIBentrySTDinterwordspacing

\bibitem{SKGPSS15}
B.~E. Schultze, P.~Karbasi, V.~Giacometti, T.~E. Plautz, K.~E. Schubert, and
  R.~W. Schulte, ``Reconstructing highly accurate relative stopping powers in
  proton computed tomography,'' in \emph{Proceedings of the IEEE Nuclear
  Science Symposium \& Medical Imaging Conference (NSS/MIC) 2015}, Oct 2015,
  pp. 1--3.

\bibitem{GBPGP17}
\BIBentryALTinterwordspacing
V.~Giacometti, V.~A. Bashkirov, P.~Piersimoni, S.~Guatelli, T.~E. Plautz,
  H.~F.-W. Sadrozinski, R.~P. Johnson, A.~Zatserklyaniy, T.~Tessonnier,
  K.~Parodi, A.~B. Rosenfeld, and R.~W. Schulte, ``Software platform for
  simulation of a prototype proton {CT} scanner,'' \emph{Medical Physics},
  vol.~44, no.~3, pp. 1002--1016, 2017. [Online]. Available:
  \url{http://dx.doi.org/10.1002/mp.12107}
\BIBentrySTDinterwordspacing

\bibitem{CEHN08}
\BIBentryALTinterwordspacing
Y.~Censor, T.~Elfving, G.~T. Herman, and T.~Nikazad, ``On diagonally relaxed
  orthogonal projection methods,'' \emph{SIAM Journal on Scientific Computing},
  vol.~30, no.~1, pp. 473--504, 2008. [Online]. Available:
  \url{https://doi.org/10.1137/050639399}
\BIBentrySTDinterwordspacing

\bibitem{imagej2}
\BIBentryALTinterwordspacing
C.~T. Rueden, J.~Schindelin, M.~C. Hiner, B.~E. DeZonia, A.~E. Walter, E.~T.
  Arena, and K.~W. Eliceiri, ``Imagej2: Imagej for the next generation of
  scientific image data,'' \emph{BMC Bioinformatics}, vol.~18, no.~1, p. 529,
  Nov 2017. [Online]. Available:
  \url{https://doi.org/10.1186/s12859-017-1934-z}
\BIBentrySTDinterwordspacing

\bibitem{DDRHB17}
\BIBentryALTinterwordspacing
G.~Dedes, L.~D. Angelis, S.~Rit, D.~Hansen, C.~Belka, V.~A. Bashkirov, R.~P.
  Johnson, G.~Coutrakon, K.~E. Schubert, R.~W. Schulte, K.~Parodi, and
  G.~Landry, ``Application of fluence field modulation to proton computed
  tomography for proton therapy imaging,'' \emph{Physics in Medicine \&
  Biology}, vol.~62, no.~15, pp. 6026--6043, 2017. [Online]. Available:
  \url{http://stacks.iop.org/0031-9155/62/i=15/a=6026}
\BIBentrySTDinterwordspacing

\bibitem{DJPDP18}
G.~Dedes, R.~P. Johnson, M.~Pankuch, N.~Detrich, W.~M.~A. Pols, S.~Rit, R.~W.
  Schulte, K.~Parodi, and G.~Landry, ``Experimental fluence modulated proton
  computed tomography by pencil beam scanning,'' \emph{Medical Physics},
  vol.~45, pp. 3287--3296, May 2018.

\bibitem{SWCSS15}
B.~E. Schultze, M.~Witt, Y.~Censor, K.~E. Schubert, and R.~W. Schulte,
  ``Performance of hull-detection algorithms for proton computed tomography
  reconstruction,'' in \emph{Infinite Products of Operators and Their
  Applications}, ser. Contemporary Mathematics, S.~Reich and A.~Zaslavski,
  Eds., vol. 636.\hskip 1em plus 0.5em minus 0.4em\relax American Mathematical
  Society, 2015, pp. 211--224.

\end{thebibliography}
\end{document}